\documentclass[11pt,a4paper]{article}
\usepackage{amsmath, amsfonts,amssymb,mathtools,nicefrac,nccmath,cases}
\usepackage{booktabs,multirow}

\usepackage{microtype}

\usepackage[usenames,dvipsnames,table]{xcolor}
\usepackage{colortbl}
\usepackage{tikz}
\usepackage{tikz-cd}
\usepackage{tkz-euclide}
\usetikzlibrary{shapes,arrows}
\usetikzlibrary{calc,shapes.callouts,shapes.arrows,bending,intersections}
\usetikzlibrary{shapes.geometric}
\usetikzlibrary{arrows.meta,arrows}
\usetikzlibrary{decorations.pathmorphing}
\usetikzlibrary{decorations.markings}
\usetikzlibrary{shapes.multipart}
\usetikzlibrary{tikzmark}
\usepackage{caption}
\usepackage{subcaption}
\usepackage{bm}
\usepackage[nosort]{cite}
\usepackage{colortbl}
\usepackage{amsthm}
\usepackage{soul}
\usepackage{dsfont}

\usepackage{multirow}
\usepackage{hhline}
\usepackage[shortlabels]{enumitem}

\usepackage{xcolor}
\usepackage[linktocpage=true]{hyperref}
\definecolor{nicecolor}{rgb}{0.1, 0.3, 0.4}
\hypersetup{colorlinks=true,linkcolor=nicecolor,citecolor=nicecolor,urlcolor=nicecolor}

\def\hybrid{\topmargin -20pt    \oddsidemargin 0pt
	\headheight 0pt \headsep 0pt
	\textwidth 6.5in        
	\textheight 9in         
	\textwidth 6.25in       
	\textheight 9 in       
	\marginparwidth .875in
	\parskip 5pt plus 1pt 
	\jot = 1.5ex
}
\hybrid
\numberwithin{equation}{section}
\numberwithin{table}{section}
\usepackage{physics}

\usepackage{float}
\usepackage{tabularx}
\newcolumntype{D}{>{\centering\arraybackslash}X}
\newcolumntype{L}{>{$}l<{$}}
\newcolumntype{R}{>{$}r<{$}}
\newcolumntype{C}{>{$}c<{$}}
\usepackage{IEEEtrantools}
\usepackage{makecell}

\newcommand{\beq}{\begin{equation}}  \newcommand{\eeq}{\end{equation}}
\newcommand{\bal}{\begin{aligned}}   \newcommand{\eal}{\end{aligned}}
\newcommand{\bea}{\begin{eqnarray}}  \newcommand{\eea}{\end{eqnarray}}
\def\beqa{\begin{eqnarray}}
	\def\eeqa{\end{eqnarray}}

\newcommand{\bmat}{\left(\begin{array}}
	\newcommand{\emat}{\end{array}\right)}

\newcommand{\cT}{\mathcal{T}}

\newcommand{\cN}{\mathcal{N}}

\newcommand{\cH}{\mathcal{H}}

\newcommand{\I}{\text{Im}\,}
\newcommand{\R}{\text{Re}\,}

\usepackage{cancel}

\newcommand{\be}{\begin{equation}}
	\newcommand{\ee}{\end{equation}}
\newcommand{\mbb}{\mathbb}

\newcommand{\mc}{\mathcal}

\newcommand{\im}{\mathbf{i}}

\definecolor{Gray}{gray}{0.95}

\begin{document}


	\baselineskip=14pt
	\parskip 5pt plus 1pt

	\vspace*{-1.5cm}
	
	\vspace*{4cm}
	\begin{center}        
		
		{\huge Global symmetry-breaking and generalized \\
		 theta-terms in Type IIB EFTs\\
			[.3cm]  }
	\end{center}
	
	\vspace{0.5cm}
	\begin{center}        
		{\large  Thomas W.~Grimm, Stefano Lanza, Thomas van Vuren}
	\end{center}
	
	\vspace{0.15cm}
	\begin{center}  
		\emph{Institute for Theoretical Physics,
			Utrecht University}\\
		\emph{Princetonplein 5, 3584 CE Utrecht, The Netherlands}
		\\[.3cm]
	\end{center}
	
	\vspace{2cm}
	
	
	\begin{abstract}

	\noindent A longstanding conjecture states that global symmetries should be absent in quantum gravity. By investigating large classes of Type IIB four-dimensional $\mathcal{N}=2$ effective field theories, we enlist the potential generalized global symmetries that could be present and explore how they are avoided. Crucial ingredients that arise in such effective field theories are generalized $\theta$-terms. These introduce non-linear couplings between axion fields and topological terms quadratic in the gauge field strengths which break a large subset of the global symmetries. Additional residual global symmetries may further be broken by assuming the existence of some charged states. However, we illustrate that the presence of generalized $\theta$-terms leads to a generalized Witten effect, which implies that the spectrum of charged states is constituted by an infinitely populated lattice. We further show that such a lattice is generated by the action of the monodromy transformation that characterizes the moduli space boundary near which the effective theory is defined. 
	\end{abstract}

	\thispagestyle{empty}
	\clearpage
	
	\setcounter{page}{1}
	
	
	\newpage

	\tableofcontents


\section{Introduction}
\label{sec:intro}

In recent years, our understanding of global symmetries has improved greatly following \cite{Gaiotto:2014kfa}, where the notion of \emph{generalized global symmetries} was introduced. 
Such global symmetries play a critical role in determining the consistency of field theories that originate from fully-fledged quantum gravity theories. In fact, quantum gravity is expected to be devoid of any global symmetries \cite{Hawking:1975vcx,Zeldovich:1976vq,Zeldovich:1977be,Banks:1988yz,Banks:2010zn}, as asserted by the \emph{No Global Symmetry Conjecture}. The latter is a foundational part of the \emph{Swampland program} which seeks to identify the criteria that any effective field theory (EFT) ought to satisfy in order to be consistent with a quantum gravity UV completion (see, for instance, \cite{Palti:2019pca,vanBeest:2021lhn} for reviews on the subject), and many of the subsequently formulated conjectures rely upon it.
The conjecture can be phrased differently as stating that, within any EFT that is valid below an energy scale $\Lambda_{\text{\tiny EFT}}$ and which admits a UV completion within quantum gravity, global symmetries are at most approximate or emergent, meaning that they are removed by mechanisms that occur at energies above $\Lambda_{\text{\tiny EFT}}$. 
For instance, in \cite{Heidenreich:2020pkc} it was shown that all generalized global symmetries of the ten-dimensional string theory EFTs, or the eleven-dimensional M-theory can be removed, either because the eventual global symmetries are gauged, or because they are broken by brane states.
However, given a lower-dimensional EFT that ought to be consistent with a quantum gravity completion, it is highly nontrivial to generically show that every generalized global symmetry is removed by invoking conventional string theory ingredients.

The aim of this work is to investigate how global symmetries are avoided in a large family of concrete string theory EFTs.
In particular, we will examine the four-dimensional EFTs that stem from compactifications of Type IIB string theory over Calabi-Yau three-folds. Within the resulting $\mathcal{N}=2$ supergravity EFTs, our attention will be focused on the generalized global symmetries associated to the bosonic fields of the vector multiplets, namely the complex scalar fields, acting as moduli fields which describe the complex structure deformations of the internal Calabi-Yau, and the supersymmetrically paired one-form gauge fields. 
Although the vector multiplets only constitute a subsector of the fields populating the more general $\mathcal{N}=2$ Type IIB EFTs, the effective description capturing their interactions is already quite rich. Indeed, the vector multiplet sector could contain some emergent generalized global symmetries, which ought to be removed in the UV-complete theory.

We will examine such Type IIB EFTs in the near-boundary regions of the moduli space. 
Remarkably, such parts of the moduli space can be investigated in detail by means of asymptotic Hodge theory \cite{MR840721}.
The latter has proven to be a powerful tool for studying the near-boundary physics in full generality. For instance, in \cite{Grimm:2018ohb,Grimm:2018cpv,Corvilain:2018lgw,Grimm:2019wtx,Grimm:2019bey} Hodge theory has been employed to show how the Distance Conjecture is realized in generic, concrete string theory EFTs, or in \cite{Grimm:2019ixq,Grimm:2020cda,Grimm:2021ckh,Grana:2022dfw} to infer general features of the scalar potential and the structure of stringy vacua, or to analyze the properties of BPS objects \cite{Font:2019cxq,Gendler:2020dfp,Lanza:2020qmt,Marchesano:2022axe}.
In this work, we will illustrate how Hodge theory can be used in order to generically study the behavior of the interactions of the gauge fields in terms of the complex structure moduli fields. In particular, in Section~\ref{sec:GGS_01mod} we will perform a detailed computation of the near-boundary effective action describing the interaction of the vector multiplets in EFTs containing one dynamical complex structure modulus.
For each of these models, we will enlist the possible global symmetries and present how these can be removed using stringy ingredients. 
These findings will be generalized in Section~\ref{sec:GGS_gen_mod} to include generic multi-moduli cases.

As we will show, a common feature of the near-boundary effective descriptions of the Type IIB EFTs is the presence of several `\emph{generalized $\theta$-terms}'. 
These terms correspond to couplings between topological terms that are quadratic in the field strengths of the gauge fields and the axions of the theory. 
We will refer to them as `\emph{generalized}' because, unlike the more standard `$\theta$-terms', of the form $\theta F \wedge F$, introduced in  \cite{Peccei:1977ur,Peccei:1977hh} which are linear in the so-called \emph{Peccei-Quinn axion} $\theta$, the `$\theta$-terms' encountered in Type IIB EFTs may be non-linear in the axionic fields of the model. 
These couplings have profound phenomenological consequences. 
In fact, as is well known in the literature, the standard $\theta$-terms are responsible for the Witten effect \cite{Witten:1979ey}: namely, the $\theta$-term guarantees that, if a magnetic monopole is present in the spectrum, then so too is an infinite tower of dyons. In this work, we will illustrate that a similar mechanism is realized in Type IIB EFTs with generalized $\theta$-terms. 
Indeed, we will show that the generalized $\theta$-terms break some of the potential generalized global symmetries, while also leading to a realization of a generalized version of the Witten effect.

The work is articulated as follows. 
In Section~\ref{sec:GGS_QG} we introduce the notion of generalized global symmetry and review some of the mechanisms that quantum gravity is expected to be endowed with that prevent the presence of global symmetries. 
We further show how the presence of $\theta$-terms implies that the breaking of global symmetries may lead to an infinitely populated charge lattice.
In Section~\ref{sec:GGS_IIB} we examine the potential generalized global symmetries that the vector multiplet sector of four-dimensional $\mathcal{N}=2$ Type EFTs may display. Therein, we additionally illustrate how a generalized version of the Witten effect is realized in such EFTs, and how this effect influences the definition of gauge-invariant conserved currents that could deliver generalized global symmetries. 
In Section~\ref{sec:GGS_01mod} we explicitly compute the gauge interactions of Type IIB EFTs with no moduli or in the near-boundary region of EFTs with one complex structure modulus. For each of the models therein presented we display the potential emerging generalized global symmetries, and how these can be broken. 
In Section~\ref{sec:GGS_gen_mod} we will then show how the Hodge theory techniques can be employed to count the global symmetries that can emerge towards any arbitrary field space boundary.

Furthermore, the Appendices contain several additional technical details that have been used throughout the main text.
In particular, Appendix~\ref{app:conv_dual} collects some conventions on gauge fields used throughout the work.
Appendix~\ref{sec:Hodge} is devoted to the introduction of asymptotic Hodge theory, the technology that is foundational for Sections~\ref{sec:GGS_01mod} and~\ref{sec:GGS_gen_mod}. 
Therein, we will show how the latter offers the tools necessary to obtain the interactions of the bosonic components of the $\mathcal{N}=2$ vector multiplets in the near-boundary region of any Calabi-Yau singularities in full generality.
Finally, in Appendix~\ref{app:Periods_One_Mod} we illustrate how to compute the periods and the gauge interactions towards generic boundaries in EFTs with a single complex structure modulus by means of the asymptotic Hodge theory.


\section{Global symmetries and the Witten effect}
\label{sec:GGS_QG}

Recently, our knowledge about how global symmetries may be realized has been greatly expanded.
Indeed, the recent \cite{Gaiotto:2014kfa} has shown that the notion of global symmetry is wider than was previously thought, leading to the notion of \emph{generalized global symmetries}, and illustrated under which conditions these can be avoided, via their breaking or gauging.

The complete identification of global symmetries within a theory plays a crucial role in Quantum Gravity.
Indeed, as captured by the \emph{No Global Symmetry Conjecture}, one of the pillars of the Swampland program, any consistent theory of quantum gravity ought to be devoid of any global symmetry. The foundations of the conjecture are grounded in the black hole physics \cite{Hawking:1975vcx,Zeldovich:1976vq,Zeldovich:1977be,Banks:1988yz,Banks:2010zn}. 
Loosely speaking, if there exist black holes carrying global charges, the Hawking evaporation process would produce an infinite number of states, with masses and sizes of the order of the Planck scale and each characterized by arbitrary values of global charges. The coexistence of an infinite number of species of such \emph{remnants} is believed to be pathological \cite{Susskind:1995da}, and in conflict with the Covariant Entropy Bound \cite{Banks:2010zn}.

However, as noticed in \cite{Heidenreich:2020pkc,Heidenreich:2021xpr} and further explained in the following sections, EFTs stemming from string theory are \emph{apparently} plagued by such generalized global symmetries. 
Therefore, in order to comply with the No Global Symmetry conjecture, their UV stringy origin has to deliver the appropriate mechanisms that avoid them. 

In this section, we set the stage for the forthcoming discussion by briefly recalling the notion of generalized global symmetries in Quantum Field Theory, and how these can be avoided.
In addition, we show that the breaking of some global symmetries may lead to the realization of a \emph{Witten effect}, which predicts properties about the composition of the allowed spectrum of states that the theory has to support.

\subsection{Generalized global symmetries, and how to avoid them}
\label{sec:GGS_avoid}

In a $D$-dimensional spacetime, a \emph{$p$-form global symmetry} is a symmetry whose action is implemented via operators supported on a $(p-1)$-codimension manifold, $\mathcal{M}^{(D-p-1)}$. 
Such global symmetries form a group, which we denote with $G$.
In general, the group $G$ may be either abelian or non-abelian; however, for $p>0$, the symmetry group $G$ needs to be abelian. Moreover, the group $G$ can be either discrete or continuous.
Given an element $g \in G$, one can then associate an operator $U_g (\mathcal{M}^{(D-p-1)})$. The operator $U_g (\mathcal{M}^{(D-p-1)})$ is assumed to be topological, namely, it is invariant under continuous deformations of the supporting manifold $\mathcal{M}^{(D-p-1)}$, provided that no charged spacetime defect is crossed.
$G$ being a group, it follows that for any two group elements $g_1, g_2 \in G$, also $g = g_1 g_2 \in G$. Such a composition law is realized at the operator level as
\begin{equation}
	\label{GGS_comp}
	U_{g_1} (\mathcal{M}^{(D-p-1)}) \times U_{g_2} (\mathcal{M}^{(D-p-1)}) = U_{g} (\mathcal{M}^{(D-p-1)}) .
\end{equation}
Observables -- such as charged operators -- can be understood as operators $V(\mathcal{C}^{(p)})$ supported on $p$-dimensional mainfolds $\mathcal{C}^{(p)}$.
Then, a symmetry operator $U_g (\mathcal{M}^{(D-p-1)})$, with the manifold $\mathcal{M}^{(D-p-1)}$ encircling $\mathcal{C}^{(p)}$, acts on $G$ as follows\footnote{Recall that, in $D$ dimensions, a $p$-dimensional object can be surrounded by a $(D-p-1)$-dimensional sphere.}
\begin{equation}
	\label{GGS_UV}
	U_{g} (\mathcal{M}^{(D-p-1)}) V(\mathcal{C}^{(p)}) = g(V) V(\mathcal{C}^{(p)})\,,
\end{equation}
where $g(V)$ denotes the action of $g$ in some convenient representation.

For concreteness, and to set the ground for the upcoming sections, let us now focus on the case where the symmetry group $G$ is the continuous, abelian $U(1)$ group. In such a case, one can single out a $(D-p-1)$-form Noether current $J_{D-p-1}$ that is conserved on-shell:\footnote{We refer to Appendix~\ref{app:conv_dual} for the conventions on the gauge fields that we use throughout the work.}
\begin{equation}
	\label{GGS_dJ}
	{\rm d} J_{D-p-1} = 0\,.
\end{equation}
In turn, given an element $g = e^{\im \alpha}$, with $\alpha \in \mathbb{R}$, one can define the following topological operator 
\begin{equation}
	\label{GGS_UQ}
	U_{g = e^{\im \alpha}}(\mathcal{M}^{(D-p-1)})  = \exp\left( \im \alpha Q(\mathcal{M}^{(D-p-1)})\right) = \exp\left( \im \alpha \int_{\mathcal{M}^{(D-p-1)}} J_{D-p-1} \right) \,,
\end{equation}
where $Q$ denotes the conserved charge. The operator \eqref{GGS_UQ} is indeed topological due to the conservation law \eqref{GGS_dJ}.
The `ordinary' global $U(1)$ symmetries correspond to the case $p=0$, for which the charge operators are supported on codimension-one manifolds and the charged observables can be viewed as localized, zero-dimensional defects.

Let us now assume that some generalized global symmetries, constituting the group $G_{\text{\tiny EFT}}$, can be identified within an EFT that is described by the action $S_{\text{\tiny EFT}}$, valid up to the energy scale cutoff $\Lambda_{\text{\tiny EFT}}$. 
Then, according to the No Global Symmetry Conjecture, these symmetries ought to be ultimately removed in a full quantum gravity theory, eventually by effects emerging at energy scales above $\Lambda_{\text{\tiny EFT}}$.
The two most common mechanisms that could remove the global symmetries are the following:
\begin{description}
    \item[Breaking the symmetries] The global symmetries can be \emph{explicitly} broken so that the group $G_{\text{\tiny EFT}}$ is not anymore a symmetry group of the theory. At the EFT level, this can be achieved by including, in the effective action $S_{\text{\tiny EFT}}$, new terms that are not invariant under the action of $G_{\text{\tiny EFT}}$.
    Indeed, these new, additional terms can be either local or supported on some higher dimensional manifold, and their inclusion leads to the explicit breaking of the conservation law \eqref{GGS_dJ}.
    The latter symmetry breaking terms may serve, for example, as operators creating higher dimensional charged states.\footnote{In this regard, it is worth mentioning that the completeness of the spectrum can be related to the breaking of all the higher form symmetries of a theory and, under some hypotheses, such a relation can be explicitly proven -- see, for instance, \cite{Rudelius:2020orz,Heidenreich:2021xpr}.}
    It is worth noticing, however, that the inclusion of charged operators typically reduces $G_{\text{\tiny EFT}}$ to a discrete subgroup thereof, $G_{\text{\tiny EFT}}^{\text{\tiny{discrete}}}$, and further assumptions on the symmetry-breaking terms may be needed in order to break also $G_{\text{\tiny EFT}}^{\text{\tiny{discrete}}}$.
	\item[Gauging the symmetries] Alternatively, the global symmetry group may be gauged in the UV-completed theory. Namely, the global symmetry group $G_{\text{\tiny EFT}}$ is promoted to a wider local symmetry group $G_{\rm gauge}$.
	In the case $G_{\text{\tiny EFT}} = U(1)$, gauging the global symmetry can be achieved by rendering the current $J_{D-p-1}$ exact so that the operator \eqref{GGS_UQ} becomes trivial. 
\end{description}

\noindent{\textbf{A simple example: the free Maxwell theory}} 

\noindent
Before closing this brief overview of generalized global symmetries, let us consider a simple, illustrative example, namely the free Maxwell theory. 
We consider the case where only a single $U(1)_{\rm g,e}$ gauge field is dynamical. The latter is represented by a gauge one-form $A_1$, subjected to the gauge transformation $A_1 \to A_1 + {\rm d}\Lambda_0$, for an arbitrary zero-form $\Lambda_0$. The free Maxwell action is
\begin{equation}
	\label{Maxw_S}
	S = -\frac1{2 g^2}  \int F_2 \wedge * F_2\,,
\end{equation}
with $F_2 = {\rm d} A_1$ being the field strength, and we will refer to $g$ as the electric gauge coupling.

The action \eqref{Maxw_S} is endowed with a $U(1)_{\rm e}$ one-form global symmetry. Such a symmetry is realized as a shift of the gauge one-form $A_1$ by a flat connection, as $A_1 \to A_1 + \Lambda_1$ with ${\rm d} \Lambda_1 = 0$. Applying Noether's theorem yields the following two-form current and topological operator:
\begin{equation}
	\label{Maxw_JQU}
	J_{2}^{\rm e} = -\frac1{g^2} * F_2\,, \qquad U_{\alpha}(\Sigma_2) = e^{\im \alpha Q^{\rm e}(\Sigma_2)} = e^{- \im \frac{\alpha}{g^2}\int_{\Sigma_2} * F_2}\,,
\end{equation}
where $\alpha \in \mathbb{R}$ and $\Sigma_2$ is a closed two-dimensional manifold.
The charged operators under such electric one-form global symmetry are the \emph{Wilson lines}
\begin{equation}
	\label{Maxw_Wline}
	W_n(\Gamma_1) = \exp \left(\im n \oint_{\Gamma_1} A_1 \right)\,,
\end{equation}
with $n \in \mathbb{N}$ denoting the elementary charge of the Wilson line, and $\Gamma_1$ the one-dimensonal closed manifold supporting the operator. 
The Wilson line transforms non-trivially under the one-form shift $A_1 \to A_1 + \Lambda_1$ for closed $\Lambda_1$. 
In fact, $W_n(\Gamma_1)$ gets mapped to $e^{\im \alpha \oint_{\Gamma_1} \Lambda_1} W_n(\Gamma_1)$, with the phase operator $e^{\im \alpha \oint_{\Gamma_1} \Lambda_1}$ depending only on the homology class of $\Gamma_1$. 
The elementary charge of the Wilson line can be measured via the relation \eqref{Maxw_JQU}, which reads
\begin{equation}
	\label{Maxw_WllineU}
	U_{\alpha}(\Sigma_2) W_n(\Gamma_1) = e^{\im n \alpha} W_n(\Gamma_1) \,,
\end{equation}
for a closed two-manifold $\Sigma_2$ encircling $\Gamma_1$.

However, the $U(1)_{\rm e}$ group is not the only one to deliver a generalized global symmetry. 
In fact, the free Maxwell theory \eqref{Maxw_S} admits a dual, magnetic description in which the degrees of freedom are described in terms of new gauge one-form $V_1$, the electromagnetic dual form of $A_1$.
By following the dualization procedure outlined in Section~\ref{app:conv_dual}, one can show that the action magnetic dual to \eqref{Maxw_S} is 
\begin{equation}
	\label{Maxw_Sd}
	S_{\rm dual} = -\frac{g^2}{2}  \int G_2 \wedge * G_2\,,
\end{equation}
with $G_2 = {\rm d} V_1$. The action \eqref{Maxw_Sd} is endowed with an additional $U(1)_{\rm m}$ one-form global symmetry that is realized via shifts of the magnetic gauge one-form $V_1$ by flat connections as $V_1 \to V_1 + \tilde{\Lambda}_1$, with ${\rm d} \tilde{\Lambda}_1 = 0$. The associated conserved two-form current and charge operator are
\begin{equation}
	\label{Maxw_JQUm}
	J_{2}^{\rm m} = - g^2 *G_2\,, \qquad U_{\beta}(\Sigma_2) = e^{\im \beta Q^{\rm m}(\Sigma_2)} = e^{-\im \beta g^2 \int_{\Sigma_2} *G_2}\,,
\end{equation}
with $\beta \in \mathbb{R}$. 
As outlined in Appendix~\ref{app:conv_dual}, the conservation of the magnetic current $J_{2}^{\rm m}$ can be equivalently read from the Bianchi identities of the electric frame, since the dualization procedure enforces the on-shell relation $F_2 = - g^2 * G_2$.
An example of observables that can be detected in the dual frame \eqref{Maxw_Sd} are the \emph{`t Hooft lines} 
\begin{equation}
	\label{Maxw_Tline}
	T_m(\Gamma_1) = \exp \left(\im m \oint_{\Gamma_1} V_1 \right)\,,
\end{equation}
with $m$ denoting the elementary, magnetic charge of the operator. 

Therefore, in summary, the free Maxwell theory is characterized by the global symmetry group $G_{\text{\tiny EFT}} = U(1)_{\rm e} \times U(1)_{\rm m}$.  But how can such symmetries be removed? The simplest possibility is that they are broken explicitly.
This can be achieved via an explicit violation of the conservation laws of both the electric and magnetic two-form currents in \eqref{Maxw_JQU} and \eqref{Maxw_JQUm} as
\begin{equation}
	\label{Maxw_break}
	{\rm d} J^{\rm e}_2 = - j^{\rm e}_3\,, \qquad {\rm d} J^{\rm m}_2 = - j^{\rm m}_3\,,
\end{equation}
where $ j^{\rm e}$( $j^{\rm m}$) denotes a localized electric (magnetic) charge distribution. For instance, in the presence of a single charged fundamental object, such currents are of the form
\begin{equation}
	 j^{\rm e} = q^{\rm e} \delta_3 (\Gamma_1) \,, \qquad j^{\rm m} = q^{\rm m} \delta_3 (\tilde{\Gamma}_1)
\end{equation}
with $ q^{\rm e}$ ($q^{\rm m}$) the electric (magnetic) elementary charge of the object and $\delta_3 (\Gamma_1)$ ($\delta_3 (\tilde\Gamma_1)$) the three-form Dirac delta distribution localized on the object's worldline $\Gamma_1$($\tilde\Gamma_1$).
At the EFT level, these charged, fundamental objects can be understood as (heavy) particles that manifest themselves via the couplings
\begin{equation}
	S^{\rm e} = q^{\rm e} \int_{\Gamma_1} A_1  = \int A_1 \wedge j^{\rm e}  \,, \qquad S^{\rm m} = q^{\rm m} \int_{\tilde{\Gamma}_1} V_1  = \int V_1 \wedge j^{\rm m}
\end{equation}
to the actions \eqref{Maxw_S} and \eqref{Maxw_Sd} respectively.

Alternatively to their explicit breaking, the electric and magnetic one-form global symmetries can be removed if they are gauged in the UV-completed theory. 
The gauging is performed by assuming that the one-form shifts, $A_1 \to A_1 + \Lambda_1$ and $V_1 \to V_1 + \tilde{\Lambda}_1$, leave the theory invariant, even for non-flat one-forms $\Lambda_1$, $\tilde{\Lambda}_1$. Clearly, under such shifts, neither of the field strengths $F_2$ and $G_2$ are invariant. In fact, constructing a gauge invariant EFT requires the presence of additional degrees of freedom that may stem from the UV completion. Such degrees of freedom can be, for instance, associated to some gauge two-forms $B_2$ or $\tilde{B}_2$, with gauge transformations  $B_2 \to B_2 + {\rm d}\Lambda_1$ and $\tilde{B}_2 \to \tilde{B}_2 + {\rm d}\tilde{\Lambda}_1$. These would allow for defining the gauge invariant field strengths $\hat{F}_2 = F_2 - B_2$ and $\hat{G}_2 = G_2 - \tilde{B}_2$, which can be employed to conveniently define the one-form kinetic terms. An example of such a gauging occurs in the so-called $BF$-theory -- see, for instance, \cite{Banks:2010zn}.

\subsection{The Witten effect and an infinitely populated charge lattice}
\label{sec:WE_mon}

As we have illustrated, breaking certain global symmetries may require the existence of particular extended objects in the spectrum.
However, whenever the theory is endowed with additional couplings, the mere existence of said states may give important information about the composition of the allowed \emph{(sub)lattice} of states that the theory must support.
In particular, if the theory is endowed with some `\emph{$\theta$-terms}', whereby an axion couples to topological terms of the form $F \wedge F$, the \emph{Witten effect}  \cite{Witten:1979ey} (see also \cite{Harvey:1996ur,Alvarez-Gaume:1997dpg,Agrawal:2022yvu} for later reviews) predicts the existence of an infinitely populated lattice of states, generated by a \emph{seed} state.

In order to illustrate how the Witten effect is realized, let us consider a simple model, namely the four-dimensional free Maxwell theory \eqref{Maxw_S} with the addition of a $\theta$-term:
\begin{equation}
	\label{WE_S}
	S = \int \left(-\frac12 F_2 \wedge * F_2 + \frac{\theta}{2}  F_2 \wedge F_2 \right)
\end{equation}
Owing to the quantization condition for the curvature $F_2$, integral `monodromy' transformations $\theta \to \theta +n$ only shift the action by an integer, as $S \to S +n$. As such, the monodromy transformation $\theta \to \theta +n$ ought to be regarded as a perturbative symmetry of the model, under which any correlation function is left invariant.

Clearly, if the field strength $F_2$ obeys the Bianchi identity ${\rm d} F_2 = 0$, the last term in \eqref{WE_S} is a topological term. 
Moreover, as discussed in the previous section, $J_2^{\rm m} = F_2$ also plays the role of a conserved current associated to a magnetic one-form global symmetry.
In order to remove such a global symmetry without introducing additional gauge fields, one can assume that a magnetic monopole with elementary charge $q_{\rm m}$ is present in the spectrum. As a result, the Bianchi identity for $F_2$ gets modified to ${\rm d} F_2 = - q_{\rm m} \delta^{(3)}(\Gamma)$, with $\Gamma$ the monopole worldline.

Here, the Witten effect comes into play. The Witten effect \cite{Witten:1979ey} asserts that the single assumption of the existence of a magnetic monopole in the spectrum is enough to guarantee that the spectrum of states also allows for the presence of dyons, namely particles both electrically and magnetically charged. 
There are several ways to see the emergence of dyons in the spectrum.
The simplest is to notice that the action \eqref{WE_S} can be recast as \cite{Coleman:1982cx} 
\begin{equation}
	\label{WE_Sb}
	S = -\frac12 \int F_2 \wedge * F_2 - q_{\rm m} \theta \int_\Gamma A_1 \,.
\end{equation}
The last term displays the coupling of the one-form gauge field $A_1$ to an electric particle with charge $q_{\rm e} = - q_{\rm m} \theta$, tracing out the same worldline as the magnetic monopole introduced above. Moreover, due to the perturbative symmetries of the action, \emph{all} the electric charges that are obtained from the one appearing in \eqref{WE_Sb} via integral shifts of the parameter $\theta$ should be present. As a result, the presence of a magnetic monopole implies an \emph{infinite} tower of dyons with elementary charges
\begin{equation}
	\label{WE_qdyon}
	{\bf q}^T = (q_{\rm m}, - q_{\rm e}) = \left( q_{\rm m}, (\theta+n) q_{\rm m} \right)\,.
\end{equation}
Moreover, due to the explicit dependence of the electric charges on the real parameter $\theta$, in general, the electric charges are not integral.\footnote{However, notice that the dyon's electric and magnetic charges are still mutually quantized, for they obey a modified version of the Dirac's quantization \cite{Witten:1979ey}.} 

Alternatively, one can see the emergence of an infinite tower of dyonic particles by examining the currents and charges of the model. In the absence of objects, the action \eqref{WE_S} exhibits a magnetic one-form global symmetry. Thus, there exists a conserved, gauge-invariant two-form current $J_2^{\rm m} = F_2$. This current is however not conserved in a monopole background, for ${\rm d} J_2^{\rm m} = - q_{\rm m} \delta^{(3)}(\Gamma)$. It is then convenient to introduce the modified current and charge:\footnote{In the language of the upcoming Section~\ref{sec:GGS_IIBsub}, these can be identified as the \emph{Page current} and the \emph{Page charge} of the model.}
\begin{equation}
    \label{WE_modQJ}
	(J_2^{\rm e})^{{\rm mod}} = -* F_2 + \theta F_2\,, \qquad (Q^{\rm e})^{{\rm mod}} = \int_{\Sigma_2} (J_2^{\rm e})^{{\rm mod}} \,.
\end{equation}
In the absence of electric sources, i.e.~$j^{\rm e} = 0$, we have  ${\rm d}(J_2^{\rm e})^{{\rm mod}} = 0$; in turn, this leads to ${\rm d} * F_2 =  - q_{\rm m} \theta \delta^{(3)}(\Gamma)$, consistently with \eqref{WE_Sb}. Still, the current and charge \eqref{WE_modQJ} are not single-valued: under shifts $\theta \to \theta + n$, they shift as
\begin{equation}
	(J_2^{\rm e})^{{\rm mod}} \to (J_2^{\rm e})^{{\rm mod}} + n F_2\,, \qquad (Q^{\rm e})^{{\rm mod}} \to (Q^{\rm e})^{{\rm mod}}  - n q_{\rm m} \,.
\end{equation}
Such shifts can be reinterpreted as electric states which source the current $j^{\rm e}$, defined in such a way that
\begin{equation}
	{\rm d} (J_2^{\rm e})^{{\rm mod}} = n q_{\rm m} \delta^{(3)}(\Gamma) =: - j^{\rm e}\,,
\end{equation}
and thus breaking the conservation of the modified current $(J_2^{\rm e})^{{\rm mod}}$.

Therefore, although the avoidance of the magnetic one-form global symmetry requires the existence of only a single magnetic monopole of elementary charge, the Witten effect predicts that the monopole is not the sole charge that has to be present in the spectrum. Rather, an \emph{infinite} tower of dyonic states has to be supported by the theory.


\section{Global symmetries and generalized \texorpdfstring{$\theta$}{tehta}-terms in Type IIB EFTs}
\label{sec:GGS_IIB}

In this section we present the class of EFTs that will be the main object of study: namely, the four-dimensional EFTs that arise from the compactification of the ten-dimensional Type IIB string theory over a Calabi-Yau three-fold. After a brief overview of the salient features of such EFTs and of the multiplet sector that we focus on, we will initiate the study of the generalized global symmetries that the 4D Type IIB EFTs may be endowed with, and how these may be broken.
We further show that the breaking of the global symmetry triggers a generalized version of the Witten effect, and we characterize the spectrum of states that this effect predicts.
The analysis carried out in this section is general, relying only on the $\mathcal{N}=2$ local supersymmetry that the Type IIB EFTs are equipped with.

\subsection{\texorpdfstring{$\mathcal{N}=2$}{N=2} Type IIB EFTs}
\label{sec:IIB_review}

In this section we review some basic aspects of $\mathcal{N}=2$ EFTs originating after compactifying Type IIB string theory over a Calabi-Yau three-fold $Y$ \cite{Louis:2002ny,Grimm:2004uq}. We will be interested in the vector multiplet sector coupled to gravity, and we will focus on the bosonic sector only. The fields entering the EFT are the graviton $g_{\mu\nu}$, the graviphoton gauge one-form $A^0$ and $h^{2,1}$ gauge one-forms $A^i$, $i =1, \ldots, h^{2,1}$, which we collectively denote $A^I$, $I = 0, 1,\ldots, h^{2,1}$, and $h^{2,1}$ complex scalar fields $t^i$.

The complex scalar fields $t^i$ parametrize deformations of the complex structure of $Y$. Indeed, by introducing a real integral basis of three-forms $\gamma_{\mathcal{I}}$, $\mathcal{I} = 1,\ldots, 2h^{2,1}+2$ of $H^3(Y)$, we can expand the Calabi-Yau holomorphic three-form as follows
\begin{equation}
	\label{N2_Omega}
	\Omega = \Pi^{\mathcal{I}} (t) \gamma_{\mathcal{I}} = {\bm \Pi}^T (t) {\bm \gamma}\,.
\end{equation}
where the functions $\Pi^{\mathcal{I}} (t)$, that are holomorphic in $t^i$, are the `periods' of the holomorphic three-form $\Omega$. The periods can be further computed as $\Pi^{\mathcal{I}} (t) = \int_{\Gamma_{\mathcal{I}}} \Omega$, given a basis $\Gamma_{\mathcal{I}}$ of three-cycles of $H_3$ such that $\int_{\Gamma_{\mathcal{I}}}  \gamma_{\mathcal{J}} = \delta_\mathcal{J}^\mathcal{I}$.
	
For the discussion in the upcoming sections, it is convenient to further specialize the choice of basis of cycles $\gamma_{\mathcal{I}}$. We introduce a symplectic basis of three-cycles $\Gamma^{\mathcal{I}} = (A^I, B_J)$ of $Y$ and, accordingly, a basis of dual three-forms $\gamma_{\mathcal{I}} = (\alpha_I, \beta^J)$, with $I, J= 1, \ldots, h^{2,1} + 1$ defined in such a way that the intersection matrix $\eta$ has the standard form
\begin{equation}
\label{Sympl_eta}
	\eta = \begin{pmatrix}
	\int \alpha_I \wedge \alpha_J & \int \alpha_I \wedge \beta^J\\  \int \beta^I \wedge \alpha_J  & \int \beta^I \wedge \beta^J
	\end{pmatrix} = \begin{pmatrix}
	0 & \mathds{1}\\ -\mathds{1} & 0
    \end{pmatrix}
\end{equation}
expressed in $(h^{2,1}+1) \times (h^{2,1}+1)$ blocks. 
	
We can then expand the holomorphic three-form $\Omega$ in the symplectic basis as
\begin{equation}
	\label{IIB_periodsymplb}
	{\bm \Pi} (t) =  \begin{pmatrix}
	\int_{B_I} \Omega \\  -\int_{A^I} \Omega
	\end{pmatrix}  =  \begin{pmatrix}
		X^I (t) \\  -\mathcal{F}_I(t)
	\end{pmatrix} \, ,
\end{equation}
with $X^I(t)$ and $\mathcal{F}_I(t)$ holomorphic functions of the complex structure moduli $t^i$.
	
The four-dimensional $\mathcal{N}=2$ effective action describing the interactions among the bosonic components of the gravity multiplet and the $h^{2,1}$ vector multiplets is
\begin{equation}
	\label{N2_action}
	S = \int \left( \frac{1}2 M^2_{\rm P} R *1 - M^2_{\rm P} K_{i\bar\jmath}^{\rm cs} {\rm d} t^i \wedge * {\rm d} \bar{t}^{\bar\jmath} + \frac12 {\rm Im} \mathcal{N}_{IJ} F^I \wedge * F^J + \frac12 {\rm Re} \mathcal{N}_{IJ} F^I \wedge F^J \right)\,.
\end{equation}
Here we have denoted with $R$ the Ricci scalar and $K_{i\bar\jmath}^{\rm cs} = \partial_{t^i} \partial_{\bar{t}^{\bar\jmath}} K^{\rm cs}$, with $\partial_{t^i} := \frac{\partial}{\partial t^i}$, is the K\"ahler metric that can be obtained out of the K\"ahler potential $K^{\rm cs}$. The latter is fully specified by the periods as
\begin{equation}
	\label{IIB_Kcs}
	K^{\rm cs}(t,\bar t) = -\log \left(  \im \int_Y \Omega \wedge \bar\Omega \right)=- \log \im\, {\bm \Pi}^T \eta \bar{\bm \Pi} = - \log \im(\bar{X}^I \mathcal{F}_I -  X^I \bar{\mathcal{F}}_I)\,.
\end{equation}
The interactions of the gauge one-forms $A^I$ are dictated by the complex matrix $\mathcal{N}_{IJ}$ computed as follows
\begin{equation}
	\label{IIB_NIJ}
	\mathcal{N}_{IJ} = (\mathcal{F}_I \quad D_{\bar\imath} \bar{\mathcal{F}}_{\bar I})(X^J \quad D_{\bar\imath} \bar{X}^{\bar J})^{-1}\,.
\end{equation}

Whenever the matrix $(D_i X^I, X^I)$ is invertible, the action \eqref{N2_action} is fully specified by a \emph{prepotential} $\mathcal{F}(X)$, which is a homogeneous function of degree two in the projective coordinates $X^I$.  Then, the quantities $\mathcal{F}_I$ that appear in \eqref{IIB_periodsymplb} can be regarded as derivatives of the prepotential $\mathcal{F}_I = \frac{\partial \mathcal{F}}{\partial X^I}$. Moreover, when a prepotential exists, one can recast the matrix $\mathcal{N}_{IJ}$ in \eqref{IIB_NIJ} in the more explicit form
\begin{equation}
	\label{IIB_NIJb}
	\mathcal{N}_{IJ} = \bar{\mathcal{F}}_{IJ} + 2 \im \frac{{\rm Im} 	\mathcal{F}_{IK} X^K {\rm Im} 	\mathcal{F}_{JL} X^L}{ {\rm Im} 	\mathcal{F}_{MN} X^M X^N }\,.
\end{equation}

A key ingredient of the analysis that follows is given by the examination of the equations of motion of the fields entering the effective description \eqref{N2_action}. The equations of motion for the gauge fields $A^I$ are:
\begin{equation}
	\label{IIB_eom_A}
	{\rm d} G_I  = 0 \,,
\end{equation}
where we have defined
\begin{equation}
	\label{IIB_eom_G}
	G_I \equiv \frac{\delta S}{\delta F_{\mu\nu}^I} {\rm d} x^\mu \wedge {\rm d} x^\nu = {\rm Im} \mathcal{N}_{IJ} *\! F^J  + {\rm Re} \mathcal{N}_{IJ} F^J \,,
\end{equation}
to which we will oftentimes refer as the field strengths \emph{dual} to $F^I$. 
	%
	%
The equations of motion \eqref{IIB_NIJb} are additionally supported by the Bianchi identities ${\rm d} F^I = 0$. 

The equations of motion for the complex structure moduli $t^i$ are
\begin{equation}
	\label{IIB_eom_phi}
	M_{\rm P}^2 \left[{\rm d} (K_{i \bar\jmath}^{\rm cs} * {\rm d} \bar{t}^{\bar\jmath} ) - K_{il \bar\jmath}^{\rm cs}{\rm d} t^l \wedge * {\rm d} \bar{t}^{\bar\jmath} \right] = -\frac{1}{2} \frac{\partial}{\partial t^i} {\rm Im} \mathcal{N}_{IJ} F^I \wedge * F^J -\frac{1}{2} \frac{\partial}{\partial t^i} {\rm Re} \mathcal{N}_{IJ} F^I \wedge F^J\,.
\end{equation}
Additionally, for the following discussion, it will be convenient to consider the special case in which the K\"ahler metric is real, and it is only saxion dependent; namely $K_{i\bar\jmath} (t, \bar t) =: K_{ij } (s)$, so that the general action \eqref{N2_action} reduces to	
\begin{equation}
\label{N2_action_realm}
	\begin{aligned}
	S &= \int \Big( \frac{1}2 M^2_{\rm P} R *1 - M^2_{\rm P} K_{ij}^{\rm cs} ({\rm d} a^i \wedge * {\rm d} a^j + {\rm d} s^i \wedge * {\rm d} s^j)  + \frac12  G_I \wedge F^I \Big)\,,
    \end{aligned}
\end{equation}
and the equations of motion for the saxions $s^i$ and the axions $a^i$ are, respectively
\begin{subequations}
\begin{equation}	
	\begin{aligned}
	\label{IIB_eom_s}
	&M_{\rm P}^2 \left[{\rm d} (K_{ij}^{\rm cs} * {\rm d} \bar{t}^{\bar\jmath} ) - \frac{\partial}{\partial s^i} K_{lj}^{\rm cs}( {\rm d} a^l \wedge * {\rm d} a^{j} + {\rm d} s^l \wedge * {\rm d} s^{j} ) \right] = -\frac{1}{2} \frac{\partial}{\partial s^i} G_I \wedge F^I\,,
	\end{aligned}
\end{equation}
\begin{equation}
	\begin{aligned}
	\label{IIB_eom_a}
	&M_{\rm P}^2\, {\rm d} (K_{ij}^{\rm cs} * {\rm d} a^{j} )   = -\frac{1}{2} \frac{\partial}{\partial a^i} G_I \wedge F^I\,.
	\end{aligned}
\end{equation}
\end{subequations}

As of now, the analysis of the Type IIB EFTs has been focused on the closed string sector only. However, the four-dimensional Type IIB EFTs can be populated by a plethora of objects that originate from higher-dimensional $p$-branes. For the upcoming discussion, we will be interested in the coupling of the vector multiplet sector to BPS particles that are electrically charged under the one-form gauge fields $A^I$ or their dual gauge fields. Such particles stem from D3-branes that wrap internal special Lagrangian three-cycles. 
The mass of the a D3-particle obtained wrapping the D3-brane on the special Lagrangian $\Sigma_{\bf q} \in H^3(Y,\mathbb{Z})$ is given by the central charge $\mathcal{Z}_{\bf q}$ as \cite{Hull:1994ys,Ceresole:1995ca}:
\begin{equation}
	\label{IIB_D3M}
	M_{\bf q} = |\mathcal{Z}_{\bf q}| = e^{\frac{K^{\rm cs}}2} \left| \int_{\Sigma_{\bf q}} \Omega \right|  = e^{\frac{K^{\rm cs}}2} \left| \int_Y q \wedge \Omega \right| \,.
\end{equation}
Here we have denoted with ${\bf q}$ the  \emph{elementary charges} of the particle, whereas $q$ denotes the three-form Poincar\'e dual to the three-cycle $\Gamma_{\bf q}$. It is convenient to expand $q = \alpha_I p^I - q_I \beta^I$, or equivalently the charge vector ${\bf q}$ as
\begin{equation}
	\label{IIB_D3q}
	{\bf q} = \begin{pmatrix}
		p^I \\ -q_I
	\end{pmatrix}\,, \qquad q_I, p^I \in \mathbb{Z}\,.
\end{equation}
We will refer to $q_I$ as the elementary \emph{electric} charges and $p^I$ as the elementary \emph{magnetic} charges of the D3-brane. With such a decomposition, the D3-particle mass \eqref{IIB_D3M} can be rewritten as
\begin{equation}
	\label{IIB_D3Mb}
	M_{\bf q} = e^{\frac{K^{\rm cs}}{2}} |{\bf q}^T\,\eta\, {\bm \Pi}(t)|\,.
\end{equation}
In addition to the elementary charges, one can define the \emph{physical charge} of the D3-particle as follows
\begin{equation}
	\label{IIB_D3Q}
	\mathcal{Q}^2_{\bf q} = \frac12  \int_Y q \wedge \star q  = - \frac12 {\bf q}^T \mathcal{T} {\bf q}\,,
\end{equation}
where
\begin{equation}
	\label{IIB_Mmat}
	 \mathcal{T}  = \begin{pmatrix}
		{\rm Im} \mathcal{N} + {\rm Re} \mathcal{N} ({\rm Im}\mathcal{N})^{-1} {\rm Re} \mathcal{N} & {\rm Re} \mathcal{N} ({\rm Im} \mathcal{N})^{-1}
		\\
		({\rm Im} \mathcal{N})^{-1} {\rm Re} \mathcal{N}  & ({\rm Im}\mathcal{N})^{-1} 
	\end{pmatrix}\,,
\end{equation}
that is expressed in the symplectic basis $\{\alpha_I,\beta^I\}$. The physical charge in \eqref{IIB_D3Q} carries information about the gauge couplings associated to the $U(1)_{\rm g}$ abelian gauge one-forms $A^I$.

In the following, we will regard the D3-particles just introduced as \emph{fundamental} objects. 
Namely, we shall assume that the EFT cutoff $\Lambda_{\text{\tiny EFT}}$ is chosen in such a way that $\Lambda_{\text{\tiny EFT}} \lesssim M_{\bf q}$, thus that additional brane worldvolume degrees of freedom can be neglected within the effective description.
Under such an assumption,  electrically charged D3-particles with elementary charges $q_I$ can be coupled to the bulk action \eqref{N2_action} via the BPS action
\begin{equation}
	\label{IIB_SD3}
	S_{\text{D3},{\bf q}_{\rm e}} = - \int_{\Gamma} {\rm d} \tau\,  M_{\bf q}  + q_I \int_{\Gamma}\,  A^I
\end{equation}
where $\tau$ the proper time parametrizing the particle worldline $\Gamma$.
Magnetically charged particles with elementary charges $p^I$ cannot be directly coupled to the EFT action \eqref{N2_action}, but rather to its electromagnetic dual action that is computed below (see \eqref{N2_action_dual}).\footnote{We stress that here, and in the following we shall always consider \emph{fundamental} objects, which are neither constituted by elementary components, nor can decay. An analysis of symmetry breaking via non-fundamental objects is left for future work.}

\subsection{The symplectic structure of Type IIB EFTs}
\label{sec:IIB_Sympl}

Within the $\mathcal{N}=2$ supergravity EFTs introduced in the previous section, the complex structure fields $t^i$ span a special K\"ahler manifold that is endowed with a flat symplectic bundle (see, for instance, \cite{Andrianopoli:1996cm,Craps:1997gp}). 
This section is devoted to the review of some basic consequences of the symplectic structure of the $\mathcal{N}=2$ supergravity EFTs. The knowledge of how the ingredients of $\mathcal{N}=2$ EFTs transform under the symplectic group will be crucial for the upcoming sections.

Preliminarily, let us recall that a matrix $\mathcal{S} \in {\rm GL}(2 h^{2,1}+2, \mathbb{R})$ is symplectic if $\mathcal{S}^T \eta \mathcal{S} = \eta$, with $\eta$ defined in \eqref{Sympl_eta}. In order to better characterize the structure of any symplectic $\mathcal{S}$ matrices, it is convenient to split the matrix $\mathcal{S}$ into $(h^{2,1}+1) \times (h^{2,1}+1)$ square matrices as follows
\begin{equation}
	\label{Sympl_S}
	\mathcal{S} = \begin{pmatrix}
		A^{I}{}_J & B^{IJ} \\ C_{IJ} & D_I{}^J
	\end{pmatrix}\,,
\end{equation}
Then, a symplectic matrix $\mathcal{S}$ obey the conditions
\begin{equation}
	\label{Sympl_prop}
	A^T D - C^T B = A D^T - B C^T =  \mathds{1}\,, \qquad AB^T, CD^T, A^T C, B^T D \quad \text{symmetric} \,,
\end{equation}
where here, and in the following, we suppress the index structure. Moreover, the inverse of any symplectic matrix $\mathcal{S}$ is given by $\mathcal{S}^{-1} = - \eta \mathcal{S}^T \eta$.

Let us now focus on the scalar sector of the Type IIB EFTs. The periods \eqref{IIB_periodsymplb} transform as sections over the symplectic bundle
\begin{equation}
	{\bm \Pi}(t) \rightarrow  {\bm \Pi}'(t) = \mathcal{S} {\bm \Pi} (t)\,.
\end{equation}
It can be easily checked that any symplectic transformation of the periods leaves the complex structure K\"ahler potential \eqref{IIB_Kcs} invariant. In fact, consistently transforming the symplectic basis $\gamma_{\mathcal{I}} = (\alpha_I, \beta^J)$ leaves \eqref{N2_Omega} invariant.\footnote{Specifically, $(-\beta^I, \alpha_I)$ transforms as $(-\beta^I, \alpha_I) \to (-\beta^I, \alpha_I) \mathcal{S}^T $. It is straightforward to check that such a transformation leaves also  \eqref{Sympl_eta} invariant.}

However, the symplectic structure of the $\mathcal{N}=2$ supergravity plays a crucial role in the study of the gauge sector. As shown by Gaillard and Zumino in the seminal work \cite{Gaillard:1981rj} (see also \cite{Andrianopoli:1996cm,Craps:1997gp,Aschieri:2008ns} for later application to $\mathcal{N}=2$ supergravities), symplectic transformations can be understood as \emph{dualities} of the action \eqref{N2_action}. With \emph{duality} we understand a linear transformation of the field strengths and the matrix $\mathcal{N}$ determining the gauge kinetic matrix and Chern-Simons couplings such that the first relation in \eqref{IIB_eom_G} is left invariant \emph{in form}. In order to see this, let us first collect the field strengths $F^I$ and their duals $G_I$ defined in \eqref{IIB_eom_G} into the single vector
\begin{equation}
	\label{Sympl_FG}
	{\bf F} = \begin{pmatrix}
		F^I \\ - G_I
	\end{pmatrix}\,.
\end{equation}
Under a duality transformation, the field strengths ${\bf F}$ transform as a section of a flat symplectic bundle that the special K\"ahler geometry supports as ${\bf F} \to {\bf F}' = \mathcal{S} {\bf F}$. Additionally, the matrix $\mathcal{N}_{IJ}$ in \eqref{IIB_NIJ} accordingly transforms as
\begin{equation}
	\label{Sympl_Ntransf}
	\mathcal{N} \rightarrow (C + D \mathcal{N}) (A + B \mathcal{N})^{-1}\,.
\end{equation}
The general action \eqref{N2_action} is not invariant under general duality transformations; indeed, it gets transformed as
\begin{equation}
	\label{Sympl_Stransf}
	S \rightarrow S + \frac12 \int \left( 2 G B^T C F - F C^T A F - G D^T B G\right)\,.
\end{equation}

As of now, we have considered the symplectic group as defined over the real field. However, as we will see momentarily, dualities do also act on the D3-particle charge vectors \eqref{IIB_D3q}.
This feature can be employed as a constraint on the allowed symplectic transformations.
In fact, let us assume that the D3-brane charges are quantized in a given frame. Then, requiring that the D3-brane charges are mutually quantized after performing a duality transformation requires that the symplectic matrix that generates the duality displays integral entries. Therefore, from now on, we shall always consider dualities as induced by matrices $\mathcal{S} \in {\rm Sp}(2 h^{2,1}+2, \mathbb{Z})$.

In the following, we will be largely interested in two families of symplectic transformations:

\noindent\textbf{Monodromies.} Monodromy transformations constitute the subset of the symplectic transformations that are induced by an integral shift of the axionic fields. These can be most readily computed by looking at how the period vectors \eqref{IIB_periodsymplb} transform under such shifts. Namely, we identify the monodromy matrix $T_i \in {\rm Sp}(2 h^{2,1}+2, \mathbb{Z})$ by how the periods \eqref{IIB_periodsymplb} transform under the sole shift $a^i \to a^i + 1$ as:\footnote{As we shall see below, such discrete shifts can be seen as the gauge subgroup of the general shift symmetries $a^i \to a^i + c$, with $c \in \mathbb{R}$.}
	\begin{equation}
		\label{Sympl_TPi}
		{\bm \Pi}(a,s) \xrightarrow{a^i \to a^i + 1}  {\bm \Pi}'(a,s) =: T_i {\bm \Pi} (a,s)\,.
	\end{equation}
	Clearly, under $k$ shifts of the axions $a^i$, the periods get mapped as ${\bm \Pi}(a,s) \to T_i^k {\bm \Pi}(a,s)$.
	Although we shall give a detailed characterization of the properties of the monodromy matrices that appear in the near-boundary EFTs in Section~\ref{sec:GGS_01mod}, for the following discussion it suffices to introduce the following matrix
	\begin{equation}
		\label{Sympl_Nlogi}
		N_i = \log T_i\,,
	\end{equation}
	to which we will refer as \emph{log-monodromy matrix}.
	In the near-boundary EFTs studied in Section~\ref{sec:GGS_01mod} such a matrix is always well-defined.
	Assuming that the periods are expressed in special coordinates, i.e.~${\bm \Pi}(t) = (1, t^i, \mathcal{F}_I (t))$, any monodromy matrix $T_i$ acquires the form
	\begin{equation}
		\label{Sympl_Ttransf}
		T_i = \begin{pmatrix}
			A & 0 \\ C & D
		\end{pmatrix}\,, \qquad \text{with} \quad D= (A^T)^{-1}\,.
	\end{equation}
	As inferred from \eqref{Sympl_Stransf}, under a generic monodromy transformation \eqref{Sympl_Ttransf}, the action \eqref{N2_action} is not invariant. However, the lack of invariance is due to the integral shift of the action
	\begin{equation}
		\label{Sympl_Ttransf_shift}
		S \to S -\frac12 \int F C^T A F\,,  \qquad \frac12 \int F C^T A F \in \mathbb{Z}\,.
	\end{equation}
	Consequently, any partition function is expected to be invariant under generic monodromy transformations. Thus, despite not being classical symmetries of the action \eqref{N2_action}, monodromies can be regarded as \emph{pertubative} symmetries of the model.

\noindent\textbf{Electromagnetic duality.} We regard as \emph{electromagnetic duality} the symplectic transformation induced by the matrix $\eta$ defined in \eqref{Sympl_eta}. Such a duality transformation exchanges the role of field strengths, mapping $F^I \to - G_I$, $G_I \to F^I$, and consistently mapping $\mathcal{N} \to -\mathcal{N}^{-1}$. Consequently, the action \eqref{N2_action} gets mapped to
	\begin{equation}
		\label{N2_action_dual}
		\begin{aligned}
		S^{\rm dual} = \int \Big( &\frac{1}2 M^2_{\rm P} R *1 - M^2_{\rm P} K_{i\bar\jmath}^{\rm cs} {\rm d} t^i \wedge * {\rm d} \bar{t}^{\bar\jmath} 
		\\& - \frac12 {\rm Im} (\mathcal{N}^{-1})^{IJ} G_I \wedge * G_J - \frac12 {\rm Re} (\mathcal{N}^{-1})^{IJ} G_I \wedge G_J \Big)\,,
		\end{aligned}
	\end{equation}
	with 
	\begin{equation}
	\small{
		\label{N2_N_inv}
		\begin{aligned}
		\mathcal{N}^{-1} &= [({\rm Re}\,\mathcal{N})^{-1} {\rm Im}\,\mathcal{N} + ({\rm Im}\,\mathcal{N})^{-1} {\rm Re}\,\mathcal{N}]^{-1} [({\rm Im}\,\mathcal{N})^{-1} - \im  ({\rm Re}\,\mathcal{N})^{-1} ]
		\\
		& = [\mathds{1} + ({\rm Im}\,\mathcal{N})^{-1} {\rm Re}\,\mathcal{N} ({\rm Im}\,\mathcal{N})^{-1} {\rm Re}\,\mathcal{N}]^{-1} [ ({\rm Im}\,\mathcal{N})^{-1} {\rm Re}\,\mathcal{N}  ({\rm Im}\,\mathcal{N})^{-1} - \im ({\rm Im}\,\mathcal{N})^{-1}]
	\end{aligned}}
	\end{equation}
	In this frame, the elementary degrees of freedom may be understood as being carried by the field strengths $G_I$. We can then set $G_I = {\rm d} V_I$ for some gauge one-forms $V_I$.
	Moreover, as is clear from \eqref{N2_action_dual}, the role of electric and magnetic gauge couplings is reversed in passing from \eqref{N2_action} to \eqref{N2_action_dual}. For this reason, we will refer to \eqref{N2_action_dual} as the \emph{electromagnetic action dual} to \eqref{N2_action}.
	 \footnote{Such dualities that relate strongly and weakly coupled regimes of a theory were dubbed $U$-dualities in \cite{Andrianopoli:2006ub} in analogy to those studied in 
	 \cite{Hull:1994ys}.}
	 In the following, we will refer to the frame where the Type IIB EFTs action acquires the form \eqref{N2_action} as the \emph{electric frame}, understanding that the electric gauge couplings $g^2_{{\rm e},I}(t) = \mathcal{Q}^2_{{\bf q}_{\rm e}^{(I)}}$ are small, namely $g^2_{{\rm el},I}(t)\lesssim 1$; instead, we will refer to the frame where the action is \eqref{N2_action_dual} as the  \emph{magnetic} one, with gauge couplings $g^2_{{\rm m},I}(t) \gtrsim 1$.

\subsection{Generalized global symmetries of Type IIB EFTs}
\label{sec:GGS_IIBsub}

In this section, we examine the generalized global symmetries that characterize the  $\mathcal{N}=2$ Type IIB EFTs reviewed in Section~\ref{sec:IIB_review}. 
The philosophy that we follow is the one outlined earlier in Section~\ref{sec:GGS_avoid}: within the Type IIB EFTs, we first single out the conserved currents associated to the global symmetries that the EFTs exhibit and then inquire whether one can identify well-defined topological operators -- i.e. the charge operators -- for said global symmetries.

To begin with, we investigate the global symmetries that are associated to the gauge sector of $\mathcal{N}=2$ Type IIB EFTs.

\noindent\textbf{Electromagnetic one-form global symmetries.} As for the simple free Maxwell theory examined in Section~\ref{sec:GGS_avoid}, the $\mathcal{N}=2$ Type IIB EFTs may be endowed with abelian $U(1)$ one-form global symmetries. Indeed, the electric-frame action \eqref{N2_action} is trivially invariant under shifts of the gauge one-forms $A^I$ by flat connections, namely $A^I \to A^I + \Lambda^I_{\rm e}$ with ${\rm d} \Lambda^I_{\rm e} = 0$. 
Analogously, the magnetic-frame action \eqref{N2_action_dual} is invariant under shifts of the magnetic gauge one-forms $V_I$, as $V_I \to V_I + \Lambda_I^{\rm m}$ with ${\rm d} \Lambda_I^{\rm m} = 0$. 
Therefore, the (maximal) global one-form symmetry group is
\begin{equation}
	\label{GGSIIB_G}
	G_{1} = U(1)^{h^{2,1}+1}_{\rm e} \times U(1)^{h^{2,1}+1}_{\rm m}\,,
\end{equation}
where, as in Section~\ref{sec:GGS_avoid}, we have introduced the subscripts to distinguish whether the symmetry subgroup is associated to an electric or a magnetic one-form symmetry.

In order to tell whether the group \eqref{GGSIIB_G} is broken to a subgroup within the Type IIB EFTs, we need to examine the structure of the associated currents. 
From the general discussion of Section~\ref{sec:GGS_QG}, the maximal global symmetry group \eqref{GGSIIB_G} may lead to, at most, $b_3(Y) = 2h^{2,1}+2$ two-form currents.
However, the simultaneous presence of multiple $p$-forms, namely the gauge one-forms $A^I$ and the axion zero-forms $a^i$, renders the identification of well-defined currents rather subtle.
Indeed, as recognized in \cite{Marolf:2000cb}, one can identify \emph{three} different kinds of currents that in \cite{Marolf:2000cb} were referred to as the  \emph{Page}, the \emph{Maxwell} and the \emph{brane currents}.  
These definitions of currents were further employed in \cite{Heidenreich:2020pkc,Heidenreich:2021xpr}, since they provide a neat classification of the available currents when the theory is endowed with couplings that involve several field strengths of the gauge fields.
In the following, we show how such three notions of charges appear in the Type IIB EFTs under examination, and illustrate their key properties.

The simplest choice of currents is the one obtained by the straightforward application of Noether's first theorem for the group \eqref{GGSIIB_G}. Applied to the electric one-form symmetries, starting from the action \eqref{N2_action}, Noether's first theorem  leads to the $h^{2,1}+1$ electric two-form currents
\begin{equation}
	\label{GGSIIB_Jel}
	(J_I^{\text{e}})^{(P)} = G_I = {\rm Im} \mathcal{N}_{IJ} *\! F^J  + {\rm Re} \mathcal{N}_{IJ} F^J\,.
\end{equation}
Instead, applying Noether's theorem to the magnetic subgroup of \eqref{GGSIIB_G} delivers the magnetic two-form currents
\begin{equation}
	\label{GGSIIB_Jm}
	(J^I_{\text{m}})^{(P)}  = F^I \,,
\end{equation}
as can be straightforwardly checked from the dual action \eqref{N2_action_dual}, with the help of \eqref{N2_N_inv}. We can collect the two-form currents in the vector ${\bm J}^{(P)} = ((J^I_{\text{m}})^{(P)}, -(J_I^{\text{e}})^{(P)})^T$. From the equations of motion for the electric one-forms $A^I$, and their Bianchi identities, the currents so defined are conserved on-shell:
\begin{equation}
	\label{GGSIIB_JPcons}
	{\rm d} {\bm J}^{(P)} = 0 \,.
\end{equation}
Are the currents ${\bm J}^{(P)}$ \emph{good} currents that lead to well-defined charge operators? In general, the currents ${\bm J}^{(P)}$ suffer from a pathology: they are not single-valued. 
In fact, recalling the discussion in Section~\ref{sec:IIB_Sympl}, the currents ${\bm J}^{(P)}$ coincides with the field strength vector \eqref{Sympl_FG}.
Therefore, the current vector ${\bm J}^{(P)}$ is a section of the flat symplectic bundle $Sp(2h^{2,1}+2, \mathbb{Z})$, and thus it does transform under general symplectic transformations.
In particular, the currents change also under isometries of the scalar manifold, such as the monodromy transformations introduced in Section~\ref{sec:IIB_Sympl} as ${\bm J}^{(P)} \to T {\bm J}^{(P)}$. 
Therefore, though conserved, the currents ${\bm J}^{(P)}$,  are \emph{not} generically gauge invariant. 
For these reasons, following \cite{Marolf:2000cb}, we will refer to the currents ${\bm J}^{(P)}$ as the \emph{Page currents}.
In the concrete examples that we will consider in Sections~\ref{sec:GGS_01mod} we will see that this is partly due to the explicit axion dependence of the currents \eqref{GGSIIB_Jel} on the axion fields entering the EFT.

Clearly, such an ambiguity in the definition of the currents is also reflected in the charge operators. Given a closed two-dimensional surface $\Sigma_2$, one can define the Page charge operators
\begin{equation}
	\label{GGSIIB_QM}
	{\bm Q}^{(P)} = \int_{\Sigma_2} {\bm J}^{(P)} \,.
\end{equation}
As we will motivate shortly, the Page charges are localized charges and are expected to be quantized \cite{Marolf:2000cb}. 
For fixed scalar fields $t^i = t^i_0$ each charge operator within ${\bm Q}^{(P)}$ is topological. However, after the axionic shifts $a^i \to a^i+1$, the charge operators nontrivially transform as ${\bm Q}^{(P)} \to T {\bm Q}^{(P)}$.

It is worth remarking that whether \emph{all} or \emph{some} of the Page currents ${\bm J}^{(P)}$ (or the corresponding Page charges ${\bm Q}^{(P)}$) are not single-valued depends on the explicit form of the monodromy transformation $T$.
Hence, among the Page currents or charges, there could be some that are both conserved and single-valued. 
These ought to be considered as stemming from genuine global symmetries of the model and, by the No Global Symmetry Conjecture, should be removed in a full quantum gravity theory. 
We will comment on how to consistently break these symmetries later in this section.

Since the Page currents and charges are generically ill-defined, one could wonder whether it is possible to define some alternative single-valued conserved currents. Single-valued currents and charges can be most readily defined out of the Page currents and charges as
\begin{equation}
	\label{GGSIIB_JQM}
	{\bm J}^{(M)}  = e^{-a^i N_i} {\bm J}^{(P)}  \,, \qquad  {\bm Q}^{(M)} = \int_{\Sigma_2} {\bm J}^{(M)} \,,
\end{equation}
where $N_i$ are the matrices defined in \eqref{Sympl_Nlogi}. We will refer to \eqref{GGSIIB_JQM} as \emph{Maxwell currents} and \emph{Maxwell charges}, respectively.
In fact, under the integral axion shift $a^i \to a^i+1$, the prefactor in the definition of the above currents gets mapped according to $e^{-a^i N_i} \to e^{-a^i N_i} T^{-1}$, undoing the monodromy transformation of the Page currents ${\bm J}^{(P)}$ and rendering \eqref{GGSIIB_JQM} single-valued.
However, the Maxwell currents do suffer from a novel issue: in a background with dynamical axions, they are not generically conserved, for in general ${\rm d} {\bm J}^{(M)} \neq 0$  and the Maxwell charges ${\bm Q}^{(M)}$ are not topological operators.\footnote{Notice that in \cite{Marolf:2000cb}, Maxwell currents are defined as non-localized, gauge-invariant conserved currents. Here we adopt the weaker definition of Maxwell charges used in \cite{Heidenreich:2020pkc}, where Maxwell currents are gauge-invariant but not generically conserved.}

To summarize, in a dynamical axion background, the sole well-defined global symmetries are those which deliver currents that are simultaneously conserved, gauge-invariant and single-valued. 
In order to break the global symmetries at the EFT level, we can employ the last type of currents, the \emph{brane currents}. 
Within the $\mathcal{N}=2$ Type IIB EFTs described by the action \eqref{N2_action}, the brane currents can originate from two sources. The first kind of brane current stems from the direct coupling of the Type IIB EFTs \eqref{N2_action} with the electric D3-particles effectively described by the action \eqref{IIB_SD3}. 
In the presence of $N_{\rm e}$ electrically charged D3-particles, the on-shell current conservation rules for the Page electric currents \eqref{GGSIIB_Jel} get modified to 
\begin{equation}
	\label{GGSIIB_Jbe}
	{\rm d} (J_I^{\text{e}})^{(P)}  = -\sum\limits_{\alpha = 1}^{N_{\rm e}} q_I^\alpha \delta^{(3)} (\Gamma_\alpha) =: - j_I^{\text{e}} 
\end{equation}
where $q_I^\alpha$ denotes the elementary electric charges of the D3-particle drawing the worldline $\Gamma_\alpha$. Additional localized sources can be included in order to break the Bianchi identities for the field strengths $F^I$ as
\begin{equation}
	\label{GGSIIB_Jbm}
	{\rm d} F^I  = {\rm d} (J^I_{\text{m}})^{(P)}  = -\sum\limits_{\beta = 1}^{N_{\rm m}} p^I_\beta \delta^{(3)} (\Gamma^\beta) =: - j^I_{\text{m}} 
\end{equation}
These localized sources can be understood as stemming from $N_{\rm m}$ magnetic D3-particles, with elementary charges $p^I_\beta$ and worldlines $\Gamma^\beta$, coupled to the dual effective action \eqref{N2_action_dual}. The relations
\eqref{GGSIIB_Jbe} and \eqref{GGSIIB_Jbm} can be more compactly written as
\begin{equation}
	\label{GGSIIB_Jb}
	{\rm d} {\bm J}^{(P)} =  - {\bm j}\,, \qquad {\bm j} := (j^I_{\text{m}}, - j_I^{\text{e}} )\,,
\end{equation}
and we will refer to ${\bm j}$ as \emph{brane currents}. These currents are localized by assumption, and consequently, the Page charges in \eqref{GGSIIB_QM} are localized. Breaking the conservation rule for the currents explicitly, the brane currents are the prime candidates that break any residual electromagnetic one-form global symmetry.
It is worth noting, however, that generic brane currents may break the continuous global symmetries down to a discrete subgroup thereof; choosing the brane currents as composed of unit charges breaks the continuous group entirely.

One could then inquire what is the minimal choice of brane currents that could lead to the breaking of a global symmetry. Specifically, let us assume that a brane with unit charge is included so as to break the conservation rule of one of the Page currents: is this configuration consistent?
As mentioned earlier, the Page current ${\bm J}^{(P)}$ is not single-valued under monodromy transformations. For instance, shifting every axion as $a^i \to a^i +1$, \eqref{GGSIIB_Jb} maps to ${\rm d} {\bm J}^{(P)} =  - T {\bm j}$. In other words, the brane currents are also required to transform under generic monodromies.  
This hints at the fact that the inclusion of single brane current ${\bm j}$ is generically inconsistent, for also the currents $T {\bm j}$, $T^2 {\bm j}$, \ldots should be present in the EFT. 
This phenomenon has profound phenomenological which will be examined in the following section.

Having exhausted the discussion of the electromagnetic one-form global symmetries, let us pass to the global symmetries that can stem from the scalar sector of the $\mathcal{N}=2$ Type IIB EFTs.
	
\noindent\textbf{Axionic zero-form global symmetries.} Let us assume that the K\"ahler metric is axion-independent, so that the $\mathcal{N}=2$ Type IIB EFTs acquires the form \eqref{N2_action_realm}. The scalar kinetic terms of the action \eqref{N2_action_realm} are then ostensibly invariant under any shift of the axion fields $a^i \to a^i + c^i$, with $c_i \in \mathbb{R}$. We further assume that the axions $a^i$ span a compact domain, namely circles of unit circumference, so that we may regard $c_i \in \mathbb{R}/\mathbb{Z}$. Then, the axion fields can be understood as zero-form gauge fields, and they can be responsible for zero-form global symmetries, that act on the axion via said shift symmetry. The maximal abelian global symmetry group that can be associated to such zero-form global symmetries is then:
\begin{equation}
	\label{GGSIIB_G0}
	G_{0} = U(1)^{h^{2,1}}_{a}\,.
\end{equation}
The following, gauge-invariant Maxwell currents can be defined:
\begin{equation}
		\label{GGSIIB_Ja}
		J_{a,i}^{(M)} = M_{\rm P}^2 K_{ij}^{\rm cs} * {\rm d} a^{j}\,.
\end{equation}
However, in general, these currents are not conserved due to the explicit dependence of the gauge interactions on the axion fields $a^i$:
\begin{equation}
	\label{GGSIIB_Ja_cons}
	{\rm d} J_{a,i}^{(M)} = -\frac{1}{2} \frac{\partial}{\partial a^i} G_I \wedge F^I\,.
\end{equation}
If ${\rm d} J_{a,i}^{(M)} = 0$ for some $i$, then the zero-form global symmetry is a genuine global symmetry of the EFT.
As we show in details in Appendix~\ref{sec:Hodge}, we should expect that these continuous axion shift symmetries are at most approximate in the EFT, for they may be broken to a discrete subgroup by some `essential' instanton effects. 
Alternatively, if those essential instantons are absent, the axion zero-form global symmetry can be explicitly broken by including some brane currents composed by $n_{i}$ fundamental instantons as:
\begin{equation}
	\label{GGSIIB_Ja_consb}
	{\rm d} J_{a,i}^{(M)} = - j_{a,i}\,, \qquad j_{a,i} = \sum\limits_{\alpha = 1}^{n_{i}} m_\alpha \delta^{(4)}(x^\alpha_0) \,,
\end{equation}
where we have denoted with $m_\alpha$ the elementary charges of the fundamental instanton localized at the spacetime point $x^\alpha_0$.

\noindent\textbf{Two-form global symmetries.} Before concluding this section, let us briefly comment on one last kind of generalized global symmetries that the Type IIB EFTs may be endowed with and should be thus removed. 
These are the two-form generalized global symmetries that can be understood as the electromagnetic duals of the axionic zero-form global symmetries discussed above.
They are associated to the one-form currents $J_{B}^i = {\rm d} a^i$, which are trivially conserved.
In order to understand why such symmetries may be also present in the models at hand, let us first consider the simple case in which the EFT is endowed with an axionic zero-form global symmetry that shifts the axion $a^i \to a^i + c$, for some $c \in \mathbb{R}$ -- namely, the K\"ahler metric is independent of the axion $a^i$ and the axion $a^i$ does not enter any generalized $\theta$-term.
Then, one could perform a dualization procedure as outlined in Appendix~\ref{app:conv_dual}, replacing the axion $a^i$ with its dual, two-form gauge field $B_{2,i}$
(see also \cite{Lanza:2020qmt,Lanza:2021udy,Lanza:2022zyg}
for details about such a dualization procedure, and how this can be performed in a manifestly supersymmetric way).
Then, proceeding as in Section~\ref{sec:GGS_avoid} for the one-form global symmetries, one could obtain the conserved one-form currents $J_{B}^i = {\rm d} a^i$ by applying the standard Noether procedure.

However, in the more general Type IIB EFTs axions participate in the generalized $\theta$-terms, and the associated continuous shift symmetries are broken down to discrete ones.
Consequently, as also highlighted in the recent \cite{Martucci:2022krl}, the standard dualization procedure summarized in Appendix~\ref{app:conv_dual} cannot be generally performed, and the Noether procedure cannot be straightforwardly applied in order to compute the eventual one-form currents.
Nevertheless, it is worth remarking that the one-form currents $J_{B}^i = {\rm d} a^i$ still appear to be conserved, and well-defined -- see \cite{Brennan:2020ehu,Heidenreich:2021yda}. This could signal that the Type IIB EFTs that we are considering are endowed with two-form global symmetries, which might be broken by including strings with elementary charges. 
However, in the presence of $\theta$-terms, the inclusion of fundamental strings comes at a price: as pointed out in \cite{Heidenreich:2021yda}, fundamental strings can only be consistently coupled provided that they carry some additional string worldsheet degrees of freedom, whose existence is hard to justify from a top-down perspective (see also \cite{Marchesano:2022axe} for a recent discussion in the context of $\mathcal{N}=2$ Type IIA EFTs).
We leave a better understanding of how these two-form global symmetries are realized in Type IIB EFTs, and how they can be avoided for future investigation.

\subsection{A generalized Witten effect for Type IIB EFTs}
\label{sec:WE_IIB}

The four-dimensional $\mathcal{N}=2$ Type IIB EFTs are plagued by the presence of various topological Chern-Simons terms, of the form $F^I \wedge F^J$, for some $I, J$, that are nontrivially coupled to the axions $a^i$, residing in ${\rm Re} \mathcal{N}_{IJ}$, via the Lagrangian term
\begin{equation}
	\label{WE_gentheta}
	\frac12 {\rm Re} \mathcal{N}_{IJ}(a) F^I \wedge F^J
\end{equation}
as appears in \eqref{N2_action}. Such terms are reminiscent of the more commonly known QCD \emph{$\theta$-terms}, of the form $\theta {\rm Tr} (F \wedge F)$, for some non-abelian field strength $F$ and $\theta \in \mathbb{R}$, that were proposed to help solve the strong $CP$-problem \cite{Peccei:1977ur,Peccei:1977hh}. For this reason, we will refer to the more general couplings \eqref{WE_gentheta} as \emph{generalized $\theta$-terms}.
As mentioned in the previous section, such couplings do not allow for a correct identification of some of the conserved currents, \emph{de facto} breaking them. 

However, this is not the only role of generalized $\theta$-terms \eqref{WE_gentheta}. 
In fact, a phenomenon similar to the Witten effect reviewed in Section~\ref{sec:GGS_gen_mod_WE} is triggered by the generalized $\theta$-terms \eqref{WE_gentheta} and, for this reason, we will refer to this as \emph{generalized Witten effect}.
In order to see in detail how this generalized Witten effect takes place in the Type IIB EFTs under examination, let us consider the equations of motion \eqref{GGSIIB_Jbe}-\eqref{GGSIIB_Jbm} in the presence of sources. After a monodromy transformation as in \eqref{Sympl_Ttransf}, these get mapped to
\begin{subequations}
\begin{align}
	&A^I{}_J {\rm d} F^J =- j^I_{\text{m}}\,, \label{WE_IIB_m}
	\\
	&D_I{}^J {\rm d} \left({\rm Im} \mathcal{N}_{JL} *\! F^L  + {\rm Re} \mathcal{N}_{JL} F^L + (D^{-1} C)_{JL} F^L \right)  =- j_I^{\text{e}}\,. \label{WE_IIB_e}
\end{align}
\end{subequations}
Let us first assume that there are no magnetic sources, i.e.~$j^I_{\text{m}} = 0$. Then, \eqref{WE_IIB_e} signals that, whenever the electric brane source $j_I^{\text{e}}$ is present in the spectrum, then also the brane source $(D^{-1})_I{}^J j_I^{\text{e}}$ has to be present. However, one can perform multiple monodromy transformations. As a result, due to monodromies induced by the Witten effect, a \emph{tower} of brane sources, namely  $j_I^{\text{e}}$, $(D^{-1}  j^{\text{e}})_I$, $(D^{-1} D^{-1}  j^{\text{e}})_I$, $\ldots$~has to be allowed by the EFT spectrum by solely assuming that one of these sources exists.

In the presence of magnetic sources, a similar argument can be performed: from the transformation law of the magnetic equations of motion \eqref{WE_IIB_m}, one finds that, if the magnetic sources $j^I_{\text{m}}$ are allowed by the spectrum, then also $(A^{-1} j_{\text{m}})$, $(A^{-1} A^{-1}  j_{\text{m}})$, etc.~are allowed. On the other hand, the electric brane sources are turned into dyons after a monodromy transformation, for they get mapped to
\begin{equation}
	j_I^{\text{e}} \to (D^{-1})_I{}^J j_J^{\text{e}} - (D^{-1})_I{}^J C_{JL} (A^{-1})^L{}_K j^K_{\text{m}}\,,
\end{equation}
thus delivering an infinite tower of dyonic brane sources.

The combined interplay of the Witten effect and nontrivial axion monodromies, therefore, has important consequences on the composition of the spectrum of any EFT.
Let us assume that a localized, dyonic brane `\emph{seed current}' ${\bm j}_{\rm seed}$ is present in the spectrum:
\begin{equation}
	\label{WE_jseed}
	{\bm j}_{\rm seed}^T = {\bf q}_{\rm seed}^T \delta^{(3)}(\Sigma) =  (q^I_{\rm m,seed}, - q^{\rm e}_{I, {\rm seed}}) \delta^{(3)}(\Sigma)\,.
\end{equation}
Such a seed charge may be given, for instance, by one of the charges that are required to break the one-form global symmetries.
For instance, the seed current may be induced by a dyonic BPS D3-particle following the worldline $\Sigma$, with elementary charge ${\bf q}_{\rm seed}$ and mass given by \eqref{IIB_D3Mb}. 
Moreover, we will assume that such a D3-particle does not form a black hole. 
Thus, we shall assume that it is characterized by small elementary charges, and cannot be resolved within the EFT; the latter requirement is achieved by imposing
\begin{equation}
	\label{WE_Mcutoff}
	M_{{\bf q}_{\rm seed}} \gtrsim \Lambda_{\text{\tiny EFT}}\,.
\end{equation}
As shown in the previous section, due to the Witten effect, the axion-monodromy induced currents $T^{-1} {\bm j}_{\rm seed}$, $T^{-1} T^{-1} {\bm j}_{\rm seed}$, $\ldots$ also ought to be allowed by the effective description.
In turn, this implies that BPS D3-particles with charges 
\begin{equation}
	\label{WE_qtower}	
	{\bf q}^{(k)} = (T^{-1})^k {\bf q}_{\rm seed}
\end{equation}
should also be allowed by the spectrum. 
Namely, the EFT charge lattice $\Gamma_{\text{\tiny EFT}}$ of D3-brane charges ought to contain \emph{at least} the monodromy-induced sublattice
\begin{equation}
	\label{WE_GammaT}
	\Gamma_{T} = \{  {\bf q} \in \mathbb{Z}^{b_3(Y)} \; | \; {\bf q} = (T^{-1})^k {\bf q}_{\rm seed} \; \text{for some $k \in \mathbb{N}$}\,  \} \subset \Gamma_{\text{\tiny EFT}}\,.
\end{equation}
In particular, if the monodromy matrix $T$ is of infinite order, the sublattice $\Gamma_{T}$ is of infinite cardinality. The masses of the BPS D3-particles constituting the monodromy-induced sublattice \eqref{WE_GammaT} can be computed from \eqref{IIB_D3Mb} as
\begin{equation}
	\label{WE_Mseedb}
	M_{{\bf q}^{(k)}} = e^{\frac{K^{\rm cs}}{2}} |( (T^{-1})^k {\bf q})^T\,\eta\, {\bm \Pi}(t)|\,.
\end{equation}
%


\section{Global symmetries in the rigid case and one-modulus asymptotic limits}
\label{sec:GGS_01mod}

The analysis of the Type IIB EFTs global symmetries carried out in the previous section was general, and mostly based on the features shared by the underlying $\mathcal{N}=2$ local supersymmetry.
However, we would like to be more specific about the global symmetries that can emerge in concrete stringy EFTs, and how these are avoided.
We will be interested in the Type IIB EFTs defined in field space regions close to singularities in the complex structure moduli space $\mathcal{M}_{\rm cs}$. Within such regions, the EFTs are expected to be weakly-coupled and well-controlled.
The physics of these near-boundary EFTs can be studied in detail by means of \emph{asymptotic Hodge Theory}. 
Established in \cite{MR840721}, asymptotic Hodge Theory has proven to be a powerful tool for examining the structure of string vacua and characteristics of the BPS spectrum in the asymptotic regions of moduli space \cite{Grimm:2018ohb,Grimm:2018cpv,Grimm:2019bey,Grimm:2019ixq,Grimm:2020cda,Grimm:2021ckh,Grana:2022dfw,Bastian:2020egp,Grimm:2021ikg,Bastian:2021eom,Bastian:2021hpc,Grimm:2021idu}. 
A brief overview of asymptotic Hodge Theory is contained in Appendix~\ref{sec:Hodge}, where we further highlight how the couplings of the vector multiplet sector can be studied in full generality exploiting the tools that the Hodge theory provides.

For ease of exposition, in this section, we limit ourselves to the investigation of the generalized global symmetries in some simple Type IIB EFTs defined in the near-boundary regime. 
Specifically, we will first examine the global symmetries of the gravity multiplet sector of $\mathcal{N}=2$ EFTs only, without any moduli, after which we will consider the global symmetries emerging in the $\mathcal{N}=2$ EFTs  characterized by $h^{2,1}=1$ towards any field space boundary.
The computations in this section rely on the general framework of asymptotic Hodge Theory reviewed in Appendix~\ref{sec:Hodge}, and on the general expressions of the one-modulus-dependent gauge interactions that are computed in Appendix~\ref{app:Periods_One_Mod}.

\subsection{The rigid case}
\label{sec:GGS_0mod}

As a warm-up for the upcoming sections, let us first consider the simplest EFT that can be described by the action \eqref{N2_action}: namely, a theory \emph{without} moduli, only describing the interaction among the components of the $\mathcal{N}=2$ gravity multiplet. The zero-moduli case can be considered as originating from the compactification of Type IIB string theory over a rigid Calabi-Yau $Y$. Formally, this case can be thought of as stemming from the prepotential $\mathcal{F} = \frac{c}2 (X^0)^2$, for some $c \in \mathbb{C}$. Without loss of generality, we will take $c = \theta - \im$, with $\theta \in \mathbb{R}$. Then, the action  \eqref{N2_action} reduces to
\begin{equation}
	\label{Zero-m_S}
	\begin{aligned}
		S &= \int \Big( \frac{1}2 M^2_{\rm P} R *1 - \frac12  F^0 \wedge * F^0 + \frac{\theta}2  F^0 \wedge * F^0 \Big)\,,
	\end{aligned}
\end{equation}
where $F^0 = {\rm d}A^0$, with $A^0$ being the graviphoton gauge one-form.
The action \eqref{Zero-m_S} might be understood as the deep infrared limit of \eqref{N2_action}, in the case that the latter is endowed with a mechanism that renders the $h^{2,1}$ vector multiplets massive. The sole equation of motion for the graviphoton field now reads ${\rm d} *\! F^0 = 0$. 

The gauge sector contributions to the action \eqref{Zero-m_S} are the same as those appearing in \eqref{WE_S}, and thus the analysis of the global symmetries of the model closely follows that of Section~\ref{sec:WE_mon}. Indeed, the action \eqref{Zero-m_S} admits two one-form global symmetries: an electric one, associated to the two-form Maxwell current $J_2^{\rm e} = * F^0$ that is conserved by the graviphoton equations of motion; and a magnetic one, with Maxwell current $J_2^{\rm m} = F^0$. In order to break both these global symmetries, it is enough to assume that there exists a D3-brane, wrapping a rigid three-cycle in $Y$, that leads to a particle magnetically charged under the graviphoton field in the external space. Namely, assuming the D3-particle has elementary magnetic charge $q_{\rm m}$ and spans a worldline $\Gamma$, we find that it breaks the conservation of the magnetic two-form current as ${\rm d} J_2^{\rm m} = - q_{\rm m} \delta^{(3)}(\Gamma)$. 
However, as stressed in Section~\ref{sec:GGS_avoid}, the generic inclusion of a particle with charge $q_{\rm m}$ would break the magnetic $U(1)_{\rm m}$ to a discrete $\mathbb{Z}_q$ subgroup. In order to ensure that the $U(1)_{\rm m}$ global symmetry group is broken completely, a particle with unit magnetic charge is required.

As illustrated in Section~\ref{sec:WE_mon}, the Witten effect induced by the $\theta$-term implies that an infinite tower of electric states, with brane charges $q_{\rm e} = - n q_{\rm m}$ with $n \in \mathbb{Z}$, has to exist. 
Indeed, such a tower of electric D3-particles breaks the conservation law of the Page two-form current as 
\begin{equation}
    {\rm d} (J_2^{\rm e})^{(P)} = {\rm d} (-* F^0 + \theta F^0) = \sum\limits_{n \in \mathbb{Z}} n q_{\rm m} \delta^{(3)}(\Gamma)\,.
\end{equation}

\subsection{The one-modulus case: the large complex structure point}
\label{sec:GGS_1mod_IV}

We now investigate the global symmetries emerging in EFTs equipped with a complex one-dimensional moduli space. 
The first class of EFTs that we examine is the familiar case of EFTs defined at large complex structure (LCS).
In the following, after first determining the most general EFT action defined at any large one-modulus complex structure point, we examine which global symmetries such EFT is endowed with, and how these can be broken.

\noindent \textbf{The LCS EFT.} In the language of Appendix~\ref{sec:Hodge_review}, the LCS point corresponds to a singularity of Type $\text{IV}_1$ within the moduli space.
Given the single complex structure modulus $t = a + \im s$ of the internal Calabi-Yau $Y$, the LCS point is reached as $s \to \infty$, and we are interested in the EFT defined in regions of very large saxion, $s \gg 1$.
In the mirror-dual Type IIA setting, the saxion $s$ can be alternatively understood as parametrizing the volume of the mirror-dual Calabi-Yau $\hat{Y}$, and the region of interest is the one where the internal volume is very large.

As outlined in Appendix~\ref{sec:Hodge_review}, any near-boundary EFT can be determined by specifying a handful of boundary data. 
In this section, we only outline the main results, and we refer to Appendix~\ref{app:ivhodge} for details. The first ingredient that we need is the nilpotent operator $N$ that realizes the weight filtration \eqref{HT_Wl}. It can be shown that the most general operator $N$ for a Type IV singularity is \cite{Green:2008}
\begin{equation}
\label{GGS_1mod_IV_N}
    N=\begin{pmatrix}
    0 & 0 & 0 & 0\\
    m & 0 & 0 & 0\\
    c & b & 0 & -m\\
    b & n & 0 & 0
    \end{pmatrix},\quad \text{with }\begin{cases}
    m,n\in\mbb{Z},\\
    b+mn/2\in\mbb{Z},\\
    c-m^2n/6\in\mbb{Z},\\
    m\neq 0, \,n>0.
    \end{cases}
\end{equation}
The above constraints on the parameters $b$, $c$, $m$, $n$ guarantee that ${\rm rk} N = 3$, as required by a Type IV singularity, and that the monodromy matrix $T = e^N$ displays integral entries. 
The matrix $N$ can be most readily understood as the lowering operator $N_- = N$ of the sole ${\rm sl}_2$-triple \eqref{HT_sl2_triples} associated to the Type IV boundary.
Secondly, we need to characterize the $F_0^3$ filtration, which allows for reconstructing the whole filtration. The most general representative thereof is \cite{Green:2008}
\begin{equation}
\label{GGS_1mod_IV_a0}
    {\bf a}_0 = \left(1,0,\xi, \frac{c}{2m}\right)^T\,,
\end{equation}
where $\xi$ is, in general, a complex parameter. For the sake of generality, we shall leave the parameters $b$, $c$, $m$, $n$, $\xi$ unspecified. However, in concrete stringy EFTs they are determined by the geometric data of the internal Calabi-Yau manifold $Y$.
Equipped with \eqref{GGS_1mod_IV_N} and \eqref{GGS_1mod_IV_a0}, it is immediate to show that the nilpotent orbit approximation of the periods \eqref{HT_Per_nil} leads to
\begin{equation}
\label{GGS_1mod_IV_period}
    {\bm \Pi}_{\rm nil}=e^{tN}\bf{a}_0=\begin{pmatrix}
    1\\
    m t\\
    -\frac16 m^2nt^3+ \frac12 ct +\xi\\
    \frac12 mnt^2 +bt+\frac{c}{2m}
    \end{pmatrix}.
\end{equation}
It is worth noticing that the periods \eqref{GGS_1mod_IV_period} can be regarded as originating from the following prepotential 
\begin{equation}
\label{GGS_1mod_IV_prepot}
    \mc{F}_{\text{\tiny{LCS}}} =-\frac{n}{6m} \frac{(X^1)^3}{X^0}-\frac{b}{2m} (X^1)^2-\frac{c}{2m} X^1X^0-\frac{1}{2}\xi (X^0)^2
\end{equation}
upon gauge-fixing $X^0 =1$, $X^1 = m t$.

Following the procedure outlined in Appendix~\ref{app:ivhodge}, it can be shown that the boundary data \eqref{GGS_1mod_IV_N} and \eqref{GGS_1mod_IV_a0} yield the following expression for matrix \eqref{IIB_NIJb} that dictates the interactions of the gauge sector:
\begin{equation}
\label{GGS_1mod_IV_cN}
    \cN_{\text{\tiny{LCS}}} =\begin{pmatrix}
    -\frac{1}{3}m^2na^3-\R\xi & \frac{1}{2}mna^2-\frac{c}{2m}\\
    \frac{1}{2}mna^2-\frac{c}{2m} & -na-\frac{b}{m}
    \end{pmatrix}+\begin{pmatrix}
        -\frac{1}{6}m^2ns(3a^2+s^2) & \frac{1}{2}mnas\\
        \frac{1}{2}mnas & -\frac{n}{2}s
    \end{pmatrix}.
\end{equation}
Therefore, the explicit expressions \eqref{GGS_1mod_IV_period} and \eqref{GGS_1mod_IV_cN} lead to the following EFT action defined towards any arbitrary LCS point: 
\begin{equation}
\label{GGS_1mod_IV_S}
\small{\begin{aligned}
    S_{\text{\tiny{LCS}}} &= \int \Big( \frac{1}2 M^2_{\rm P} R *1 - M^2_{\rm P} \frac{3}{4 s^2} \frac{1 - \frac{3 {\rm Im} \xi}{m^2 n s^3}}{ \left(1 - \frac{3 {\rm Im} \xi}{2 m^2 n s^3}\right)^2} ({\rm d} a \wedge * {\rm d} a + {\rm d} s \wedge * {\rm d} s)  
    \\
    &\qquad\quad - \frac1{12} m^2 n s (3 a^2 + s^2) F^0 \wedge * F^0 + \frac1{2} m n a s F^0 \wedge * F^1 - \frac1{4} n s  F^1 \wedge * F^1
    \\
    &\qquad\quad - \frac1{6} (m^2 n a^3 + 3 {\rm Re} \xi) F^0 \wedge F^0 + \frac1{2 m} \left( m^2 n a^2 - c \right) F^0 \wedge F^1 - \frac1{2m} (b + mn a)   F^1 \wedge F^1
    \Big)\,.
\end{aligned}}
\end{equation}

Remarkably, the EFT action above exhibits generalized $\theta$-terms couplings of the form as in \eqref{WE_gentheta} with the axion appearing with a \emph{cubic} power, rather than via the more standard linear axion coupling.\footnote{Notice that, in order to get generalized $\theta$-terms with couplings that are at most quadratic in the axions, the one-complex structure modulus models examined in this work are not enough.
One should rather consider EFT defined towards boundaries where multiple saxions acquire large vevs.}

In the following, we assume that the saxion $s$ is not dynamical and fixed at a given vev $s \gg 1$.

\noindent \textbf{The LCS one-form global symmetries.} The LCS effective action \eqref{GGS_1mod_IV_S} is endowed with, at most, four one-form global symmetries. These are associated to the invariance of the effective action \eqref{GGS_1mod_IV_S} under transformations of the electric gauge one-forms as $A^I \to A^I + \Lambda^I$, with ${\rm d}\Lambda^I = 0$, and its magnetic counterpart. The corresponding two-form Page currents are the following
\begin{equation}
\label{GGS_1mod_IV_PageJ}
\small{\begin{aligned}
    (J^0_{\rm m})^{(P)} &= F^0\,,
    \\
    (J^1_{\rm m})^{(P)} &= F^1\,,
    \\
    (J_0^{\rm e})^{(P)} &= G_0 := - \frac{m^2 n s}{6} (3 a^2 + s^2) * F^0 + \frac{m n a s}2  * F^1 - \frac{m^2 n a^3 + 3 {\rm Re} \xi}{3} F^0 + \frac{m^2 n a^2 - c}{2 m} F^1\,,
    \\
    (J_1^{\rm e})^{(P)} &= G_1 := \frac{m n a s}{2}  * F^0 - \frac{n s}{2}   * F^1 + \frac{ m^2 n a^2 - c }{2 m} F^0 - \frac{b + mn a}{m}  F^1\,,
\end{aligned}}
\end{equation}
which are conserved by virtue of the Bianchi identities and the equations of motion of the gauge fields $A^I$.

Indeed, as shown in Section~\ref{sec:WE_mon}, the Page currents \eqref{GGS_1mod_IV_PageJ} are not in general single-valued, for they transform as
\begin{equation}
\label{GGS_1mod_IV_PageJtr}
\begin{aligned}
    \begin{pmatrix} (J^I_{\rm m})^{(P)} \\ -(J_I^{\rm e})^{(P)} \end{pmatrix} \quad \xrightarrow{a \to a + 1} \quad e^N \begin{pmatrix} (J^I_{\rm m})^{(P)} \\ -(J_I^{\rm e})^{(P)}\end{pmatrix}\,,
\end{aligned}
\end{equation}
under a unitary axion shift $a \to a + 1$. 
In particular, from the expression \eqref{GGS_1mod_IV_N} of the nilpotent matrix $N$, one can show that the Page currents $(J^1_{\rm m})^{(P)}$ and $(J_I^{\rm e})^{(P)}$ are not invariant under axion shifts and they are therefore ill-defined. 
The sole Page current that is single-valued is $(J^0_{\rm m})^{(P)} = F^0$,\footnote{Notice that this is the sole Page current among those in \eqref{GGS_1mod_IV_PageJ} that can also be regarded as a Maxwell current.} and can thus lead to a genuine, well-defined one-form $U(1)_{\rm m}$ global symmetry.
Such a symmetry can be broken explicitly by including D3-particles magnetically charged under the graviphoton field $A^0$. Minimally, it is enough to include a single of such particles, so that the Bianchi identity for the graviphoton breaks down to ${\rm d} F^0 = p^0 \delta^3(\Gamma)$, with $\Gamma$ the worldline the D3-particle describes and $p^0$ its charge. As shown in detail in Appendix~\ref{app:ivhodge}, in the vector basis in which the nilpotent operator is as in \eqref{GGS_1mod_IV_N}, one can choose $p^0 \in \mathbb{Z}$. For a general $p^0$, the $U(1)_{\rm m}$ group is broken to a discrete subgroup $\mathbb{Z}_{p^0}$; breaking the symmetry completely requires that the charge $p^0$ has to be the unit charge $p^0 =1$.

\noindent \textbf{The generalized Witten effect in LCS EFTs.} 
The effective action \eqref{GGS_1mod_IV_S} displays generalized $\theta$-terms that mix the graviphoton gauge one-form $A^0$ with the vector multiplet gauge one-form $A^1$.
It is worth remarking that such generalized $\theta$-terms appearing in \eqref{GGS_1mod_IV_S} are \emph{unavoidable} in the field space region where the action \eqref{GGS_1mod_IV_S} is valid. 
In fact, it can be shown that, after canonically normalizing both the axion $a$ and the gauge fields $A^0$ and $A^1$, the gauge kinetic terms for the one-form gauge fields $A^I$ and the generalized $\theta$-terms scale in the same fashion with the saxion $s$.
Thus, as explained in Section~\ref{sec:WE_IIB}, a generalized Witten effect takes place, also towards the limit $s\to \infty$.

Indeed, as we have seen above, the avoidance of the one-form global symmetries requires the existence of (at least) a `\emph{seed charge}' carrying elementary charge ${\bf q}_{\rm seed} = (1,0,0,0)^T$. 
According to the generalized Witten effect, the spectrum has to additionally contain all the states characterized by the elementary charges
\begin{equation}
\label{GGS_1mod_IV_qk}
    {\bf q}^{(k)} = e^{-kN} {\bf q}_{\rm seed} = \left(1, -km, \frac16(k^3 n m^2- 6kc), -\frac12 k(2b- kmn) \right)^T, \qquad \text{with $k\in \mathbb{Z}$}\,.
\end{equation}
These are dyonic states and, in particular, BPS D3-particle states with elementary charges \eqref{GGS_1mod_IV_qk} can be constructed. 
As can be checked from the general expression for the masses and physical charges \eqref{IIB_D3Mb} and \eqref{IIB_D3Q}, the associated BPS D3-particles are heavy and strongly coupled, with both masses and physical charges diverging as $s \to \infty$.

\noindent \textbf{The LCS axion global symmetries.} Before concluding the analysis of the LCS EFTs, let us briefly comment on the axion zero-form global symmetries. The Maxwell three-form current
\begin{equation}
\label{GGS_1mod_IV_Jax}
    J^{(M)}_a = - M_{\rm P}^2 \frac{3}{4 s^2} \frac{1 - \frac{3 {\rm Im} \xi}{m^2 n s^3}}{ \left(1 - \frac{3 {\rm Im} \xi}{2 m^2 n s^3}\right)^2} * {\rm d} a\,,
\end{equation}
associated to the continuous axion shift symmetry $a \to a + \chi$, $\chi \in \mathbb{R}$, is not conserved. The lack of conservation of \eqref{GGS_1mod_IV_Jax} is due to the axion-dependent gauge interactions appearing in the effective action \eqref{GGS_1mod_IV_S}. 
As such, the global zero-form axion $U(1)_{\rm a}$ is naturally broken to a discrete subgroup $\mathbb{Z}_N$, for some $N$.  Fully breaking the axion shift symmetry can be achieved by including an instanton of unit charge.

\subsection{The one-modulus case: Tyurin degeneration}
\label{sec:GGS_1mod_II}

The LCS point is not the only infinite-distance boundary that can be reached within a complex one-dimensional moduli space. Indeed, in the language of Appendix~\ref{sec:Hodge_review}, one finds that type II boundaries are similarly located at infinite geodesic distance. In analogy with \cite{Bastian:2021eom}, we shall call this class of boundaries `\emph{Tyurin degeneration}' because of their analogy with the Calabi-Yau degenerations first studied in \cite{arxiv.math/0302101} (see also \cite{Doran:2016uea,Joshi:2019nzi,Lee:2019wij} for later works that consider such classes of field space boundaries).
Following in the footsteps of the previous section, we illustrate which global symmetries emerge in the EFTs defined close to any Tyurin degeneration, and how these can be avoided.

\noindent \textbf{The Tyurin degeneration EFT.} We first determine the EFT describing the vector multiplets interactions close to a Tyurin degeneration, by means of the boundary data characterizing the field space boundary. 
We will again be brief and refer to Appendix~\ref{app:iihodge} for details.
As in the previous section, we split the single complex structure modulus as $t = a + \im s$, and we assume that the Tyurin degeneration point is reached as $s\to\infty$.
The first ingredient that we need to construct the near-boundary EFT is the nilpotent operator associated to the Tyurin degeneration; the nilpotent operator that we shall consider is \cite{Green:2008}
\begin{equation}
\label{GGS_1mod_II_N}
    N=\begin{pmatrix}
    0 & 0 & 0 & 0\\
    0 & 0 & 0 & 0\\
    m & 0 & 0 & 0\\
    0 & n & 0 & 0\\
    \end{pmatrix},\quad \text{with } m,n\in \mathbb{N}\,.
\end{equation}
Secondly, the most general representative ${\bf a}_0$ of the set $F_0^3$ is \cite{Green:2008}
\begin{equation}
\label{GGS_1mod_II_a0}
    {\bf a}_0 = \left(1,0,\im\, \alpha, \im\,\alpha c + d\right)^T\,,\quad \text{with } \alpha = \sqrt{m/n}\quad \text{and}\quad c,d \in \mathbb{R}.
\end{equation}
Then, Schmid's nilpotent orbit theorem \eqref{HT_Per_nil} yields the following approximated periods
\begin{equation}
\label{GGS_1mod_II_period}
    {\bm \Pi}_{\rm nil}=e^{tN}\bf{a}_0=\begin{pmatrix}
    1\\
    \im\, \alpha \\
    m t\\
    d + \im\, c \alpha + \im\, n \alpha t
    \end{pmatrix}.
\end{equation}
The periods \eqref{GGS_1mod_II_period} do not have a clear origin in terms of a prepotential. However, performing the symplectic transformation
\begin{equation}
\label{GGS_1mod_II_period_S}
    \mathcal{S}= 
    \begin{pmatrix} 
    1 & 0 & 0 & 0 \\
    -\alpha d & - \alpha c & 0 & \alpha \\
    0 & -d & 1 & 0 \\
    0 & -\alpha & 0 & 0
    \end{pmatrix},
\end{equation}
the periods \eqref{GGS_1mod_II_period} get mapped to
\begin{equation}
\label{GGS_1mod_II_period_b}
    \widetilde{\bm \Pi}_{\rm nil} = \mathcal{S} {\bm \Pi}_{\rm nil} =\begin{pmatrix}
    1\\
    \im\, m t \\
    m t - \im \alpha d\\
    - \im
    \end{pmatrix}.
\end{equation}
The latter periods can be most readily computed out of the prepotential
\begin{equation}
\label{GGS_1mod_II_prepot}
    \mathcal{F}_{\text{\tiny{Tyurin}}} = \im X^0 X^1 + \im \frac{\alpha d}{2} (X^0)^2\,,
\end{equation}
upon gauge-fixing $X^0 =1$, $X^1 = \im\, m t$.

Following the steps described in Appendix~\ref{app:iihodge}, one can show that the $\mathfrak{sl}(2)$-approximated matrix $\mathcal{N}$ is
\begin{equation}
\label{GGS_1mod_II_cN}
    \cN_{\text{\tiny{Tyurin}}} = -\begin{pmatrix}
    ma & \frac{d}{2}\\
    \frac{d}{2} & na+c
    \end{pmatrix}-\im \begin{pmatrix}
    ms & 0\\
    0 & ns
    \end{pmatrix}.
\end{equation}
Thus, the EFT defined for large saxionic vevs, near the Tyurin degeneration is:
\begin{equation}
\label{GGS_1mod_II_S}
\begin{aligned}
    S_{\text{\tiny{Tyurin}}} &= \int \Big( \frac{1}2 M^2_{\rm P} R *1 - M^2_{\rm P} \frac{m^2}{(\alpha d - 2 m s)^2} ({\rm d} a \wedge * {\rm d} a + {\rm d} s \wedge * {\rm d} s)  
    \\
    &\qquad\quad - \frac1{2} m s F^0 \wedge * F^0 - \frac1{2} n s  F^1 \wedge * F^1
    \\
    &\qquad\quad - \frac12 m a F^0 \wedge F^0 - \frac{d}{2} F^0 \wedge F^1  - \frac12 (n a + c) F^1 \wedge F^1
    \Big)\,.
\end{aligned}
\end{equation}

\noindent \textbf{The Tyurin degeneration one-form global symmetries.} As in Section~\ref{sec:GGS_1mod_IV}, the gauge sector can lead to four one-form global symmetries, associated to shifts of the one-forms $A^I$ by flat connections and their magnetic counterparts. Employing the Bianchi identities for the one-form gauge fields $A^I$ and the equations of motions stemming from the EFT action \eqref{GGS_1mod_II_S}, one can show that the putative one-form global symmetries are related to the following, conserved two-form Page currents:
\begin{equation}
\label{GGS_1mod_II_PageJ}
\begin{aligned}
    (J^0_{\rm m})^{(P)} &= F^0\,,
    \\
    (J^1_{\rm m})^{(P)} &= F^1\,,
    \\
    (J_0^{\rm e})^{(P)} &= G_0 := - m s * F^0 - m a F^0- \frac{d}{2} F^1 \,,
    \\
    (J_1^{\rm e})^{(P)} &= G_1 :=  - n s * F^1 - \frac{d}{2} F^0- (n a + c) F^1\,.
\end{aligned}
\end{equation}
Recalling that the Page currents ${\bm J}^{(P)} =( (J^I_{\rm m})^{(P)}, -(J_I^{\rm e})^{(P)} )^T$ transform as ${\bm J}^{(P)} \to e^N {\bm J}^{(P)}$ under a unit axion shift $a \to a +1$, it can be shown that the electric Page currents nontrivially transform as
\begin{equation}
\label{GGS_1mod_II_PageJ_eltr}
\begin{pmatrix}
 - (J_0^{\rm e})^{(P)} \\ - (J_1^{\rm e})^{(P)}
\end{pmatrix} \quad \xrightarrow{a \to a + 1} \quad \begin{pmatrix}
- (J_0^{\rm e})^{(P)} + m (J^0_{\rm m})^{(P)} \\ - (J_1^{\rm e})^{(P)} + n (J^1_{\rm m})^{(P)}
\end{pmatrix}\,.
\end{equation}
As such, the electric currents $(J_I^{\rm e})^{(P)}$ are ill-defined. On the other hand, the magnetic Page currents $(J^I_{\rm m})^{(P)}$ are single-valued and thus deliver genuine one-form global symmetries.
These global symmetries can be minimally broken by assuming the presence of a D3-particle with nontrivial magnetic charges $p^I = (p^0, p^1)$ so that the two-form Page currents $(J^I_{\rm m})^{(P)}$ are not conserved as
\begin{equation}
\label{GGS_1mod_II_PageJ_mag_break}
    {\rm d} (J^I_{\rm m})^{(P)} = {\rm d} F^I = - p^I \delta^{(3)}(\Gamma)\,.
\end{equation}
A generic choice of magnetic charges $p^I$ breaks the global group $(U(1)_{\rm m})^2$ to the global discrete subgroup $\mathbb{Z}_{p^0} \times \mathbb{Z}_{p^1}$. Choosing the magnetic charges $p^I$ as unit-charges, the one-form global symmetries are fully broken.

\noindent \textbf{The generalized Witten effect in EFTs defined close to the Tyurin degeneration.} 
As for the EFTs defined close to an LCS point, also EFTs defined towards a Tyurin degeneration point are characterized by generalized $\theta$-terms and, as such, are subjected to a nontrivial Witten effect.  
In fact, as we have seen, requiring the breaking of the genuine one-form global symmetries $(U(1)_{\rm m})^2$ requires the presence of a D3-particle with the seed charge
\begin{equation}
\label{GGS_1mod_IV_qseedm}
    {\bf q}_{\rm seed} = (p^0,p^1,0,0)^T\,.
\end{equation}
Then, the generalized Witten effect then predicts the following, infinite tower of dyonic states
\begin{equation}
\label{GGS_1mod_II_qtower}
    {\bf q}^{(k)}_{\rm tower} = (p^0,p^1,-k m p^0,-k n p^1)^T\,, \quad \text{with} \quad k \in \mathbb{Z}.
\end{equation}
It can be checked that any state belonging to the tower \eqref{GGS_1mod_II_qtower} has super-Planckian mass towards the Tyurin degeneration, as $M_{{\bf q}_{\rm tower}} \to \infty$ as $s\to \infty$.

\noindent \textbf{The Tyurin degeneration axion global symmetries.} Finally, let us consider the zero-form axion global symmetry. The three-form Maxwell current 
\begin{equation}
\label{GGS_1mod_II_Jax}
    J^{(M)}_a = - M_{\rm P}^2\frac{m^2}{(\alpha d - 2 m s)^2}  * {\rm d} a
\end{equation}
is not conserved, as can be checked from the equations of motion for the axion, and the lack of conservation is due to the generalized $\theta$-terms appearing in \eqref{GGS_1mod_II_S}. Therefore, the continuous axion shift symmetry is unavoidably broken to a discrete subgroup.

\subsection{An example of finite distance boundary: the conifold singularity}
\label{sec:GGS_1mod_I}

The final class of EFTs endowed with a single dynamical complex structure modulus are those encountered close to a conifold singularity. 
In contrast with the LCS point examined in Section~\ref{sec:GGS_1mod_IV} and the Tyurin degeneration studied in Section~\ref{sec:GGS_1mod_II}, the conifold singularity is reached at \emph{finite} distance in field space.
As such, we will see that EFTs defined close to conifold point exhibit crucial differences compared to the EFTs investigated in the previous sections.

\noindent \textbf{The conifold EFT.} We first determine the general form of the $\mathcal{N}=2$ Type IIB EFTs defined close to a conifold singularity. In the terminology of Appendix~\ref{sec:Hodge_review}, a conifold singularity corresponds to a Type I field space boundary. The nilpotent operator $N = N_-$ that describes a Type I boundary and delivers the lowering operator of the single ${\rm sl}_2$-triple is given by \cite{Green:2008}
\begin{equation}
\label{GGS_1mod_I_N}
    N=\begin{pmatrix}
    0 & 0 & 0 & 0\\
    0 & 0 & 0 & 0\\
    n & 0 & 0 & 0\\
    0 & 0 & 0 & 0\\
    \end{pmatrix},\quad \text{with } 0>n\in \mathbb{Z}\,,
\end{equation}
such that $N^2 = 0$. A general representative ${\bf a}_0 \in F_0^3$ is \cite{Green:2008}
\begin{equation}
\label{GGS_1mod_I_a0}
    {\bf a}_0 = \left(0,1,\beta - \gamma\tau, \tau\right)^T\,,\quad \text{with } \quad \beta, \gamma \in \mathbb{R},\quad \tau\in\mathbb{C}.
\end{equation}
Moreover, we shall additionally assume $d>0$. 
The nilpotent operator \eqref{GGS_1mod_II_N} acts trivially on $F_0^3$ as $N {\bf a}_0 = 0$. Therefore, the leading term in the nilpotent orbit theorem \eqref{HT_Per_nil} is moduli-independent, and leads to a constant K\"ahler potential \eqref{IIB_Kcs}. 
In order to get a nontrivial K\"ahler potential one needs to consider subleading terms in the nilpotent orbit \eqref{HT_Per_nil}, which are generated by the instanton map. Its general form is given in \cite{Bastian:2021eom}, which, transformed to our basis, leads to the following expression:
\begin{equation}
\label{GGS_1mod_I_a1}
    {\bf a}_1 = \Gamma'(0){\bf a}_0=\alpha \left(1,\gamma, 0, \beta\right)^T,
\end{equation}
with $\alpha \in \mathbb{R}$ a model-dependent parameter appearing in the instanton map.
Then, employing \eqref{GGS_1mod_I_a0}-\eqref{GGS_1mod_I_a1} along with the explicit expression for the nilpotent operator $N$ \eqref{GGS_1mod_I_N}, the nilpotent orbit theorem \eqref{HT_Per_nil} yields the following approximated periods:
\begin{equation}
\label{GGS_1mod_I_period}
    {\bm \Pi} = e^{tN}({\bf a}_0 + e^{2\pi\im\,t} {\bf a}_1  + \ldots)  \simeq \begin{pmatrix}
    \alpha e^{2\pi \im\, t}\\
    1 + \alpha \gamma e^{2\pi \im\, t}  \\ 
    \beta -\gamma \tau  + \alpha n e^{2\pi \im\, t} t \\
    \tau + \alpha\beta e^{2\pi \im\, t}
    \end{pmatrix}.
\end{equation}
The approximate K\"ahler potential \eqref{IIB_Kcs} computed out of the periods \eqref{GGS_1mod_I_period} is
\begin{equation}
\label{GGS_1mod_I_Kcs}
    e^{-K^{\rm cs}_{\text{\tiny{conifold}}}} \simeq 2 d + 2  \alpha^2 n s e^{-4\pi s}\,.
\end{equation}
It is worth mentioning that, by performing the symplectic transformation
\begin{equation}
\label{GGS_1mod_I_period_S}
    \mathcal{S}= 
    \begin{pmatrix} 
    0 & 1 & 0 & 0 \\
    1 & 0 & 0 & 0 \\
    0 & -c & 0 & 1 \\
    0 & 0 & 1 & 0
    \end{pmatrix},
\end{equation}
the periods \eqref{GGS_1mod_I_period} can be recast as
\begin{equation}
\label{GGS_1mod_I_period_b}
    \widetilde{\bm \Pi}_{\rm nil} \simeq \mathcal{S} {\bm \Pi}_{\rm nil} =\begin{pmatrix}
    1\\
    e^{2\pi \im t}\\
    \im d\\
    \beta-\gamma\tau+\alpha n t e^{2\pi \im t}\\
    \end{pmatrix},
\end{equation}
where we have neglected subleading corrections.
In turn, the periods \eqref{GGS_1mod_I_period_b} can be regarded as originating from the superpotential
\begin{equation}
\label{GGS_1mod_I_prepot}
    \mathcal{F}_{\text{\tiny{conifold}}} = - \frac{\im d}{2} (X^0)^2 - (\beta-\gamma\tau)X^0X^1+ \frac{\im \alpha n}{4 \pi} (X^1)^2 \log \frac{X^1}{X^0}\,,
\end{equation}
upon gauge-fixing $X^0 =1$, $X^1 = e^{2\pi \im t}$. Indeed, it can be checked that computing the periods \eqref{IIB_periodsymplb} via the prepotential \eqref{GGS_1mod_I_prepot} and plugging them into the general \eqref{IIB_Kcs} deliver the same K\"ahler potential as \eqref{GGS_1mod_I_Kcs}.

In Appendix~\ref{app:ihodge} the following $\mathfrak{sl}(2)$-approximated matrix $\mathcal{N}$ is obtained:\footnote{Notice that, for $s\gg 1$ and $n < 0$, the eigenvalues of ${\rm Im} \mathcal{N}_{\text{\tiny{conifold}}}$ are both negative. \label{footnote:conifold_signs}}
\begin{equation}
\label{GGS_1mod_I_cN}
    \cN_{\text{\tiny{conifold}}} \simeq - \begin{pmatrix} na-\gamma(\beta-\gamma c) & \beta-\gamma c\\
    \beta-\gamma c & c
    \end{pmatrix}+\im \begin{pmatrix}
    ns & \gamma d\\
    \gamma d & - d\end{pmatrix},
\end{equation}
where we have defined
\begin{equation}
    \tau=c+\im\,d,\qquad c,d\in\mathbb{R}.
\end{equation}
Hence, using the general periods \eqref{GGS_1mod_I_period} and the matrix \eqref{GGS_1mod_I_cN}, the EFT defined close to a conifold singularity, for large value of saxion $s \gg 1$, is\footnote{In order for the kinetic terms to get the correct signs, one has to set $n < 0$, $d < 0$ -- see also footnote~\ref{footnote:conifold_signs}.}
\begin{equation}
\label{GGS_1mod_I_S}
\begin{aligned}
    S_{\text{\tiny{conifold}}} &\simeq \int \Big( \frac{1}2 M^2_{\rm P} R *1 - M^2_{\rm P} \frac{2\pi n}{d}\alpha^2 e^{-4\pi s}(2\pi s -1) ({\rm d} a \wedge * {\rm d} a + {\rm d} s \wedge * {\rm d} s)  
    \\
    &\qquad\quad + \frac1{2} ns F^0 \wedge * F^0 +\gamma d F^0 \wedge * F^1 - \frac{d}{2} F^1 \wedge * F^1
    \\
    &\qquad\quad - \frac12(\gamma \beta - c\gamma^2 - an) F^0 \wedge F^0 - (\beta -\gamma c) F^0 \wedge F^1  - \frac{c}2 F^1 \wedge F^1
    \Big)\,.
\end{aligned}
\end{equation}

\noindent \textbf{The conifold one-form global symmetries.} The four one-form global symmetries that the conifold EFT may at most support are associated to the following two-form Page currents
\begin{equation}
\label{GGS_1mod_I_PageJ}
\begin{aligned}
    (J^0_{\rm m})^{(P)} &= F^0\,,
    \\
    (J^1_{\rm m})^{(P)} &= F^1\,,
    \\
    (J_0^{\rm e})^{(P)} &= G_0 := (ns - d\gamma^2) * F^0 + \gamma d * F^1 + (\gamma \beta - c\gamma^2 - an) F^0- (\beta - \gamma c) F^1 \,,
    \\
    (J_1^{\rm e})^{(P)} &= G_1 :=  \gamma d * F^0 - d * F^1 - (\beta - \gamma c) F^0 -c F^1\,,
\end{aligned}
\end{equation}
which are conserved due to the Bianchi identities and the equations of motion for the one-forms $A^I$.
The Page currents ${\bm J}^{(P)} =( (J^I_{\rm m})^{(P)}, -(J_I^{\rm e})^{(P)} )^T$ transform as ${\bm J}^{(P)} \to e^N {\bm J}^{(P)}$ under a unit-axion shift. 
From the explicit expression of the nilpotent matrix \eqref{GGS_1mod_I_N}, one recognizes that the electric Page current $(J_0^{\rm e})^{(P)} = G_0$ is subjected to the nontrivial shift 
\begin{equation}
\label{GGS_1mod_I_G0}
G_0 \quad \xrightarrow{a \to a + 1} \quad G_0 - n F^0
\end{equation}
under a monodromy transformation.
Thus, the Page current $(J_0^{\rm e})^{(P)}$ is ill-defined.
On the other hand, the remaining Page currents, $(J^I_{\rm m})^{(P)}$ and $(J_1^{\rm e})^{(P)}$ are single-valued and lead to unbroken one-form global symmetries.
In order to break them, it is enough to assume the existence of a dyonic D3-particle with charge ${\bf q} = (p^0,p^1,0,-q_1)^T$, that breaks the conservation laws of the well-defined Page currents as
\begin{equation}
\label{GGS_1mod_I_PageJ_break}
    {\rm d} (J^I_{\rm m})^{(P)} = {\rm d} F^I = - p^I \delta^{(3)}(\Gamma)\,, \qquad {\rm d} (J_1^{\rm e})^{(P)} = {\rm d} G_1 = - q_1 \delta^{(3)}(\Gamma)\,.
\end{equation}

\noindent \textbf{The generalized Witten effect in EFTs defined close to a conifold singularity.} 
The $\theta$-terms populating the near-conifold EFT \eqref{GGS_1mod_I_S} induce a nontrivial Witten effect.
Here, the Witten effect manifest in the generation of an infinite tower of dyonic, massive states.
In fact, consider a seed charge of the form ${\bf q}_{\rm seed} = (p^0,p^1,0,-q_1)^T$, as is required in order to break the three genuine one-form global symmetries as in \eqref{GGS_1mod_I_PageJ_break}.
Then, an infinite tower of states with charge
\begin{equation}
\label{GGS_1mod_I_qtower}
    {\bf q}^{(k)}_{\rm tower} = e^{-k N}{\bf q}_{\rm seed} = (p^0,p^1,-k m p^0,-q_1)^T\,, \quad \text{with} \quad k \in \mathbb{Z}\,,
\end{equation}
is also present in the spectrum. Thus, as can be inferred from the BPS mass formula \eqref{IIB_D3Mb} employing the periods \eqref{GGS_1mod_I_period}, the tower of states \eqref{GGS_1mod_I_qtower} is constituted by massive, super-Planckian states.

\noindent \textbf{The conifold axion global symmetries.} Lastly, let us comment on the eventual symmetries that involve the scalar field $a$. First, we remark that oftentimes, in the literature (see, for instance, \cite{Reece:2018zvv}), the scalar field $a$ is referred to as pseudo-Nambu-Goldstone boson, whenever the field space boundary is located at finite geodesic distance.
The reason is rooted in the fact that, at the conifold singularity, the continuous axion shift symmetry is restored. Nevertheless, here, we shall still refer to the scalar field $a$ as `axion'.
Indeed, in analogy with the EFTs defined towards boundaries located at infinite field distance studied above, integral shifts of the field $a$ generate monodromy transformations as in \eqref{Sympl_TPi}. 
Furthermore, at leading order in the saxion, the field space metric appearing in \eqref{GGS_1mod_I_S} is invariant under shifts of the field $a$.

As such, the EFT action might be endowed with global shift symmetries for the field $a$, which have to be avoided.
However, the Maxwell three-form axion current 
\begin{equation}
\label{GGS_1mod_I_Jax}
    J^{(M)}_a = M_{\rm P}^2 \frac{2\pi n}{d}e^{-4\pi s}(2\pi s -1)   * {\rm d} a\,,
\end{equation}
which can be computed from the effective action \eqref{GGS_1mod_I_S}, is not conserved, due to the $\theta$-term appearing in \eqref{GGS_1mod_I_S} that involves the graviphoton:
\begin{equation}
\label{GGS_1mod_I_Jaxb}
    {\rm d} J^{(M)}_a = - \frac{n}2 F^0 \wedge F^0\,.
\end{equation}
Therefore, the continuous axion shift symmetry is certainly broken in the field space region close to a conifold singularity, similarly to what happens towards the infinite field distance boundaries studied above.


\section{Global symmetries in general asymptotic limits}
\label{sec:GGS_gen_mod}

Inferring the global symmetries that emerge towards a generic field space boundary might appear to be a daunting task. 
Indeed, it is far from trivial to extend the analysis carried in Section~\ref{sec:GGS_01mod} in the one-modulus limits to multi-moduli limits.
However, although the explicit derivation of the most general EFTs that describe the dynamics towards general field space boundaries remains out of reach, the Hodge theory technology reviewed in Section~\ref{sec:Hodge} delivers valuable information about the asymptotic symmetries that emerge in such limits.

In order to set the stage for the upcoming discussion, let us state the setup on which we focus.
For notational simplicity, we consider generic field space boundaries that are reached when the first $k$ saxions acquire large values, as $s^1, \ldots, s^k \to \infty$. Limits that involve a different set of saxions can be described analogously by a relabeling of the coordinates.
Let us assume that such a boundary is reached by sending the saxions to large values one by one. For instance, let us first consider the limit in which only the first saxion $s^1$ acquires large vev, as $s^1 \to \infty$. The locus towards which $s^1 \to \infty$ is a boundary on its own, and following the discussion in Appendix~\ref{sec:Hodge_review} it is characterized by a type, that we generically denote `Type ${\sf A}_1$', with ${\sf A}_1 = {\rm I}_a, {\rm II}_b, {\rm III}_c, {\rm IV}_d$.
We then further send $s^2 \to \infty$, thus reaching a new boundary, with type ${\sf A}_2$, and analogously proceed so that we ultimately reach a Type ${\sf A}_k$ boundary towards which all the first $k$ saxions acquire large vev.
This chain of successive `\emph{enhancements}' of the boundary type can be schematically written as
\begin{equation}
	\label{GGS_gen_enh}
	\text{Type ${\sf A}_1$} \rightarrow \text{Type ${\sf A}_2$} \rightarrow \ldots \rightarrow \text{Type ${\sf A}_k$}\,. 
\end{equation}
In the following, employing the general considerations of Appendix~\ref{sec:Hodge}, we will investigate which global symmetries may emerge at each step of the enhancement chain \eqref{GGS_gen_enh}, and how these are inter-related.

\subsection{The generalized Witten Effect and global symmetries towards a generic boundary}
\label{sec:GGS_gen_mod_WE}

First, we consider the generalized global symmetries that can emerge at a given step along the enhancement chain \eqref{GGS_gen_enh}.
Without loss of generality, we shall focus on the last step of the chain represented in \eqref{GGS_gen_enh}.
Additionally, we shall regard the residual $(n-k)$-complex moduli fields $t^i$ as non-dynamical, fixed at given vevs.

Our general investigation starts with the analysis of the allowed electromagnetic one-form global symmetries that the EFT defined close to the Type ${\sf A}_k$ may be equipped with.
As illustrated in Section~\ref{sec:GGS_IIBsub} on general grounds, the maximal $2(h^{2,1}+1)$ one-form global symmetries may be ill-defined, since the conserved, gauge-invariant Page currents \eqref{GGSIIB_Jel}--\eqref{GGSIIB_Jm} non-trivially transform as ${\bm J}^{(P)} \to T {\bm J}^{(P)}$ under the axion monodromies $a^i \to a^i + 1$ for all $i = 1,\ldots, k$.
The monodromy matrix $T$ is here determined by the nilpotent matrix $N_{(k)}$ defined above \eqref{HT_Wl}  as $T = e^{N_{(k)}}$. Therefore, the knowledge of the generic properties of the operator $N_{(k)}$ deliver pivotal information about the transformation that the Page currents are subjected to. 
The relevant properties of the nilpotent matrix $N_{(k)}$ are indeed related to the boundary type close to which the EFT is defined, and are summarized in Table~\ref{table:HDclass}.

In particular, the Page currents that get nontrivially transformed are those which lie in the image of the operator $N_{(k)}$. 
Therefore, the currents that can deliver genuine, well-defined Page charges are those which lie in the complement space $H^3(Y) \smallsetminus {\rm Im}\, N_{(k)}$.
Interestingly, one can give an estimation on the number of independent Page currents towards any field space boundary based on the general considerations of Appendix~\ref{sec:Hodge_review}.
In fact, the independent, well-defined Page currents are counted by the dimension of the space $H^3(Y) \smallsetminus {\rm Im}\, N_{(k)}$, and the latter can be most readily inferred from the rank of the operator $N_{(k)}$, that depends on the boundary type as given in Table~\ref{table:HDclass}. 
Thus, the number of independent, electromagnetic one-form global symmetries $N_{\rm em}$ depends on the boundary type as follows:
\begin{equation}
	\label{GGS_gm_Nem}
	N_{\rm em} = {\rm dim}_{\mathbb{R}} \left(H^3(Y) \smallsetminus {\rm Im}\, N_{(k)}\right) =
	\begin{cases}
		2 h^{2,1} + 2 - a & \text{Type I$_{a}$}\,,
		\\
		2 h^{2,1} - b & \text{Type II$_{b}$}\,,
		\\
		2 h^{2,1} - 2 - c & \text{Type III$_{c}$}\,,
		\\
		2 h^{2,1}  - d & \text{Type IV$_{d}$}\,.
	\end{cases}
\end{equation}
Employing the bounds on the parameters $a$, $b$, $c$, $d$ that distinguish the boundary type listed in Table~\ref{table:HDclass}, the following bounds on the number $N_{\rm em}$ of genuine, conserved one-form global symmetries can be obtained:\footnote{Recall that boundaries of Type III$_{c}$ can appear only for $h^{2,1} \geq 2$.}
\begin{equation}
	\label{GGS_gm_Nem_bounds}
	N_{\rm em} =
	\begin{cases}
		2 \leq N_{\rm em} \leq 2 h^{2,1} + 2 & \text{Type I$_{a}$}\,,
		\\
		h^{2,1} +1 \leq N_{\rm em} \leq 2 h^{2,1} & \text{Type II$_{b}$}\,,
		\\
		h^{2,1}  \leq N_{\rm em} \leq 2 h^{2,1} - 2 & \text{Type III$_{c}$}\,,
		\\
		h^{2,1} \leq N_{\rm em} \leq 2 h^{2,1} -1 & \text{Type IV$_{d}$}\,.
	\end{cases}
\end{equation}

Clearly, the bounds in \eqref{GGS_gm_Nem_bounds} are very lax, and their range widens as the number of moduli increases.
However, it is interesting to notice that \emph{for every moduli space boundary, there is always at least one well-defined one-form global symmetry}, leading to a global abelian $U(1)_{\rm em}^{N_{\rm em}}$ continuous symmetry group.
Such a feature is just a rephrasing of the fact that a nilpotent matrix $N_{(k)}$ can never be full-rank.

The immediate consequence is that, regardless of the boundary the EFT describes, one needs to assume the existence of at least one fundamental object that breaks the residual continuous one-form global symmetries. 
With a convenient choice of the basis, the fundamental charges ${\bf q}$ of the required D3-particles can be picked in the set $(H^3(Y) \smallsetminus {\rm Im}\, N_{(k)}) \cap \mathbb{Z}^{N_{\rm em}}$.
A generic choice of charges would break the continuous group $U(1)_{\rm em}^{N_{\rm em}}$ down to a discrete subgroup thereof. 
However, choosing the charges as being unit charges, the $U(1)_{\rm em}^{N_{\rm em}}$ group is entirely broken.

In general, the properties of the fundamental objects that ought to be included to break the residual one-form global symmetries depend on the specific geometrical data of the compactification. 
Nevertheless, some of their crucial features can be understood in general terms. 
First, notice that the elementary charges ${\bf q}$ cannot be generated as the image of any of the operators $N_{(i)}$, $i =1, \ldots, k$.
Along each of the enhancement steps in \eqref{GGS_gen_enh} that leads to the Type ${\sf A}_k$ boundary that is of interest to us, one can draw a Deligne diagram as the ones in Table~\ref{table:HDclass}.
The requirement that ${\bf q} \notin {\rm Im}\, N_{(i)}$, for every $i =1, \ldots, k$, can then be understood as the fact that ${\bf q}$ has to lie in the upper-part of each of these Deligne diagrams, namely ${\bf q} \in I^{p,q}_{(i)}$, with $p+q \geq 3$.
Recalling that the physical charge of the D3-particles \eqref{IIB_D3Q} are Hodge norms, this implies that $\mathcal{Q}_{\bf q} \gtrsim 1$. 
Namely, the objects required to break the residual one-form global symmetries are not expected to be weakly coupled, as observed in one-modulus examples of Section~\ref{sec:GGS_01mod}.

Finally, let us comment on the zero-form axion global symmetries that the EFT defined towards a generic boundary can be endowed with.
A necessary condition in order to have a continuous axion shift symmetry is that the EFT action~\eqref{N2_action} is invariant under arbitrary shifts of the axions $a^i \to a^i + c^i$, for some $c^i \in \mathbb{R}$.
In particular, the K\"ahler metric ought to be invariant under such shifts.
However, recall that the periods are generically subjected to corrections of order $\mathcal{O}(e^{2\pi \im t})$ as captured by the nilpotent orbit theorem \eqref{HT_Per_nil}.
Consequently, the K\"ahler potential \eqref{IIB_Kcs} acquires the general form
\begin{equation}
	\label{GGS_gm_KP}
	e^{-K} = \mathcal{P}(s) + \mathcal{O}(e^{-2\pi s}, \cos (2\pi a), \sin (2\pi a))\,,
\end{equation}
where $\mathcal{P}(s)$ is a polynomial in the saxions and $\mathcal{O}(e^{-2\pi s}, \cos (2\pi a), \sin (2\pi a))$ denotes additional corrections that may be either axion-dependent or exponentially suppressed in the saxions.
As noticed in \cite{Bastian:2021eom,Bastian:2021hpc}, the axion-dependent corrections that appear in \eqref{GGS_gm_KP} may be necessary for consistency of the Hodge decomposition near boundaries different from the Type IV ones, namely the large complex structure boundaries.
Hence, in general, only EFTs defined close to a Type IV boundaries may lead to an axion-independent K\"ahler potential, and thus they might exhibit the axion-shift symmetry.
However, the analysis of the K\"ahler metric is not enough to conclude that some continuous axionic zero-form symmetries are unbroken. 
In fact, Type IV boundaries are associated to nilpotent matrices of the highest rank, and the generalized $\theta$-terms necessarily depend non-trivially on the axions, which appear within either linear, quadratic or cubic couplings with the topological four-form terms $F^I \wedge F^J$, which explicitly break the axion shift symmetries.
Whether the continuous axion zero-form symmetry is entirely broken or broken down to a discrete subgroup towards any boundary depends on the specific EFT data. Requiring that the discrete shift symmetry is broken completely may additionally require the presence of instantons with unit charges.

\subsection{Avoiding global symmetries along boundary enhancement chains}
\label{sec:GGS_gen_mod_enh}

\begin{table}[!t]
	\begin{center}
		\begin{tabular}{ | c | l  | } 
			\hline
			\rowcolor{orange!20} Starting Type ${\sf A}_{k}$ boundary & Enhanced Type ${\sf A}_{k+1}$ boundary 
			\\
			\Xhline{1pt} 
			\multirow{4}{*}{Type I$_{a}$} & Type I$_{\hat a}$ for $a \leq \hat{a}$
			\\
			& Type II$_{\hat b}$ for $a \leq \hat{b}$, $a < h^{2,1}$
			\\
			& Type III$_{\hat c}$ for $a \leq \hat{c}$, $a < h^{2,1}$
			\\
			& Type IV$_{\hat d}$ for $a \leq \hat{d}$, $a < h^{2,1}$
			\\[.05cm]
			\hline
			\hline
			\multirow{3}{*}{Type II$_{b}$} & Type II$_{\hat b}$ for $b \leq \hat{b}$
			\\
			& Type III$_{\hat c}$ for $2 \leq b \leq \hat{c}+2$
			\\
			& Type IV$_{\hat d}$ for $1 \leq b \leq \hat{d}-1$
			\\[.05cm]
			\hline
			\hline
			\multirow{2}{*}{Type III$_{c}$} & Type III$_{\hat c}$ for $c \leq \hat{c}$
			\\
			& Type IV$_{\hat d}$ for $c+2 \leq \hat{d}$
			\\[.05cm]
			\hline
			\hline
			Type IV$_{d}$ & Type IV$_{\hat d}$ for $d \leq \hat{d}$
			\\[.05cm]
			\hline
		\end{tabular}
		\caption{The allowed enhancements from a boundary of Type ${\sf A}_k$ reached as the first $k$ saxions $s^1,\ldots,s^k \to \infty$ to a boundary of Type ${\sf A}_{k+1}$ reached when also the $(k+1)$-th saxion acquires large vev as $s^{k+1} \to \infty$ \cite{KPR,Grimm:2018cpv}.  \label{tab:enh}}
	\end{center}
\end{table}

Effective theories defined close to different boundaries display a different number of global symmetries, as the nilpotent operator that determines the nontrivial transformation of the Page currents changes.
The general framework of Hodge theory allows one to keep track of how the number of one-form global symmetries changes as the boundary is enhanced along the chain \eqref{GGS_gen_enh}. 
For concreteness, consider a boundary of Type ${\sf A}_k$, which -- as in the previous section -- is reached as the first $k$ saxions $s^1, \ldots, s^k \to \infty$. 
Then, we consider the boundary that is reached when one of the residual $(n-k)$-saxions is very large -- say, when the $(k+1)$-th saxion $s^{k+1} \to \infty$.
The boundary type is then enhanced to the different type, that we generically denote ${\sf A}_{k+1}$.

Crucially, there exist restrictions on the Type ${\sf A}_{k+1}$ a Type ${\sf A}_k$ boundary can enhance to, as is explained in \cite{KPR,Grimm:2018cpv}. 
For any boundary of Type ${\sf A}_k$, the allowed types ${\sf A}_{k+1}$ of the enhanced boundary are collected in Table~\ref{tab:enh}.

\begin{table}[!t]
	\begin{center}
		\begin{tabular}{ | c  l | c l | } 
			\hline
			\rowcolor{orange!20} Starting boundary & Starting $N_{\rm em}^{\rm st}$ & Enhanced boundary & Enhanced $N_{\rm em}^{\rm enh}$
			\\
			\Xhline{1pt} 
			& \cellcolor{cyan!10}  & Type I$_{\hat a}$ & \cellcolor{lime!10} $2 h^{2,1} + 2 - {\hat a} \leq 2 h^{2,1} + 2 - a$
			\\
			& \cellcolor{cyan!10} & Type II$_{\hat b}$ & \cellcolor{lime!10} $2 h^{2,1} - \hat{b}  \leq 2 h^{2,1} - a $
			\\ 
			& \cellcolor{cyan!10} & Type III$_{\hat c}$ & \cellcolor{lime!10} $2 h^{2,1} - 2 - \hat{c}  \leq 2 h^{2,1} - 2 - a$
			\\
			\multirow{-4}{*}{Type I$_{a}$}  & \cellcolor{cyan!10} \multirow{-4}{*}{$2 h^{2,1} + 2 - a$} & Type IV$_{\hat d}$ & \cellcolor{lime!10} $2 h^{2,1} -  \hat{d}  \leq 2 h^{2,1}  - a$
			\\[.05cm]
			\hline
			\hline
			& \cellcolor{cyan!10}  & Type II$_{\hat b}$ & \cellcolor{lime!10} $2 h^{2,1} - \hat{b} \leq 2 h^{2,1} - b$
			\\
			& \cellcolor{cyan!10} & Type III$_{\hat c}$ & \cellcolor{lime!10} $2 h^{2,1} - 2 - \hat{c}  \leq 2 h^{2,1}  - b$
			\\
			\multirow{-3}{*}{Type II$_{b}$} & \cellcolor{cyan!10} \multirow{-3}{*}{$2 h^{2,1} - b$}& Type IV$_{\hat d}$ & \cellcolor{lime!10} $2 h^{2,1} -  \hat{d}  \leq 2 h^{2,1}  -1 - b$
			\\[.05cm]
			\hline
			\hline
			 & \cellcolor{cyan!10}  & Type III$_{\hat c}$ & \cellcolor{lime!10} $2 h^{2,1} - 2 - \hat{c} \leq 2 h^{2,1} - 2 - c$
			\\
			\multirow{-2}{*}{Type III$_{c}$}& \cellcolor{cyan!10} \multirow{-2}{*}{$2 h^{2,1} - 2 - c$} & Type IV$_{\hat d}$ & \cellcolor{lime!10} $2 h^{2,1} -  \hat{d} \leq 2 h^{2,1} - 2 - c$
			\\[.05cm]
			\hline
			\hline
			Type IV$_{d}$ & \cellcolor{cyan!10} $2 h^{2,1} - d$ & Type IV$_{\hat d}$ & \cellcolor{lime!10} $2 h^{2,1} -  \hat{d} \leq 2 h^{2,1} - d$
			\\[.05cm]
			\hline
		\end{tabular}
		\caption{The one-form global symmetries o a boundary of Type ${\sf A}_{k+1}$ reached when also the $(k+1)$-th saxion acquires large vev as $s^{k+1} \to \infty$.  \label{tab:enh_ggs}}
	\end{center}
\end{table}

We denote by $N_{\rm em}^{\rm st}$ the maximal number of independent, continuous one-form global symmetries of the EFT towards the starting boundary of Type ${\sf A}_{k}$, as given by \eqref{GGS_gm_Nem}. Likewise, we let $N_{\rm em}^{\rm enh}$ denote the maximal number of symmetries towards the enhanced Type ${\sf A}_{k+1}$ boundary. Table~\ref{tab:enh_ggs} lists these numbers for each of the allowed enhancements from Table~\ref{tab:enh}, and one finds that
\begin{equation}
	N_{\rm em}^{\rm enh} \leq N_{\rm em}^{\rm st} \qquad \text{for every allowed boundary enhancement}.
\end{equation}
Therefore, enhancing the boundary type, by sending additional saxions toards large vevs, generically decreases the number of generalized global symmetries that one has to avoid.

\section{Conclusions}
	
In this work we have investigated which generalized global symmetries appear in the near-boundary regimes of four-dimensional $\mathcal{N}=2$ EFTs obtained from compactifying Type IIB string theory on a Calabi-Yau three-fold.
We have focused on the global symmetries associated to the vector multiplet sector, specifically on the electromagnetic one-form and the axion zero-form global symmetries.
As in \cite{Heidenreich:2020pkc} we have analyzed whether these global symmetries are broken by studying the associated conserved currents, and inquiring whether they are well-defined.

A key role in the analysis is played by certain `generalized $\theta$-terms’ that non-linearly couple the axions to the topological $F\wedge F$-terms of the theory.
In particular, for the $\mathcal{N}=2$ Type IIB EFTs examined in this work, these couplings are at most cubic in the axions. 
However, although a more standard linear coupling of the axion to the topological $F \wedge F$-term has been extensively studied in the physics literature, these generalized $\theta$-terms deserve further study, specifically in light of their effect on the symmetry properties of the theory.
In fact, our observations illustrate that these generalized $\theta$-terms are ubiquitous and unavoidable in stringy EFTs.
Indeed, in the EFTs studied in this work, their presence is due to the underlying $\mathcal{N}=2$ supersymmetry: the vector multiplets contain one-form gauge fields and axions among their components and a supergravity action describing their interactions mixes such components.

Moreover, the generalized $\theta$-terms may lead to a rich phenomenology.
Indeed we have shown that, on the one hand, such terms break the conservation laws of some generalized global symmetries; on the other hand, they trigger a generalized Witten effect, that renders the charge lattice infinitely populated provided that a handful of seed charges exist. 
However, as we have seen in the one-modulus cases scrutinized in Section~\ref{sec:GGS_01mod}, the infinite towers of states that the generalized Witten effect delivers are constituted by \emph{dyonic} states, with masses becoming superplanckian as the field space boundary is reached.

Thus, such towers of states -- that are obtained via the combined effect of breaking the generalized global symmetries and the generalized Witten effect -- are different from the infinite towers of massless states that realize the Distance Conjecture.
Indeed, as showed in \cite{Grimm:2018ohb,Grimm:2018cpv,Corvilain:2018lgw}, towers of states that are candidates for realizing the Distance Conjecture in Type IIB EFTs should rather be constituted by \emph{electric} D3-particles.
It is then tantalizing to relate the infinite tower of states found in this work with those of \cite{Grimm:2018ohb,Grimm:2018cpv,Corvilain:2018lgw}, and an investigation of a tight relationship between these two kinds of towers is left for future work.
Indeed, if the generalized Witten effect studied in this work could also justify the existence of the infinite towers of massless states found in \cite{Grimm:2018ohb,Grimm:2018cpv,Corvilain:2018lgw}, this could give a neat bottom-up argument for the realization of the Distance Conjecture.

It is worth stressing that the analysis carried out here does not cover \emph{all} the possible generalized global symmetries related to the vector multiplet sector.
In particular, as stressed in Section~\ref{sec:GGS_IIBsub}, alongside the zero-form axion symmetries, their electromagnetic dual, two-form global symmetries may be present.
The difficulties in examining the two-form global symmetries in Type IIB EFTs are rooted in the obstructions to applying the standard dualization procedure of Appendix~\ref{app:conv_dual} in a straightforward way, due to the presence of generalized $\theta$-terms.
As outlined in \cite{Heidenreich:2021yda}, understanding how these two-form global symmetries are avoided may deliver valuable information about the worldsheet theory of the eventual axion strings that ought to be present in the spectrum.
Moreover, higher-form generalized global symmetries associated to the Chern-Weil currents studied in \cite{Heidenreich:2020pkc,Heidenreich:2021xpr} might be present.
We leave the systematic analysis of these additional global symmetries for future work.

It is also tempting to extend our findings to EFTs endowed with higher supersymmetry, eventually in dimension greater than four. 
In the $\mathcal{N}=2$ Type IIB EFTs examined in this work, the monodromy group, which is part of the isometry group of the scalar manifold, determines the structure of the generalized $\theta$-terms and dictates the realization of the Witten effect.
In theories with higher supersymmetry or diverse dimension, the role here played by the monodromy group is expected to be played by a proper subgroup of the isometries of the scalar manifold.
A detailed investigation along this direction is left for future work.

\vspace{2em}

\noindent{\textbf{Acknowledgments}}

\noindent We would like to thank Rigers Aliaj, Luca Martucci, Irene Valenzuela for insightful discussions and we are especially grateful to Brice Bastian for fruitful, intensive discussions throughout the completion of this work. This research is partly supported by the Dutch Research Council (NWO) via a Start-Up grant and a Vici grant.


\appendix


\section{Charge quantization and dualization}
\label{app:conv_dual}

In this section we collect the main conventions that we have employed in this work on gauge fields and their quantization properties. 

For concreteness, let us focus on four-dimensional theories. Consider a free abelian $U_{\rm g}(1)$ gauge theory, whose gauge field is represented by a $(p-1)$-form $A_{p-1}$. We regard $A_{p-1}$ as the `electric' gauge field. The kinetic terms of such an $A_{p-1}$ gauge field are
\footnote{In our conventions, in $D$-dimensions, the Hodge dual of a $p$-form $\omega_p$ is the $(D-p)$-form
\begin{equation}
	\label{conv_pform}
	\omega_p = \frac{1}{p!} \omega_{M_1 \ldots M_p} {\rm d}x^{M_1} \wedge \ldots \wedge {\rm d}x^{M_p}\,,
\end{equation}
with $\varepsilon^{12\ldots D} = 1$. 
This implies that $*_D(*_D\omega_p) = \sigma (-)^{p(D-p)} \omega_p$ with $\sigma = +1$ ($\sigma=-1$) if the space has Euclidean (Minkowskian) signature.}
\begin{equation}
	\label{Qu_La}
	L = - \frac{1}{2 g^2} F_p \wedge * F_p
\end{equation}
with $F_p = {\rm d} A_{p-1}$ and $g$ is the gauge coupling. The $p$-form field strength is assumed to obey the quantization rule
\begin{equation}
	\label{Qu_Fp}
	\int_{\Sigma_p} F_p \in \mathbb{Z}\,,
\end{equation}
for any closed $p$-dimensional surface $\Sigma_p$.
	
As recalled in Section~\ref{sec:GGS_avoid}, the pure Maxwell theory exhibits two $U(1)$ one-form global symmetries. The first, electric $U_{\rm e}(1)$ one-form global symmetry is manifest in the electric frame. Indeed, $U_{\rm e}(1)$ acts by shifting $A_{p-1}$ by a closed connection, namely $A_{p-1} \to A_{p-1} + \Lambda_{p-1}$ with ${\rm d} \Lambda_{p-1} = 0 $, under which the Lagrangian \eqref{Qu_La} is invariant. 
	
The conserved current associated to the $U_{\rm e}(1)$ global symmetry can be computed via the standard Noether's procedure, leading to 
\begin{equation}
	\label{Qu_Jel}
	J^{\rm e}_{D-p} = \frac{(-1)^{p-1}}{g^2} * F_{p}\,,\qquad  {\rm d} J^{\rm e}_{D-p} = 0\,. 
\end{equation}
Given a closed $(D-p)$-surface $\Sigma_{D-p}$, we identify the quantized charges
\begin{equation}
	\label{Qu_Qel}
	Q^{\rm e} = \frac{1}{g^2} \int_{\Sigma_{D-p}} * F_{p} \in \mathbb{Z}\,.
\end{equation}

However, the free Maxwell theory described by the Lagrangian \eqref{Qu_La} is endowed with an additional, `magnetic' $U_{\rm m}(1)$ global symmetry, which we call $U_{\rm m}(1)$. Such a symmetry is made manifest once the electric gauge field $A_{p-1}$ is dualized to its magnetic dual gauge field. The dualization can be performed by considering a `master Lagrangian' that is obtained from \eqref{Qu_La} by promoting $F_p$ to an arbitrary $p$-form $\Theta_p$ and adding a dualizing term that contains an generic $(D-p-1)$-form $V_{D-p-1}$:
\begin{equation}
	\label{Qu_Lb}
	L_{\text{master}} = - \frac{1}{2 g^2} \Theta_p \wedge * \Theta_p + \Theta_p \wedge {\rm d}  V_{D-p-1}\,.
\end{equation}
Starting from \eqref{Qu_Lb}, one can follow two paths. First, integrating out $V_{D-p-1}$ from \eqref{Qu_Lb} yields ${\rm d} \Theta_p = 0$. The latter can be locally solved by setting $\Theta_p = {\rm d} A_{p-1}$ for a generic $(p-1)$-form $A_{p-1}$. Plugging such a solution in \eqref{Qu_Lb} gives back the electric-frame Lagrangian \eqref{Qu_La}. Alternatively, one can integrate out $\Theta_p$ from \eqref{Qu_Lb}, obtaining
\begin{equation}
	\label{Qu_V}
	\frac{1}{g^2} * \Theta_p = {\rm d} V_{D-p-1} \equiv G_{D-p} \,.
\end{equation}
Plugging this solution in the Lagragian \eqref{Qu_Lb} delivers the `dual' Lagrangian
\begin{equation}
	\label{Qu_Lc}
	\begin{aligned}
	L_{\rm dual} &= - \frac{g^2}{2} G_{D-p} \wedge * G_{D-p}\,,
	\end{aligned}
\end{equation}
that solely depends on the field strength of the magnetic field $V_{D-p-1}$, dual to $A_{p-1}$.
	
Analogously to the electric $U_{\rm e}(1)$, here the action of the global $U_{\rm m}(1)$ is manifest, for it acts by shifting $V_{D-p-1} \to V_{D-p-1} + \Lambda_{D-p-1}$, with ${\rm d} \Lambda_{D-p-1} = 0$. Accordingly, Noether's first theorem delivers the conserved $p$-form current
\begin{equation}
	\label{Qu_Jmag}
	J^{\rm m}_{p} =  (-1)^{D-p-1} g^2 * G_{D-p}\,,\qquad  {\rm d} J^{\rm m}_{p}  = 0\,. 
\end{equation}
Thus, taking a closed $p$-surface $\Sigma_p$, we the magnetic charges obey the quantization condition
\begin{equation}
	\label{Qu_Qm}
	Q^{\rm m} = (-1)^{D-p-1} g^2 \int_{\Sigma_p} * G_{D-p} = (-1)^{(D-p)(p+1)} \int_{\Sigma_p} F_p\in \mathbb{Z}\,.
\end{equation}
%


\section{Gauge interactions in asymptotic limits}
\label{sec:Hodge}

In this section, we review some basic facts about asymptotic Hodge theory and we show how the tools that such a framework provides may be helpful to study the vector multiplet gauge interactions of any near-boundary Type $\mathcal{N}=2$ IIB EFT.
The technology reviewed here is used extensively throughout Sections~\ref{sec:GGS_01mod} and~\ref{sec:GGS_gen_mod} of the main text.

\subsection{A brief overview of asymptotic Hodge Theory}
\label{sec:Hodge_review}

We start by collecting some of the basic concepts of asymptotic Hodge Theory. Our exposition will be brief, and we refer to \cite{Grimm:2018cpv,Grimm:2021ckh} for detailed reviews on the subject. In general, the complex structure moduli space $\mathcal{M}_{\rm cs}$ is noncompact, and it may be characterized by singularities that serve as \emph{boundaries} for the moduli space. These singularities are distributed along intersecting divisors. We can choose local coordinates $z^\alpha$ such that each boundary component, corresponding to $k \leq h^{2,1}$ intersecting divisors, is described as the vanishing locus $z^\alpha = 0$, with $\alpha =1, \ldots, k$. Each boundary component is further characterized by a set of monodromy transformations $T_\alpha \in {\rm Sp}(2h^{2,1}+2,\mathbb{R})$ of the kind introduced in Section~\ref{sec:IIB_Sympl}. Upon encircling the singularity $z^{\alpha}=0$ along one of the $m$ transverse directions by drawing a path that sends $z^\alpha \to e^{2 \pi \im} z^\alpha$, some quantities may non-trivially transform, and their transformation laws are dictated by appropriate representation of the monodromy group.

Given the singularity $z^\alpha = 0$, we will focus on the near-boundary region $\mathcal{E}$ that we characterize as the punctured polydisk $\mathcal{E} = (\Delta^*)^k \times (\Delta)^{h^{2,1}-k}$ where, $\Delta \coloneqq \{ |z| < 1\}$ and $\Delta^* \coloneqq \{ 0< |z| < 1\}$. To emphasize this split we similarly decompose our local coordinates as $(z^{\alpha},\zeta^i)$. The coordinates $\zeta^i$ of the disk $\Delta$ are referred to as spectator moduli. They will not play an important role in the following and we suppress any explicit dependence on these spectator moduli. It is convenient to introduce the following parametrization for the $k$ punctured components
\begin{equation}
	\label{HT_phi}
	t^\alpha \equiv a^\alpha+ \im s^\alpha = \frac{1}{2 \pi \im} \log z^\alpha\,,
\end{equation}
in which the singularities are reached towards large values of the saxions $s^\alpha$ and encircling the locus $z^{\alpha} = 0$ once corresponds to shifting the axions $a^\alpha$ by one period as $a^\alpha \to a^\alpha +1$. 
The near boundary region of interest is then described by the upper-half plane $\cH^m = \{t^\alpha = a^\alpha + \im s^\alpha \mid 0 \leq a^\alpha \leq 1, s^\alpha > 0 \}$, with the identification $a^\alpha \simeq a^\alpha + 1$. 

Let us now better characterize the properties of the monodromy matrices $T_\alpha$.
Around each singularity, one can decompose each of the monodromy matrices $T_\alpha$ into a \emph{semi-simple}, finite-order part $T_\alpha^{(s)}$ and into a \emph{unipotent}, infinite-order $T_\alpha^{(u)}$ as $T_\alpha = T_\alpha^{(s)} T_\alpha^{(u)}$. These satisfy the following properties \cite{Grimm:2018ohb}:
\begin{equation}
	\label{HT_Tus}
	\begin{aligned}
		&(T_\alpha^{(s)})^{m_\alpha-1} \neq \mathds{1} \,, &\qquad&  (T_\alpha^{(s)})^{m_\alpha} = \mathds{1}\,,
		\\
		&(T_\alpha^{(u)} - \mathds{1})^{n_\alpha} \neq 0\,, &\qquad&  (T_\alpha^{(u)} - \mathds{1})^{n_\alpha} = 0\,,
	\end{aligned}
\end{equation}
for some $m_\alpha, n_\alpha \in \mathbb{N}$. For the upcoming discussion, it will be convenient to extract the unipotent part of the monodromy matrices by defining the \emph{log-monodromy matrices}
\begin{equation}
	\label{HT_Nalpha}
	N_\alpha := \frac{1}{m_\alpha} \log T^{m_\alpha} = \log T_\alpha^{(u)} \,,
\end{equation}
which are nilpotent matrices, obeying $N_\alpha^{n_\alpha} \neq 0$, $N_\alpha^{n_\alpha+1}=0$, for some $n_\alpha \in \mathbb{N}$. In particular, for Calabi-Yau three-folds, $0 \leq n_\alpha \leq 3$.
Moreover, the log-monodromy matrices associated to different axion shifts around the same boundary component commute, as $[N_\alpha, N_\beta] = 0$. The log-monodromy matrices just introduced will play a critical role in the analysis that follows, and in the following sections we shall see that some physical quantities are closely related to their properties.

One of the core points of asymptotic Hodge Theory resides in the identification of the structures that characterize the near-boundary regime of the complex structure moduli space.
To begin with, let us recall that the middle cohomology of a Calabi-Yau three-fold splits as
\begin{equation}
	\label{HT_Hdec}
	H^3(Y, \mathbb{C}) = \bigoplus\limits_{p=0}^3 H^{3-p,p} = H^{3,0} \oplus H^{2,1} \oplus H^{1,2} \oplus H^{0,3}\,,
\end{equation}
such that $\overline{H^{p,q}} = H^{q,p}$. Such a decomposition identifies a \emph{pure} Hodge structure, namely it exhibits a fixed \textit{weight}, given by $w = p+q=3$. 
The complex structure moduli space is then characterized by how the Calabi-Yau holomorphic three-form $\Omega$ varies inside $H^3(Y, \mathbb{C})$ as a function of the moduli.
In order to describe such variations, it will prove more convenient to introduce the \emph{Hodge filtration}:
\begin{equation}
	\label{HT_Fdec}
	\begin{aligned}
		&F^3 = H^{3,0}\,,  &\qquad&  F^2 = H^{3,0} \oplus H^{2,1}\,,
		\\
		&F^1 = H^{3,0} \oplus H^{2,1} \oplus H^{1,2}\,,  &\qquad&  F^0 = H^{3,0} \oplus H^{2,1} \oplus H^{1,2} \oplus H^{0,3}\,,
	\end{aligned}	
\end{equation}
from which one retrieves the Hodge decomposition \eqref{HT_Hdec} by setting $H^{p,q} = F^p \cap \overline{F^q}$.
Given the Gauss-Manin connection $\nabla_{\alpha} \equiv \nabla_{\partial/\partial z^\alpha}$, one has $\nabla_{\alpha} F^p \subset F^{p-1}$. For our purposes, the advantage of the Hodge filtration \eqref{HT_Fdec} is that it can be fully spanned by the Calabi-Yau three-form and its derivatives.

In general, it is very hard to construct the filtration \eqref{HT_Fdec} at arbitrary points in the moduli space. One should indeed compute the periods \eqref{IIB_periodsymplb} explicitly, which can be achieved by solving a set of Picard-Fuchs equations (see, for instance,~\cite{Morrison:1991cd}), computing its Gauss-Manin derivatives and then reconstructing the filtration \eqref{HT_Fdec}.
Nevertheless, in the near-boundary region $s^\alpha \gg 1$, the problem greatly simplifies.
Near any moduli space boundary, the filtration $F^p$ can be approximated by means of \emph{Schimd's nilpotent orbit theorem} \cite{MR0382272}, which states that, in the asymptotic region characterized by large values of the saxions, the Hodge filtration takes the following asymptotic form
\begin{equation}
	\label{HT_Fnil}
	F^p  = e^{t^\alpha N_\alpha} e^{\Gamma(z)} F^p_0\,.
\end{equation}
Here, $F^p_0$ denotes the (moduli-independent) \emph{limiting Hodge filtration}, defined via the relation $ F^p_0 : = \lim\limits_{t^\alpha \to \im \infty} e^{- t^\alpha N_\alpha}  F^p$;
moreover, $\Gamma(z)$ is a holomorphic $\mathfrak{sp}(2h^{2,1}+1)$-valued map, such that $\Gamma(0) = 0$, to which we will refer as \emph{instanton map}. Any singular behavior of the Hodge filtration near the singularity is thus completely characterized by the log-monodromy matrices. 

It is worth stressing that equation \eqref{HT_Fnil} is a vector space relation. However, evaluating it for $p=3$ leads to an asymptotic form for the period vector ${\bm \Pi}$, appearing in the expansion of the Calabi-Yau three-form in \eqref{N2_Omega}, as 
\begin{equation}
	\label{HT_Per_nil}
	{\bm \Pi}  = e^{t^\alpha N_\alpha} e^{\Gamma(z)} {\bf a}_0 = e^{t^\alpha N_\alpha} \left({\bf a}_0 + z^{\alpha} {\bf a}_{\alpha}  + z^{\alpha}z^{\beta} {\bf a}_{\alpha\beta} + \ldots \right) \simeq e^{t^\alpha N_\alpha} {\bf a}_0\equiv {\bm \Pi}_{\rm nil} .
\end{equation}
Here, ${\bf a}_0 \in F^p_0$ does not carry any dependence on the moduli. Remarkably, due to the nilpotency of the matrices $N_\alpha$, the leading contribution to the periods \eqref{HT_Per_nil} is polynomial in the complex structure moduli $t^\alpha$. Eventual corrections to the leading contributions are suppressed exponentially in the saxionic fields $s^\alpha$.

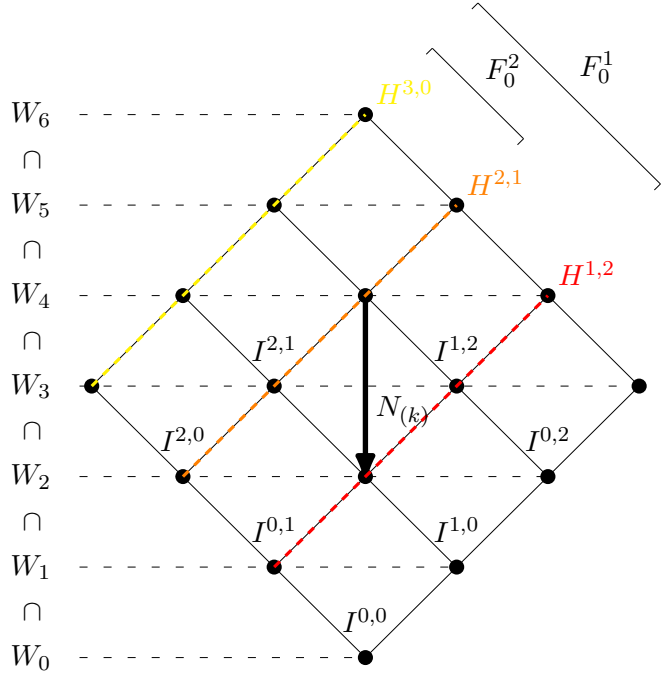
\begin{figure}
    \centering
    \begin{tikzpicture}[scale=0.8]
        \draw[step=2.12132034354,black,rotate=45, very thin] (0,0) grid (6.363961032,6.363961032);
        \tkzDefPoint(0,9){I33}
        \tkzDefPoint(1.5,7.5){I23}
        \tkzDefPoint(-1.5,7.5){I32}
        \tkzDefPoint(-3,6){I31}
        \tkzDefPoint(0,6){I22}
        \tkzDefPoint(3,6){I13}
        \tkzDefPoint(-4.5,4.5){I30}
        \tkzDefPoint(-1.5,4.5){I21}
        \tkzDefPoint(1.5,4.5){I12}
        \tkzDefPoint(4.5,4.5){I03}
        \tkzDefPoint(-3,3){I20}
        \tkzDefPoint(0,3){I11}
        \tkzDefPoint(3,3){I02}
        \tkzDefPoint(-1.5,1.5){I10}
        \tkzDefPoint(1.5,1.5){I01}
        \tkzDefPoint(0,0){I00}
        \foreach \n in {I33,I32,I23,I31,I22,I13,I30,I21,I12,I03,I20,I11,I02,I10,I01,I00}
          \node at (\n)[circle,fill,inner sep=2pt]{};
        \draw[black,line width=2pt,-{Latex[round]},] (I22) -- (I11);
        \node[anchor=east] at (-5,0){$W_0$};  \node[anchor=east] at (-5.15,.75){$\cap$};
        \node[anchor=east] at (-5,1.5){$W_1$}; \node[anchor=east] at (-5.15,2.25){$\cap$};
        \node[anchor=east] at (-5,3){$W_2$};  \node[anchor=east] at (-5.15,3.75){$\cap$};
        \node[anchor=east] at (-5,4.5){$W_3$};  \node[anchor=east] at (-5.15,5.25){$\cap$};
        \node[anchor=east] at (-5,6){$W_4$};  \node[anchor=east] at (-5.15,6.75){$\cap$};
        \node[anchor=east] at (-5,7.5){$W_5$};  \node[anchor=east] at (-5.15,8.25){$\cap$};
        \node[anchor=east] at (-5,9){$W_6$};
        \draw[loosely dashed, very thin] (-4.7,0) -- (0,0);
        \draw[loosely dashed, very thin] (-4.7,1.5) -- (1.5,1.5);
        \draw[loosely dashed, very thin] (-4.7,3) -- (3,3);
        \draw[loosely dashed, very thin] (-4.7,4.5) -- (4.5,4.5);
        \draw[loosely dashed, very thin] (-4.7,6) -- (3,6);
        \draw[loosely dashed, very thin] (-4.7,7.5) -- (1.5,7.5);
        \draw[loosely dashed, very thin] (-4.7,9) -- (0,9);
        \node[anchor=south west, yellow] at (0,9){$H^{3,0}$};
        \draw[dashed, very thick, yellow] (-4.5,4.5)--(0,9);
        \node[anchor=south west, orange] at (1.5,7.5){$H^{2,1}$};
        \draw[dashed, very thick, orange] (-3,3)--(1.5,7.5);
        \node[anchor=south west, red] at (3,6){$H^{1,2}$};
        \draw[dashed, very thick, red] (-1.5,1.5)--(3,6);
        \node[anchor=west] at (0,4.08){$N_{(k)}$};
        \node[anchor=south] at (0,0.25){$I^{0,0}$};
        \node[anchor=south] at (-1.5,1.75){$I^{0,1}$};
        \node[anchor=south] at (1.5,1.75){$I^{1,0}$};
        \node[anchor=south] at (-3,3.25){$I^{2,0}$};
        \node[anchor=south] at (3,3.25){$I^{0,2}$};
        \node[anchor=south] at (-1.5,4.75){$I^{2,1}$};
        \node[anchor=south] at (1.5,4.75){$I^{1,2}$};
        \draw[very thin] (1,10) -- (1.1,10.1);
        \draw[very thin] (2.5,8.5) -- (2.6,8.6);
        \draw[very thin] (1.1,10.1) -- (2.6,8.6);
        \node[anchor=south west] at (1.8,9.3){$F_0^2$};
        \draw[very thin] (1.75,10.75) -- (1.85,10.85);
        \draw[very thin] (4.75,7.75) -- (4.85,7.85);
        \draw[very thin] (1.85,10.85) -- (4.85,7.85);
        \node[anchor=south west] at (3.35,9.35){$F_0^1$};
    \end{tikzpicture}
    \caption{\label{fig:hddiamond}A Hodge-Deligne diamond illustrating how the spaces $F_0^p$, $W_l$, $H^{p,q}$ and $I^{p,q}$ are related. The arrow indicates a generic action of the log-monodromy matrix acting on these spaces. }
\end{figure}

As is clear from \eqref{HT_Per_nil}, the approximation of the periods achieved via the nilpotent orbit theorem relies on two data: the nilpotent matrices $N_\alpha$ and the limiting filtration $F^p_0$. This data can succinctly be captured by associating to it a \emph{mixed Hodge structure} whose properties are particularly well-understood. 
We consider a boundary reached as $s^\alpha \to \infty$, for $\alpha = 1,\ldots, k$ and set $N_{(k)} := N_1 + \ldots N_k$ and, in order to keep track of the saxions that acquire large vevs towards such a boundary, we denote as $F^p_{(k)}$ the limiting filtration $F^p_0$.
Then, we first define the \emph{monodromy weight filtration}:
\begin{equation}
	\label{HT_Wl}
	W^{(k)}_\ell (N_{(k)}) = \sum\limits_{j \geq {\rm max}(-1,\ell-3)} \ker N_{(k)}^{j+1} \cap {\rm Im} N_{(k)}^{j- \ell +3}\,.
\end{equation}
While the filtration $W^{(k)}_\ell (N_{(k)})$ involves only the nilpotent matrices $N_{(k)}$, we can use them to encode the limiting filtration $F_0^p$ in a \emph{Deligne splitting}:  
\begin{equation}
	\label{HT_Ipq}
	I^{p,q}_{(k)} = F^p_{(k)} \cap W^{(k)}_{p+q} \cap \Big(\bar{F}^q_{(k)} \cap W^{(k)}_{p+q} + \sum\limits_{j \geq 1} \bar{F}^{q-j}_{(k)} \cap W^{(k)}_{p+q-j-1}\Big)\,.
\end{equation}
This splitting identifies a \emph{mixed} Hodge structure, namely a Hodge structure involving sub-spaces $I^{p,q}$ of various weights $w=p+q$. Moreover, the limiting filtration $F_{(k)}^p$ and the monodromy weight filtration $W_\ell^{(k)}$ can be recovered from the Deligne splitting \eqref{HT_Ipq} as follows:
\begin{equation}
	\label{HT_F0Wl_Ipq}
	F_{(k)}^p = \bigoplus_{r\geq p} \bigoplus_s I^{r,s}_{(k)}, \qquad W^{(k)}_\ell = \bigoplus_{p+q =\ell} I^{p,q}_{(k)}\,.
\end{equation}
It will be convenient to organize the subspaces $I^{p,q}_{(k)}$ in a diagram as in Figure~\ref{fig:hddiamond}, on which the action of the nilpotent matrices $N_{(k)}$ can neatly be depicted. In particular, we have that $N_{(k)} I^{p,q}_{(k)} \subset I^{p-1,q-1}_{(k)}$ so that the nilpotent matrices $N_{(k)}^n$ map a subspace $I^{p,q}_{(k)}$ into another along the same vertical lines, as depicted in Figure~\ref{fig:hddiamond}. 
It can be shown that the instanton map can relate different $I^{p,q}_{(k)}$ spaces that do not lie on the same vertical lines -- we refer to \cite{Bastian:2021eom} for a detailed discussion.

It is worth mentioning that, in general, the spaces $I^{p,q}_{(k)}$ and $I^{q,p}_{(k)}$ are not related via a complex conjugation, rather $\overline{I^{p,q}}_{(k)} = I^{q,p}_{(k)} {\rm mod} \bigoplus\limits_{r < q,s<p} I^{r,s}_{(k)}$. However, in this work we will focus on $\mathbb{R}$-split decompositions $\{{\hat I}^{p,q}\}_{p,q = 0, \ldots, 3}$, for which $\overline{{\hat I}^{p,q}} = {\hat I}^{q,p}$ and \eqref{HT_Ipq} reduces to the simpler 
\begin{equation}
	\label{HT_Ipq_real}
	{\hat I}^{p,q}_{(k)} = F^p_{(k)} \cap \bar{F}^q_{(k)} \cap W^{(k)}_{p+q}\,.
\end{equation}

Altogether, the mixed Hodge structures introduced above can be used to give a coarse but useful classification of singular Calabi-Yau three-folds. In particular, we may use classify them by the Hodge-Deligne numbers of their associated Deligne splitting, defined as
\begin{equation}
    i^{p,q}\coloneqq \text{dim}_{\mathbb{C}}I^{p,q}.
\end{equation}
Many physically relevant quantities in the asymptotic regime depend only on the boundary type in this classification. Moreover, the Calabi-Yau condition and the properties of the middle cohomology place various constraints on what these Hodge-Deligne numbers can be. In particular, only one of $i^{3,q}$ for $q=0,1,2,3$ can be non-zero. We label these possibilities by roman numerals $\rm{I},\rm{II},\rm{III},\rm{IV}$. For a given $h^{2,1}$, the remaining $i^{p,q}$ are fixed completely by the value of $i^{1,1}$, so that we add this number as a subscript. This classification is summarized in Table~\ref{table:HDclass}, and we refer to \cite{Bastian:2021eom} for further details.

\begin{table}[h!]
\centering
\renewcommand*{\arraystretch}{2.0}
\begin{tabular}{| c| c | c | c | c |}
\hline boundary & $\mathrm{I}_a$ & $\mathrm{II}_b$ & $\mathrm{III}_c$ & $\mathrm{IV}_d$ \\ \hline \hline 
\begin{minipage}{0.15\textwidth}
\vspace{-1.3cm}
HD diamond
\vspace{1.3cm}
\end{minipage} &
\rule[-0.25cm]{.0cm}{3.5cm} \begin{tikzpicture}[scale=0.65,cm={cos(45),sin(45),-sin(45),cos(45),(15,0)}]
  \draw[step = 1, gray, ultra thin] (0, 0) grid (3, 3);

  \draw[fill] (0, 3) circle[radius=0.05];
  \draw[fill] (1, 2) circle[radius=0.05] node[above]{$a'$};
  \draw[fill] (2, 1) circle[radius=0.05] node[above]{$a'$};
  \draw[fill] (1, 1) circle[radius=0.05] node[above]{$a$};
  \draw[fill] (2, 2) circle[radius=0.05] node[above]{$a$};
  \draw[fill] (3, 0) circle[radius=0.05];
\end{tikzpicture} &
\begin{tikzpicture}[scale=0.65,cm={cos(45),sin(45),-sin(45),cos(45),(15,0)}]
  \draw[step = 1, gray, ultra thin] (0, 0) grid (3, 3);

  \draw[fill] (0, 2) circle[radius=0.05];
  \draw[fill] (1, 3) circle[radius=0.05];
  \draw[fill] (1, 2) circle[radius=0.05] node[above]{$b'$};
  \draw[fill] (1, 1) circle[radius=0.05] node[above]{$b$};
  \draw[fill] (2, 1) circle[radius=0.05] node[above]{$b'$};
  \draw[fill] (2, 2) circle[radius=0.05] node[above]{$b$};
  \draw[fill] (2, 0) circle[radius=0.05];
  \draw[fill] (3, 1) circle[radius=0.05];
\end{tikzpicture} &
\begin{tikzpicture}[scale=0.65,cm={cos(45),sin(45),-sin(45),cos(45),(15,0)}]
  \draw[step = 1, gray, ultra thin] (0, 0) grid (3, 3);
  \draw[fill] (0, 1) circle[radius=0.05];
  \draw[fill] (1, 0) circle[radius=0.05];
  \draw[fill] (1, 2) circle[radius=0.05] node[above]{$c'$};
  \draw[fill] (2, 1) circle[radius=0.05] node[above]{$c'$};
  \draw[fill] (2, 3) circle[radius=0.05];
  \draw[fill] (1, 1) circle[radius=0.05] node[above]{$c$};
  \draw[fill] (2, 2) circle[radius=0.05] node[above]{$c$};
  \draw[fill] (3, 2) circle[radius=0.05];
\end{tikzpicture} &
\begin{tikzpicture}[scale=0.65,cm={cos(45),sin(45),-sin(45),cos(45),(15,0)}]
  \draw[step = 1, gray, ultra thin] (0, 0) grid (3, 3);

  \draw[fill] (0, 0) circle[radius=0.05];
  \draw[fill] (1, 1) circle[radius=0.05] node[above]{$d$};
  \draw[fill] (1, 2) circle[radius=0.05] node[above]{$d'$};
  \draw[fill] (2, 1) circle[radius=0.05] node[above]{$d'$};
  \draw[fill] (2, 2) circle[radius=0.05] node[above]{$d$};
  \draw[fill] (3, 3) circle[radius=0.05];
\end{tikzpicture} \\ \hline
index &  \begin{minipage}{.15\textwidth}\centering\vspace*{0.3cm} \begin{equation*}\begin{aligned} a+a'&=h^{2,1} \\ 0\leq a &\leq h^{2,1} \end{aligned}\end{equation*}  \vspace*{-0.2cm} \end{minipage} & \begin{minipage}{.18\textwidth}\centering\vspace*{0.3cm} \begin{equation*}\begin{aligned}b+b'&=h^{2,1}-1 \\ 0\leq b &\leq h^{2,1}-1 \end{aligned}\end{equation*}  \vspace*{-0.2cm} \end{minipage}& \begin{minipage}{.18\textwidth}\centering\vspace*{0.3cm} \begin{equation*}\begin{aligned} c+c'&=h^{2,1}-1 \\ 0 \leq c &\leq h^{2,1}-2 \end{aligned}\end{equation*}  \vspace*{-0.2cm}\end{minipage}&\begin{minipage}{.15\textwidth}\centering\vspace*{0.3cm} \begin{equation*}\begin{aligned} d+d'&=h^{2,1} \\ 1 \leq d &\leq h^{2,1} \end{aligned}\end{equation*}  \vspace*{-0.2cm} \end{minipage}\\ \hline
$\text{rk}(N,N^2,N^3)$ & $(a,\, 0,\, 0)$ & $(2+b,\, 0,\, 0)$ & $(4+c,\, 2,\, 0)$ & $(2+d,\, 2,\, 1)$ \\ \hline
eigvals $\eta N$ & $a$ negative & \begin{minipage}{0.15\textwidth}\centering\vspace*{0.2cm}
$b$ negative \\
2 positive  \vspace*{0.15cm}
\end{minipage} & not needed & not needed \\ \hline
\end{tabular}
\caption{\label{table:HDclass} Classification of boundary types in complex structure moduli space based on the $4h^{2,1}$ possible different Hodge-Deligne diamonds. In each Hodge-Deligne diamond we indicated non-vanishing $i^{p,q}$ by a dot on the roster, where the dimension has been given explicitly when $i^{p,q} >1$. In the last two rows we listed the characteristic properties of the log-monodromy matrix $N$ and the symplectic pairing $\eta$ that are sufficient to make a distinction between the types. The table is taken from \cite{Bastian:2021eom}.}
\end{table}

While the classification introduced above is sufficient for many purposes, studying the details of the gauge kinetic sector will require a more refined approximation scheme than that afforded by the nilpotent orbit theorem. In particular, we will make use of certain elements of the \textit{SL(2)-orbit theorem}, due in full generality to Cattani, Kaplan and Schmid \cite{MR840721}. While the full power of this result becomes apparent only near intersecting divisors in higher-dimensional moduli spaces, for the purposes of this work we will focus on the case $h^{2,1}=1$.

The SL(2)-orbit theorem introduces several additional structures. In particular, it states that there is a special $\mathbb{R}$-split Deligne splitting, denoted $\tilde{I}^{p,q}$, associated to the one obtained from the pair $(W_l(N), F_0^p)$, which makes manifest the relationship to certain $\mathfrak{sl}(2)$-triples. This special splitting is obtained from the original by rotating the limiting Hodge filtration using a pair $\delta,\zeta$ of $\mathfrak{sp}(2h^{2,1}+2,\mathbb{R})$-valued operators. The new filtration is then defined as
\begin{equation}
    \tilde{F}_0^p=e^{\zeta}\hat{F}_0^p=e^{\zeta}e^{-\im\delta}F_0^p.
\end{equation}
We refer to \cite{Bastian:2021eom} for the explicit definition of these operators. One then evaluates Equation~\ref{HT_Ipq} (or, because it is $\mathbb{R}$-split, Equation~\ref{HT_Ipq_real}) to obtain the so-called $\mathfrak{sl}(2)$-splitting.

Next, we note that any nilpotent element of $\mathfrak{sp}(2h^{2,1}+2,\mbb{R})$ can be completed to a triple
\begin{equation}
	\label{HT_sl2_triples}
	\{N\equiv N^-, N^+, N^0\}\in\mathfrak{sp}(2h^{2,1}+2,\mbb{R})\,,
\end{equation}
satisfying the $\mathfrak{sl}(2)$ algebra
\begin{equation}
	\label{HT_sl2_rel}
	[N^0, N^+] = 2 N^+\,, \qquad [N^0, N^-] = 2 N^-\,, \qquad [N^+, N^-] = N^0\,.
\end{equation}
The choice of triple is not unique, however, but the $\mathfrak{sl}(2)$-splitting picks out a particular such choice. Indeed, there now exist a unique operator $N^0$ which acts on the $\mathfrak{sl}(2)$-splitting as a weight operator
\begin{equation}
    N^0\tilde{I}^{p,q}=(p+q-3)\tilde{I}^{p,q}.
\end{equation}
The $\mathfrak{sl}(2)$-splitting, along with the triple \ref{HT_sl2_triples}, constitutes the $\mathfrak{sl}(2)$-data associated to a given boundary component which enters in the $\mathfrak{sl}(2)$-approximated Hodge star, to be introduced in the following section. Before moving on, however, let us briefly comment on the general case, i.e. when $1<k\leq h^{2,1}\neq 1$ are sent to the boundary. In this case one constructs a set of $k$ mutually commuting $\mathfrak{sl}(2)$-triples. However, the identification $N_i=N_i^-$ is not immediate, and one should perform an iterative procedure to construct these triples. We refer to \cite{Bastian:2021eom} for further details.

\subsection{The asymptotic form of the gauge-kinetic matrix and the generalized \texorpdfstring{$\theta$}{Theta}-terms}
\label{sec:Hodge_gauge}

In this section we will illustrate how the knowledge of the boundary data can be used in order to deliver the leading asymptotic terms of the matrix \eqref{IIB_Mmat}, from which both the gauge kinetic terms and generalized $\theta$-terms appearing in the generic action \eqref{N2_action} can be retrieved.

Let us preliminarily identify the operators that microscopically characterize the matrix \eqref{IIB_Mmat}.
As is clear from \eqref{IIB_D3q}, the matrix \eqref{IIB_Mmat} captures the moduli dependence introduced by the `Hodge star' operator, $\star$. 
Thus, let us preliminarily introduce the \emph{Weil operator} $C$, that realizes the action of the Hodge-$\star$ in the middle cohomology $H_3(Y)$. For any given $\omega^{p,q} \in H^{p,q}(Y)$ of the Dolbeault cohomology, the $C$-operator acts as a simple multiplication as
\begin{equation}
	\label{HT_Csymdef}
	C \omega^{p,q}  = \im^{p-q} \omega^{p,q}\,.
\end{equation}
In the real, symplectic basis $\{\alpha_I, \beta^J\}$ introduced in Section~\ref{sec:IIB_review}, the action of Hodge $\star$-operator is rather expressed in terms of a matrix $C$ as 
\begin{equation}
	\label{HT_Csymb}
	\star \begin{pmatrix}
		\alpha_I \\ \beta^I 
	\end{pmatrix} = C \begin{pmatrix}
	\alpha_I \\ \beta^I
\end{pmatrix}\,,
\end{equation}
and the matrix $C$ can be fully expressed in terms of the matrix $\mathcal{N}_{IJ}$ defined in \eqref{IIB_NIJ} as \cite{Ceresole:1995ca}
\begin{equation}
	\label{HT_Csym}
	C = \begin{pmatrix}
		{\rm Re} \mathcal{N} ({\rm Im} \mathcal{N})^{-1} & -  {\rm Im} \mathcal{N} - {\rm Re} \mathcal{N} ({\rm Im} \mathcal{N})^{-1} {\rm Re} \mathcal{N} 
		\\
		({\rm Im} \mathcal{N})^{-1}  & - ({\rm Im} \mathcal{N})^{-1} {\rm Re} \mathcal{N} 
	\end{pmatrix} 
\end{equation}
In such a symplectic basis, the matrix \eqref{IIB_Mmat} can be straightforwardly obtained from the matrix \eqref{HT_Csym} as:
\begin{equation}
	\label{HT_Msym}
	 \mathcal{T}  = -\eta C\,.
\end{equation}

Now, our aim is to obtain an approximation to the matrix \eqref{HT_Msym}. This requires the introduction of some additional boundary data, whose existence is guaranteed by the SL(2)-orbit theorem. In particular, we have that given the mixed Hodge structure underlying the $\mathfrak{sl}(2)$-splitting, we can associate to it a pure Hodge structure. The latter is defined as follows
\begin{equation}
    F_{\infty}^p\coloneqq e^{\im N^-}\tilde{F}_0^p\,,\qquad H^{p,q}_{\infty}\coloneqq F_{\infty}^p\cap \bar{F}_{\infty}^q.
\end{equation}
Just as for the pure Hodge structure present away from the singularity, this pure Hodge structure comes with its own Weil operator, denoted $C_{\infty}$, which acts on elements of the cohomology in the usual way:
\begin{equation}\label{HT_Cinf}
    C_{\infty}\omega^{p,q}=\im^{p-q}\omega^{p,q}\,,\qquad \omega^{p,q}\in H_{\infty}^{p,q}.
\end{equation}
%
One then extends this (moduli-independent) boundary Weil operator into the near-boundary region using the $\mathfrak{sl}(2)$-data as follows \cite{MR840721}:
\begin{equation}
	\label{HT_Csl2}
	C_{\mathfrak{sl}(2)} = e^{a^k N_k^-} e(s)^{-1} C_\infty e(s) e^{-a^kN_k^-}  \,, \qquad {\rm with}\quad e(s) := \exp \left[\frac12 \sum\limits_{k} \log (s^k) N_k^0 \right]\,,
\end{equation}
Consequently, the matrix \eqref{HT_Msym} gets approximated as
\begin{equation}
	\label{HT_Msl2}
	 \mathcal{T} _{\mathfrak{sl}(2)} = -\eta C_{\mathfrak{sl}(2)} \,.
\end{equation}

Among the advantages of working with an ${\mathfrak{sl}}(2)$-basis is the simplicity of estimating the growth of some physical quantities, namely those that can be recast as \emph{Hodge norms}.
The Hodge norm of any given $\omega \in H^3(Y,\mathbb{C})$ is defined as 
\begin{equation}
\label{HT_norm}
    \| \omega \|^2 := \int_Y \omega \wedge \star \,\bar{\omega}\,.
\end{equation}
Then, on the set
\begin{equation}
    \Sigma_{m} = \{ |a^\alpha| < \delta,\ \  s^1 >  s^2 >\cdots > s^m > 1 \}\,,
\end{equation}
the norm of any given $\mathfrak{sl}(2)$-eigenvector $\omega \in V_{\ell_1 \cdots \ell_m}$ can be estimated as \cite{MR840721}
\begin{equation}
\label{HT_norm_growth}
    \| \omega \|^2 \sim \left( \frac{s^1}{s^2} \right)^{\ell_1} \cdots \left( \frac{s^{m - 1}}{s^m} \right)^{\ell_{m - 1}} (s^m)^{\ell_m}
\end{equation}
The D3-particle physical charge \eqref{IIB_D3Q} is an example of such a Hodge norm, being given by
\begin{equation}
\label{HT_Qsl2}
    \mathcal{Q}_{\bf q}^2 = \frac12 \| {\bf q} \|^2\,,
\end{equation}
and, as such, the charge of any D3-particle residing in an $\mathfrak{sl}(2)$-eigenspace $V_{\ell_1 \cdots \ell_m}$ can be estimated via \eqref{HT_norm_growth}.

\subsection{Outline of the procedure}\label{sec:Hodge_outline}

Having summarized the basic knowledge about asymptotic Hodge Theory, we now outline the procedure that can be followed in order to grasp the approximate, asymptotic generalized global symmetries that emerge towards any generic moduli space boundary reached as $s^1, \ldots s^m \to \infty$. For simplicity, let us first consider the case where the boundary is reached when only a single modulus is sent to large vev, namely $s^1 \to \infty$.
Our strategy consists of the following steps:
\begin{enumerate}
    \item  \label{Proc_1} The first step is to fix a basis $\{e_{\mathcal{I}}\}$, $\mathcal{I} = 1, \ldots, 2h^{2,1}+2$, for the space $H_{\mbb{C}}$ with respect to which we will cast our results. This basis cannot be arbitrary however, and should satisfy three important properties:
    \begin{enumerate}[(a)] 
        \item \textbf{Symplectic:} The first property is that we wish for it to be symplectic with respect to the Mukai pairing
        \begin{equation}
            \eta_{\mathcal{I}\mathcal{J}} = \int_Y e_\mathcal{I} \wedge e_\mathcal{J}
        \end{equation}
        on $H_{\mbb{C}}$. While it is not strictly necessary that this pairing take the canonical form \eqref{Sympl_eta}, it will be convenient for physical applications. This means concretely that our basis elements can be identified with the symplectic basis elements as
        \begin{equation}
        \begin{split}
            &e_1\cong \alpha_0,
            \; \ldots,\; e_{h^{2,1}+1} \cong \alpha_{h^{2,1}},\; e_{h^{2,1}+2}\cong\beta^0,\; \ldots,\; e_{2h^{2,1}+2}\cong\beta^{h^{2,1}},
        \end{split}
        \end{equation}
        \item \textbf{Integral:} The second property is that it should be integral, which is to say that it should constitute a $\mbb{Z}$-basis\footnote{A $\mbb{Z}$-basis is a basis $\{e_{\mathcal{I}}\}$ for $H_{\mbb{R}}$ such that any point in $H_{\mbb{Z}}$ is a linear combination of the $e_{\mathcal{I}}$ with integer coefficients. It follows that the elements $\{e_{\mathcal{I}}\}$ are primitive, i.e. $\nexists n\in\mbb{Z}, |n|\geq 2: [e_{\mathcal{I}}]=0\in H_{\mbb{Z}}/nH_{\mbb{Z}}$ (intuitively, $e_i$ is not an integer multiple of some other element in the lattice). This is always possible when the lattice $H_{\mbb{Z}}$ is torsion-free \cite{Green:2008}.}  for the lattice $H_{\mbb{Z}}\subset H_{\mbb{C}}$. What we mean by this is that the basis should be primitive with respect to the lattice $H_{\mbb{Z}}\subset H_{\mbb{C}}$. Recall that this lattice is given by $H^3(X,\mbb{Z})$, which is Poincar\'e dual to $H_3(X,\mbb{Z})$. In turn, this implies that D3-particles have \emph{integral} charge vectors and that the associated field strengths have quantized fluxes, i.e.
        \begin{equation}
            \int_{S^2}\begin{pmatrix}
            F^I\\
            -G_I
            \end{pmatrix}\in\mbb{Z}^{2h^{2,1}+2}.
        \end{equation}
        \item \textbf{Asymptotic Weak Coupling:} 
        The third property is that the chosen basis should correspond to a weakly coupled frame for our gauge fields. This means that given a symplectic basis $(\alpha_I,\beta^J)$, we demand that charges which couple to the corresponding $F^I$ (i.e. the electric fields of the given frame) have vanishing physical charge in the limit $s^1\rightarrow \infty$. These correspond to states with charge vector ${\bf q}=(0,-q_I)^T$ so that we demand that the basis elements $\beta^I$ satisfy $\lVert \beta^I\rVert\lesssim\mc{O}(1)$. Recalling the definition of the D3-physical charge \eqref{HT_Qsl2} and the behavior of the Hodge norm \eqref{HT_norm_growth}, this can be arranged by choosing a basis so that the $\{\beta^I\}\subset W_{l<3}$.
    \end{enumerate}
    \item \label{Proc_2} Once a basis is fixed, one can derive the most general form of the log-monodromy matrix, making sure to comply with its defining conditions, namely:
    \begin{equation}\label{eqn:Nconstraints}
        NW_l\subseteq W_{l-2},\quad N^d=0,\quad N^T\eta+\eta N=0,\quad e^N\in Sp(4,\mbb{Z}).
    \end{equation}
    In addition to these, there exist so-called polarization conditions which further constrain the unfixed entries of the matrix $N$. For our purposes, it suffices to note that these appear as positivity conditions that ensure that the kinetic terms will exhibit the correct sign, and we refer to \cite{Bastian:2021eom} for further details. As a caveat, we remark that our method does not allow us to recover a possible semi-simple piece (cf. Equation~\ref{HT_Tus}) of the monodromy matrix $T$.
    \item \label{Proc_3} Finally, we write down the most general basis vectors for the limiting filtration $F^p_0$. These too should satisfy various compatibility conditions, such as 
    $N F_0^p \subset F^{p-1}_0$; these are listed, for instance, in \cite{Green:2008}.

    It is worth remarking that, although $N F^p_0 \subset F^{p-1}_0$, one cannot in general generate the full, lower filtrations via the action of the log-monodromy matrices only. 
    The instanton map $\Gamma(z)$ may in fact be needed to determine the lower filtrations only starting from the highest one, $F^3_0$.
    It is beyond the scope of this work to deliver a generic recipe to compute the instanton map, and we refer to \cite{Bastian:2021eom} for details.
    \item \label{Proc_4} Having assembled the most general nilpotent orbit data which can arise in a single-variable degeneration, we can apply procedure from the Appendices~\ref{sec:Hodge_review} and~\ref{sec:Hodge_gauge} to obtain approximate expressions for the period vector and gauge kinetic functions. The former follows directly from the nilpotent orbit theorem \eqref{HT_Per_nil}: one takes a representative ${\bf a}_0$ of $F_0^3$ constructed above and then recovers the periods from \eqref{HT_Per_nil}.
    The kinetic functions are constructed from the $\mathfrak{sl}(2)$-approximated Hodge star. Following the procedure of Section~\ref{sec:Hodge_gauge}, we start by translating the mixed Hodge structure $(F_0^p,W_l)$ to the more convenient Deligne splitting $I^{p,q}$. We then construct the rotation operators that take us to the $\mathfrak{sl}(2)$-splitting where we have control over the action of the Hodge star. By identifying the pure Hodge structure associated to the boundary, we can extract the limiting form of the Weil operator \eqref{HT_Cinf}. Finally, we recover the gauge kinetic functions from the matrix $\mathcal{T}$ in \eqref{HT_Msl2}.
    \item \label{Proc_5} With the approximate expressions of the periods and gauge kinetic functions in hands, one can write down the near-boundary effective action for the vector multiplet bosonic fields \eqref{N2_action}. By examining the interactions of the resulting EFT, one can follow the steps of Section~\ref{sec:GGS_IIBsub} and compute the currents and charge operators for the generalized global symmetries the EFT may possess, and study their conservation.
\end{enumerate}

If the near-boundary regime under examination is reached whenever multiple saxions acquire large vevs, as $s^1, \ldots, s^m \gg 1$, then the steps from~\ref{Proc_1} to~\ref{Proc_3} can be iteratively applied. Having then at hands a set of $m$ nilpotent matrices, the approximate periods can be computed using the nilpotent orbit theorem \eqref{HT_Per_nil}. Furthermore, one can then proceed to determining the $m$ $\mathfrak{sl}(2)$-triples, and from those the gauge kinetic interactions can be computed out of the general formula \eqref{HT_Msl2}, hence the EFT plugging the latter and the expression of the approximate periods in \eqref{N2_action}.


\section{Reconstructing the Hodge-star in the one-modulus cases}
\label{app:Periods_One_Mod}

In this appendix we will perform the procedure laid out in Section~\ref{sec:Hodge_outline} for each of the three one-modulus cases discussed in the main text. To further streamline the discussion, we group the five steps into two parts. The first consists of writing down the most general nilpotent orbit data in an appropriate integral basis for each type of limit, which covers the first three steps of the procedure. Here, we will use the results of \cite{Green:2008}. From this starting point, we then compute the approximate expressions for the kinetic matrix and period vector, covering steps four and five.

\subsection{The Large Complex Structure Point}\label{app:ivhodge}
We start by discussing the Type $\text{IV}_1$ case. In order to illustrate the method, we will be quite detailed for this first case. We emphasize however, that for each of the three cases, the construction of the nilpotent orbit data was done in \cite{Green:2008}, so that we are relatively brief about this part.

\noindent\textbf{Nilpotent Orbit Data}

\noindent Following the steps outlined in the main text, we start by fixing our integral symplectic basis. In \cite{Green:2008} it was shown that this may always be done such that the weight filtration is spanned as follows
\begin{equation}\label{eqn:weightIV}
    \text{Type } \text{IV}_1: \quad\begin{cases}W_6=\text{span}\begin{pmatrix}
            1 & 0 & 0 & 0
        \end{pmatrix}\bigoplus W_4,\\
        W_4=\text{span}\begin{pmatrix}
            0 & 1 & 0 & 0
        \end{pmatrix}\bigoplus W_2,\\
        W_2=\text{span}\begin{pmatrix}
            0 & 0 & 0 & 1
        \end{pmatrix}\bigoplus W_0,\\
        W_0=\text{span}\begin{pmatrix}
            0 & 0 & 1 & 0
        \end{pmatrix}.
        \end{cases}
\end{equation}
We emphasize that this is a non-trivial statement whose proof can be found in \cite{Green:2008}. In this basis one may then construct the most general log-monodromy matrix compatible with this filtration, as well as the constraints \eqref{eqn:Nconstraints}. The result is given by:
\begin{equation}\label{eqn:NmatIV}
    N=\begin{pmatrix}
    0 & 0 & 0 & 0\\
    m & 0 & 0 & 0\\
    c & b & 0 & -m\\
    b & n & 0 & 0
    \end{pmatrix},\quad \text{with }\begin{cases}
    m,n\in\mbb{Z},\\
    b+mn/2\in\mbb{Z},\\
    c-m^2n/6\in\mbb{Z},\\
    m\neq 0, \,n>0.
    \end{cases}
\end{equation}
We note that the property \eqref{eqn:weightIV} does not uniquely fix a basis. In particular, there remain integral symplectic transformations that preserve the weight filtration, whereby lower-weight pieces get rotated into the higher-weight pieces. These can be used to simplify some of the coefficients appearing in the log-monodromy matrix, but we do not do so here. 

The final piece of nilpotent orbit data is the limiting filtration $F_0^p$. It will be convenient to express this filtration in terms of a \textit{period matrix} whose columns successively span the limiting filtration
    \begin{equation}\label{eqn:periodmatdef}
        \Omega=\begin{pmatrix}
        F_0^3 & \subset & F_0^2 & \subset & F_0^1 & \subset & F_0^0
        \end{pmatrix}.
    \end{equation}
It can then shown that the most general limiting filtration $F_0^p$ for this case can be spanned by the following period matrix \cite{Green:2008}:
\begin{equation}\label{eqn:periodmatIV}
    \Omega=\begin{pmatrix}
    1 & 0 & 0 & 0\\
    0 & 1 & 0 & 0\\
    \xi & c/(2m) & 0 & 1\\
    c/(2m) & b/m & 1 & 0
    \end{pmatrix},\quad \text{with }\xi\in\mbb{C}.
\end{equation}

\noindent\textbf{$\mathfrak{sl}(2)$-orbit Data}

\noindent With the nilpotent orbit in hand, we can now begin to compute the associated Deligne splitting by evaluating Equation~\ref{HT_Ipq}. It turns out that for this particular case the $I^{p,q}$ simply correspond to the columns of the period matrix above, so that we do not give their explicit expression here. Instead, we proceed right away to constructing the rotation operator that renders this splitting $\mbb{R}$-split. Following the procedure from \cite{Bastian:2021eom}, we find the following operator:
\begin{equation}
    \delta=\begin{pmatrix}
    0 & 0 & 0 & 0\\
    0 & 0 & 0 & 0\\
    \I \xi & 0 & 0 & 0\\
    0 & 0 & 0 & 0
    \end{pmatrix},
\end{equation}
which we then use to rotate to the $\mbb{R}$-split Deligne splitting
\begin{equation}\label{eqn:rsplitlcs}
    \begin{pmatrix}
    \hat{I}^{3,3} & \hat{I}^{2,2} & \hat{I}^{1,1} & \hat{I}^{0,0}\\
    1 & 0 & 0 & 0\\
    0 & 1 & 0 & 0\\
    \R \xi & c/(2m) & 0 & 1\\
    c/(2m) & b/m & 1 & 0
    \end{pmatrix}.
\end{equation}
The second rotation matrix is trivial so that we immediately arrive at the final $\mathfrak{sl}(2)$-splitting. In the following we will need the weight operator of the $\mathfrak{sl}(2)$-triple associated to this splitting. This is readily obtained by solving
\begin{equation}\label{eqn:weightopdef}
    N^0\hat{I}^{p,q}=(p+q-3)\hat{I}^{p,q}.
\end{equation}
Note that because the $\hat{I}^{p,q}$ span the full $H_{\mbb{C}}$, this equation fixes $N^0$ unambiguously. The result is given by
\begin{equation}\label{eqn:weightopIV}
    N^0=\begin{pmatrix}
    3 & 0 & 0 & 0\\
    0 & 1 & 0 & 0\\
    6\R \xi & 2c/m & -3 & 0\\
    2c/m & 2b/m & 0 & -1
    \end{pmatrix}.
\end{equation}

\noindent\textbf{Boundary Data}

\noindent Next, we compute the limiting pure Hodge structure and associated Weil operator by applying the log-monodromy matrix. The former is given by a limiting Hodge filtration with associated period matrix given by
\begin{equation}
    F_{\infty}^p=e^{\im N}F^p_0\quad\Leftrightarrow\quad\Omega_{\infty}^p=e^{\im N}\Omega^p,
\end{equation}
which evaluates to
\begin{equation}
    \Omega_{\infty}=\begin{pmatrix}
    1 & 0 & 0 & 0\\
    \im m & 1 & 0 & 0\\
    \frac{1}{6}\im (m^2n+3c)+\R \xi & c/(2m)+mn/2 & b/m+\im n & 1\\
    c/(2m)-mn/2+\im b & b/m+\im n & 1 & 0
    \end{pmatrix}.
\end{equation}
From it, one evaluates
\begin{equation}
    H^{p,q}_{\infty}=F^p_{\infty}\cap \bar{F}^q_{\infty},
\end{equation}
which yields
\begin{equation}
    \begin{split}
        &H^{3,0}_{\infty}:\quad\begin{pmatrix}1, & \im m, & \hspace{4.5mm}\frac{1}{6}\im(m^2n+3c)+\R\xi, & c/(2m)-mn/2+\im b \end{pmatrix},\\
        &H^{2,1}_{\infty}:\quad\begin{pmatrix}1, & \frac{1}{3} \im m, & -\frac{1}{6}\im (m^2n-c)+\R\xi, & c/(2m)+mn/6+\im b/3 \end{pmatrix},
    \end{split}
\end{equation}
with the other two spaces being related to these by complex conjugation. One sees that the pure Hodge structure at the boundary is related to the mixed Hodge structure in a rather non-trivial way. One can now solve for the limiting Weil operator using
\begin{equation}\label{eqn:limweil}
    C_{\infty}v=\im^{p-q}v,\qquad v\in H^{p,q}_{\infty},
\end{equation}
which like \eqref{eqn:weightopdef} unambiguously defines $C_{\infty}$. The result is given by
\begin{equation}\label{eqn:limweilIV}
    C_{\infty}=\frac{1}{m^2n}\begin{pmatrix}
    6\R\xi & 3c/m & -6 & 0\\
    mc & 2mb & 0 & -2m^2\\
    (m^4n^2+3c^2+36(\R\xi)^2)/6 & c(b+3\R\xi/m) & -6\R\xi&-mc\\
    c(b+3\R\xi/m) & m^2n^2/2+2b^2+3c^2/(2m^2) & -3c/m & -2mb
    \end{pmatrix}.
\end{equation}

\noindent\textbf{Final Result}

\noindent Finally, we can put everything together to compute the $\mathfrak{sl}(2)$-approximated Hodge star. Recall that the $\mathfrak{sl}(2)$-approximated Weil operator is obtained by evaluating
\begin{equation}\label{eqn:csl2def2}
    C_{\mathfrak{sl}(2)}(a,s)=e^{ aN^{-}}e(s)^{-1}C_{\infty}e(s)e^{-aN^-},\qquad e(s)=\exp\left(\frac{1}{2}\log(s)N^0 \right)
\end{equation}
where $N^-$ and $N^0$ are given by \eqref{eqn:NmatIV} and \eqref{eqn:weightopIV}, respectively. By multiplying from the left with $\eta$, we recover the Hodge star matrix
\begin{equation}\label{eqn:hsmat2}
    \cT=-\eta C_{\mathfrak{sl}(2)}=-\begin{pmatrix}
    \langle \alpha_I,*\alpha_J \rangle& \langle \alpha_I,*\beta^J \rangle\\
    \langle \beta^I,*\alpha_J \rangle & \langle \beta^I,*\beta^J \rangle
    \end{pmatrix}.
\end{equation}
For this case in particular, the expression obtained from \eqref{eqn:hsmat2} is too large to present explicitly here, but we will do so for the following two cases. Instead, we note that this matrix determines the kinetic matrix $\cN$ via the identification
\begin{equation}
    \cT=-\eta C_{\mathfrak{sl}(2)}=\begin{pmatrix}
    \I \cN+\R\cN(\I \cN)^{-1}\R\cN & \R\cN (\I \cN)^{-1}\\
    (\I \cN)^{-1}\R\cN & (\I \cN)^{-1}
    \end{pmatrix}.
\end{equation}
Thus, by inverting the bottom-right block of the resulting matrix, we recover the imaginary part of $\cN$, while multiplying the former by the bottom-left block yields the real part of $\cN$. Rather than present the full matrix $\cT$, we give here the bottom two blocks 
\begin{equation}
\begin{split}
    &\langle \alpha_I,*\beta^J\rangle = -(\I \cN)^{-1}\R\cN=\\\vspace{4mm}
    &\frac{-1}{m^3ns^3}\begin{pmatrix}
    -mna^3+3(c/m)a+6\,\R\xi/m & 3a(2b/m+na)+3c/m^2\\
    c(3a^2+s^2)-m^2na^2(a^2+s^2)+6a\,\R\xi & 3a(c/m+a(2b+mna))+2(b+mna)s^2
    \end{pmatrix},\\ \\
    &\langle \beta^I,*\beta^J\rangle=-(\I \cN)^{-1}=\frac{1}{m^2ns^3}\begin{pmatrix}
    6 & 6ma\\
    6ma & 2m^2(3a^2+s^2)
    \end{pmatrix}.
\end{split}
\end{equation}
From these, we compute the real and imaginary parts of $\cN$ to obtain our final result for the kinetic matrix
\begin{equation}
    \cN=\begin{pmatrix}
    -\frac{1}{3}m^2na^3-\R\xi & \frac{1}{2}mna^2-\frac{c}{2m}\\
    \frac{1}{2}mna^2-\frac{c}{2m} & -na-\frac{b}{m}
    \end{pmatrix}+\im \begin{pmatrix}
        -\frac{1}{6}m^2ns(3a^2+s^2) & \frac{1}{2}mnas\\
        \frac{1}{2}mnas & -\frac{n}{2}s
    \end{pmatrix}.
\end{equation}
The corresponding period vector (in the nilpotent orbit approximation) given by
\begin{equation}\label{eqn:periodIV}
    \Pi_{\rm nil}=e^{tN}\bf{a}_0=\begin{pmatrix}
    1\\
    m t\\
    -m^2nt^3/6+ ct/2+\xi\\
    mnt^2/2+bt+c/(2m)
    \end{pmatrix},
\end{equation}
where $\bf{a}_0$ is a representative of $F_0^3$, for which we have used the first column of \eqref{eqn:periodmatIV}.

\subsection{Type \texorpdfstring{$\text{II}_0$}{II0}}
\label{app:iihodge}

Next, we consider the type $\text{II}_0$ case. We will not be quite as detailed as above but do present intermediate results.

\noindent\textbf{Nilpotent Orbit Data}

\noindent For this case, it can be shown that one can always find an integral symplectic basis such that the weight filtration (here consisting only of $W_2\subset W_4$) is spanned by \cite{Green:2008}:
\begin{equation}\label{eqn:weightII}
    \text{Type } \text{II}_0: \quad\begin{cases}W_4=\text{span}\left\{\begin{pmatrix}
            1 & 0 & 0 & 0
        \end{pmatrix},\begin{pmatrix}
            0 & 1 & 0 & 0
        \end{pmatrix}\right\}\bigoplus W_2 \\
        W_2=\text{span}\left\{\begin{pmatrix}
            0 & 0 & 1 & 0
        \end{pmatrix},\begin{pmatrix}
            0 & 0 & 0 & 1
        \end{pmatrix}\right\}
        \end{cases}.
\end{equation}
Again, we have that Equation~\ref{eqn:weightII} does not uniquely specify the symplectic basis. In particular, integral symplectic transformations of the form
\begin{equation}\label{eqn:symppresweightII}
    \mc{S}=\begin{pmatrix}
     A^{-1} & 0\\
     C & A^T
    \end{pmatrix},
\end{equation}
with $AC$ symmetric preserve both the symplectic pairing $\eta$ as well as the form of the weight filtration \eqref{eqn:weightII}. We will use this freedom to simplify some of the unfixed coefficients appearing below. The most general log-monodromy matrix compatible with the weight filtration above is given by
\begin{equation}
    N=\begin{pmatrix}
    0 & 0 & 0 & 0\\
    0 & 0 & 0 & 0\\
    m & k & 0 & 0\\
    k & n & 0 & 0\\
    \end{pmatrix},\quad \text{with }\begin{cases}
    m,n,k\in\mbb{Z},\\
    m,n>0.
    \end{cases}
\end{equation}
Following \cite{Green:2008}, we will assume that $n/k$ is integer, in which case we may use a symplectic transformation of the kind \eqref{eqn:symppresweightII} to set $k=0$ and assume $m\geq n>0$. We emphasize, however, that this is not quite the most general choice, although we clearly cover a large subset of this class of limits. In \cite{Green:2008} it was then shown that the most general period matrix compatible with \eqref{eqn:weightII} and this choice of log-monodromy, is given by
\begin{equation}\label{eqn:periodmatII}
    \Omega=\begin{pmatrix}
    1 & 0 & 0 & 0\\
    \im \alpha & 0 & 1 & 0\\
    \beta & 1 & 0 & 0\\
    \im \alpha\delta & \im /\alpha & \delta & 1
    \end{pmatrix},\quad \text{with }\begin{cases}
    \alpha\coloneqq\sqrt{m/n}\in\mbb{R},\\
    \beta,\delta\in\mbb{C}.
    \end{cases}
\end{equation}
Under rescalings $z\rightarrow e^{-2\pi \im  \lambda}z$ we have that 
\begin{equation}
    \beta\rightarrow \beta+m\lambda,\quad \delta\rightarrow \delta+n\lambda,
\end{equation}
which we use to set $\beta=0$ in the following. For ease of notation, we decompose the remaining parameter $\delta$ into real and imaginary parts as follows
\begin{equation}
    \delta=c-\im d/\alpha.
\end{equation}

\noindent\textbf{$\mathfrak{sl}(2)$-Orbit Data}

\noindent Using Equation~\ref{HT_Ipq} we can compute the associated Deligne splitting, which is now non-trivially related to the period matrix above
\begin{equation}
    \begin{pmatrix}
    I^{3,1} & I^{1,3} & I^{2,0} & I^{0,2}\\
    1 & 1 & 0 & 0\\
    \im \alpha & -\im \alpha & 0 & 0\\
    0 & -\im \alpha d & 1 & 1\\
    \im \alpha\delta & -\im \alpha c & \im \alpha & -\im \alpha
    \end{pmatrix}.
\end{equation}
Following the procedure from \cite{Bastian:2021eom} we construct the relevant rotation operators to obtain the $\mbb{R}$-split Deligne splitting. The first of these is given by
\begin{equation}
    \delta=-\frac{d}{2}\begin{pmatrix}
    0 & 0 & 0 & 0\\
    0 & 0 & 0 & 0\\
    \alpha & 0 & 0 & 0\\
    0 & \alpha^{-1} & 0 & 0
    \end{pmatrix},
\end{equation}
while the rotation matrix $\zeta$ is again trivial. The resulting $\mbb{R}$-split Deligne splitting is given by
\begin{equation}
    \begin{pmatrix}
    \hat{I}^{3,1} & \hat{I}^{1,3} & \hat{I}^{2,0} & \hat{I}^{0,2}\\
    1 & 1 & 0 & 0\\
    \im \alpha & -\im \alpha & 0 & 0\\
    \frac{1}{2}\alpha d & -\frac{1}{2} \im \alpha d & 1 & 1\\
    \frac{d}{2}+\im \alpha c& \frac{d}{2}-\im \alpha c & \im \alpha & -\im \alpha
    \end{pmatrix}.
\end{equation}
Moreover, the associated weight operator is given by
\begin{equation}
    N^0=\begin{pmatrix}
    1 & 0 & 0 & 0\\
    0 & 1 & 0 & 0\\
    0 & d & -1 & 0\\
    d & 2c & 0 & -1
    \end{pmatrix}.
\end{equation}

\noindent\textbf{Boundary Data}

\noindent Rotating the associated filtration according to $F_{\infty}^p=e^{\im N}\hat{F}_0^p$, we find the following period matrix
\begin{equation}
    \Omega_{\infty}=\begin{pmatrix}
    1 & 0 & 0 & 0\\
    \im \alpha & 0 & 1 & 0\\
    \im (m+d\alpha/2) & 1 & 0 & 0\\
    -n\alpha+d/2+\im c\alpha & \im\alpha & c-\im d/(2\alpha)+\im n & 1
    \end{pmatrix}.
\end{equation}
The Hodge decomposition associated to this filtration is given by
\begin{equation}
    \begin{split}
        &H^{3,0}_{\infty}:\quad\begin{pmatrix}1, & \im \alpha, & \im (m+d\alpha/2), & -n\alpha+d/2+\im c\alpha \end{pmatrix},\\
        &H^{2,1}_{\infty}:\quad\begin{pmatrix}1, & \im \alpha, & \im (-m+d\alpha/2), & n\alpha+d/2+\im c\alpha \end{pmatrix},
    \end{split}
\end{equation}
with the remaining two spaces following by complex conjugation. As before, we can solve for the limiting Weil operator to find
\begin{equation}
    C_{\infty}=\begin{pmatrix}
    0 & d/(2m) & -1/m & 0\\
    d/(2n) & c/n & 0 & -1/n\\
    m+d^2/(4n) & cd/(2n)&0&-d/(2n)\\
    cd/(2n) & n+c^2/n+d^2/(4m)&-d/(2m)&-c/n
    \end{pmatrix},
\end{equation}
which one checks to have the correct action on the $H_{\infty}^{p,q}$ given above.

\noindent\textbf{Final Result}

\noindent With these results, we can again construct the $\mathfrak{sl}(2)$-approximated Weil operator, now given by
\begin{equation}
    \eta C_{\mathfrak{sl}(2)}=\frac{1}{s}\begin{pmatrix}
    d^2/(4n)+m(a^2+s^2) & ad+cd/(2n) & -a & -d/(2n)\\
    ad+cd/(2n) & d^2/(4m)+(an+c)^2/n+ns^2 & -d/(2m) & -a-c/n\\
    -a & -d/(2m) & 1/m & 0\\
    -d/(2n) & -a-c/n & 0 & 1/n
    \end{pmatrix}.
\end{equation}
From the bottom two blocks, we can then obtain the final result for the kinetic matrix, given by
\begin{equation}
    \cN=-\begin{pmatrix}
    ma & \frac{d}{2}\\
    \frac{d}{2} & na-c
    \end{pmatrix}-\im \begin{pmatrix}
    ms & 0\\
    0 & ns
    \end{pmatrix}.
\end{equation}
Moreover, the nilpotent orbit approximation to the period vector is given by
\begin{equation}
    \Pi_{\rm nil}=e^{tN}\bf{a}_0=\begin{pmatrix}
    1\\
    \im \alpha\\
    mt\\
    \im \alpha(\delta+n t)
    \end{pmatrix},
\end{equation}
where, for $\bf{a}_0$, we again use the first column in equation \ref{eqn:periodmatII} (recall that we set $\beta=0$). We remark that this period vector is not in the special coordinates of Section~\ref{sec:IIB_Sympl} (although this can always be achieved by a symplectic rotation).

\subsection{Type \texorpdfstring{$\text{I}_1$}{I1}}
\label{app:ihodge}
This final case is slightly more involved because it lies at finite distance. In particular, what this means is that the period vector in the nilpotent orbit approximation will not span the full Hodge filtration, which practically implies that to leading approximation the Kahler metric vanishes. We will address this issue shortly but for now we begin as before by writing down the most general nilpotent orbit data for this case.

\noindent\textbf{Nilpotent Orbit Data}

\noindent For this case, it is always possible to pick an integral symplectic basis such that the weight filtration $W_2\subset W_3\subset W_4$ is spanned by \cite{Green:2008}
\begin{equation}
    \text{Type } \text{I}_1: \quad\begin{cases}W_4=\text{span}\left\{\begin{pmatrix}
            1 & 0 & 0 & 0
        \end{pmatrix}\right\}\bigoplus W_3 \\
        W_3=\text{span}\left\{\begin{pmatrix}
            0 & 1 & 0 & 0
        \end{pmatrix},\begin{pmatrix}
            0 & 0 & 0 & 1
        \end{pmatrix}\right\}\bigoplus W_2\\
        W_2=\text{span}\left\{\begin{pmatrix}
            0 & 0 & 1 & 0
        \end{pmatrix}\right\}.
        \end{cases}
\end{equation}
Similarly to the previous cases compatibility with the weight filtration does not uniquely fix our basis, as the former is invariant under integer symplectic transformations of the form
\begin{equation}
    \begin{pmatrix}
        1 & 0 & 0 & 0\\
        b_1 & 1 & 0 & 0\\
        0 & b_2 & 1 & -b_1\\
        b_2 & 0 & 0 & 1
    \end{pmatrix},\quad \begin{pmatrix}
        b & 0 & 0 & 0\\
        0 & c_{11} & 0 & c_{12} \\
        0 & 0 & b^{-1}& 0\\
        0 & c_{21} & 0 & c_{22}
    \end{pmatrix},\quad\begin{cases}
    0<b\in\mbb{Z},\\
    \{c_{ij}\}\in SL(2,\mbb{Z}).
    \end{cases}
\end{equation}
We will not make use of this freedom, however. The most general log-monodromy matrix associated to this weight filtration is given by 
\begin{equation}
    N=\begin{pmatrix}
    0 & 0 & 0& 0\\
    0 & 0 & 0& 0\\
    n & 0 & 0& 0\\
    0 & 0 & 0& 0
    \end{pmatrix},\quad 0<n\in\mbb{Z}.
\end{equation}
Meanwhile, we can choose our period matrix to be of the form
\begin{equation}
    \Omega=\begin{pmatrix}
    0 & 1 & 0 & 0\\
    1 & \gamma & 0 & 0\\
    \beta-\tau\gamma & 0 & 1 & 0\\
    \tau & \beta & 0 & 1
    \end{pmatrix},\quad \begin{cases}
    \gamma,\beta\in\mbb{R},\\
    \I \tau\neq 0.
    \end{cases}
\end{equation}

\noindent\textbf{$\mathfrak{sl}(2)$-Orbit Data}

\noindent The Deligne-splitting obtained from this data is given by
\begin{equation}\label{eqn:delignesplit}
    \begin{pmatrix}
    I^{2,2} & I^{3,0} & I^{0,3} & I^{1,1}\\
    1 & 0 & 0 & 0\\
    \gamma & 1 & 1 & 0\\
    0 & \beta-\gamma\tau & \beta-\gamma\bar{\tau} & 1 \\
    \beta & \tau & \bar{\tau} & 0
    \end{pmatrix}.
\end{equation}
It follows that the resulting Deligne-splitting is already $\mbb{R}$-split and both rotation operators are trivial. The weight operator associated to this Deligne splitting is then given by
\begin{equation}
    N^0=\begin{pmatrix}
    1 & 0 & 0 & 0\\
    \gamma & 0 & 0 & 0\\
    0 & \beta & -1 & -\gamma\\
    \beta & 0 & 0 & 0
    \end{pmatrix}.
\end{equation}

\noindent\textbf{Boundary Data}

\noindent Evaluating $F_{\infty}^p=e^{\im N}F_0^p$ we find the following period matrix
\begin{equation}
    \Omega_{\infty}=\begin{pmatrix}
    0 & 1 & 0 & 0\\
    1 & \gamma & 0 & 0\\
    \beta - \gamma\tau & \im n & 1 & 0\\
    \tau & \beta & 0 & 1
    \end{pmatrix},
\end{equation}
with associated Hodge decomposition given by
\begin{equation}
    \begin{split}
        &H^{3,0}_{\infty}:\quad\begin{pmatrix}0, & 1, & \beta-\gamma\tau, & \tau \end{pmatrix},\\
        &H^{2,1}_{\infty}:\quad\begin{pmatrix}1, & \gamma, & \im n, & \beta \end{pmatrix}.
    \end{split}
\end{equation}
Next, we introduce the following short-hands
\begin{equation}
    \tau=c+\im d,
\end{equation}
keeping in mind that $\I \tau=d\neq 0$. We can then solve for the limiting Weil operator acting on the $H_{\infty}^{p,q}$, which now reads
\begin{equation}
    C_{\infty}=\frac{1}{nd}\begin{pmatrix}
    0 & -\beta d & d & \gamma d\\
    n(\beta-\gamma c) & nc-\beta \gamma d & \gamma d & -n +\gamma^2 d\\
    -n^2 d+n|\beta-\gamma\tau|^2 & n(\beta c-\gamma |\tau|^2) & 0 & -n(\beta-\gamma c)\\
    n(\beta c-\gamma|\tau|^2) & -\beta^2d+n|\tau|^2 & \beta d & -nc+\beta\gamma d
    \end{pmatrix}.
\end{equation}

\noindent\textbf{Final Result}

\noindent Finally, we compute the $\mathfrak{sl}(2)$-approximated Hodge star matrix, which is given by
\begin{equation}
    \eta C_{\mathfrak{sl}(2)}=\frac{1}{s}\begin{pmatrix}
    \frac{s|\beta-\gamma \tau|^2}{d} -n(a^2+s^2)& \frac{\beta (sc-ad)-\gamma s(c^2+d^2)}{d} & a & \frac{a\gamma d+s\gamma c-\beta s}{d}\\
    \frac{\beta(sc-ad)-\gamma s(c^2+d^2)}{d} & -\frac{\beta^2}{n}+\frac{s(c^2+d^2)}{d} & \frac{\beta}{n} & \frac{\beta \gamma }{n}-\frac{sc}{d}\\
    a & \frac{\beta}{n} & -\frac{1}{n} & -\frac{\gamma}{n}\\
    \frac{-\beta s+\gamma (sc +a d)}{d} & \frac{\beta \gamma }{n}-\frac{sc}{d} & -\frac{\gamma}{n} & -\frac{\gamma^2}{n}+\frac{s}{d}
    \end{pmatrix}.
\end{equation}
Extracting the relevant blocks, we then find that the gauge kinetic matrix is given by
\begin{equation}
    \cN=-\begin{pmatrix}
    na-\gamma(\beta-\gamma c) & \beta-\gamma c\\
    \beta-\gamma c & c
    \end{pmatrix}+\im \begin{pmatrix}
    ns-\gamma^2d & \gamma d\\
    \gamma d & - d
    \end{pmatrix},
\end{equation}
where we can perform a shift $a\rightarrow a+\gamma(\beta-\gamma c)/n$ to simplify the real part (this could also be done at the level of the mixed Hodge structure, but its purpose is more clear here). The associated nilpotent orbit is given by
\begin{equation}
    \Pi_{\rm nil}=\begin{pmatrix}
        0\\
        1\\
        \beta-\gamma\tau\\
        \tau
    \end{pmatrix}.
\end{equation}
It is clear that the vanishing of $N\bf{a}_0$ implies that the associated nilpotent orbit is moduli-independent. As such it leads to a constant Kahler potential and a vanishing Kahler metric. Indeed, this was exactly what we used to argue that these limits are located at finite distance. While this means that we cannot naively use ${\bf{a}_0}\in F_0^3$ to recover the full Hodge filtration, we have access to the full limiting filtration $F_0^p$. One expects that this should contain sufficient information to reconstruct an approximate period vector that does span the full filtration. Indeed, this was the strategy used in \cite{Bastian:2021eom}, where limiting expressions for the period vector were recovered that include the first sub-leading corrections. Although the authors there do not work in an integral basis, we can rotate their expressions to our basis to recover the necessary exponential corrections to the period vector (see also the discussion in the main text)
\begin{equation}
    {\bm \Pi}  \simeq \begin{pmatrix}
    \alpha e^{2\pi \im\, t}\\
    1 + \alpha \gamma e^{2\pi \im\, t}  \\ 
    \beta -\gamma \tau  + \alpha n e^{2\pi \im\, t} t \\
    \tau + \alpha\beta e^{2\pi \im\, t}
    \end{pmatrix}.
\end{equation}


\providecommand{\href}[2]{#2}\begingroup\raggedright\endgroup

\end{document}